%% file: Theory_of_Diffraction_Tomography.tex
\documentclass[fontsize=11pt, paper=letter, pagesize, oneside]{scrartcl}

\usepackage{amsmath}
\usepackage{amssymb}

\usepackage[utf8]{inputenc} 

\usepackage{graphicx} 
\usepackage{wrapfig}

\usepackage{array} 
\usepackage{paralist} 
\usepackage{verbatim} 
\usepackage{subfig} 
\DeclareCaptionLabelFormat{mycaption}{#1 #2}
\DeclareCaptionLabelSeparator{comma}{, }
\setcapindent{0pt} 

\newcommand{\mycaption}[2]{\caption[~~#1]{\textbf{#1.} #2}}
\usepackage{sistyle}
\usepackage{calc}

\usepackage{makeidx}
\makeindex

\usepackage{nicefrac}

\usepackage{float}

\usepackage{pdflscape}
     
\usepackage[percent]{overpic}

\usepackage{placeins}

\usepackage{titletoc}

\usepackage{titlesec}

\usepackage{chngcntr}

\usepackage{csquotes}
\usepackage[
    backend=bibtex, 
    style=numeric-comp,
    natbib=true,
	sorting=none,
    url=false, 
    doi=true,
    eprint=false
]{biblatex}
\usepackage[svgnames]{xcolor} 

\usepackage[includeheadfoot, headsep=.1in, margin=1in]{geometry} 
\usepackage{rerunfilecheck}

\numberwithin{equation}{section}
\numberwithin{figure}{section}
\numberwithin{table}{section}

\usepackage{fancyhdr}
\fancypagestyle{myfrontbackstyle}{%
	\rhead{}
	\lhead{}
    \lfoot{\small The Theory of Diffraction Tomography}
	\rfoot{\small \thepage}
	\cfoot{}
}

\usepackage{sectsty}
\allsectionsfont{\sffamily\mdseries\upshape} 

\usepackage[nonumberlist, nomain]{glossaries}
\newglossary{symbols}{sym}{sbl}{List of Symbols}
\usepackage{glossary-superragged} 
\setglossarystyle{superragged} 
\loadglsentries[symbols]{glossary_symbols}
\makeglossaries

\newcommand{\myincludegraphics}[2]{
  \begin{overpic}[]{#1.#2}
     \put(0,0){\includegraphics[]{#1_text.pdf}}
  \end{overpic}
}

\newcommand{\vc}[1]{\ensuremath{\mathbf{#1}}} 

\titleformat{\section}{\large\bfseries}{\thesection}{0.5em}{}
\titleformat{\subsection}{\bfseries}{\thesubsection}{0.5em}{}

\titlecontents{table}
[0pt]                                  
{\addvspace{.5cm}}
{\contentsmargin{0pt}                  
    Table~\thecontentslabel%
    \large}
{\contentsmargin{0pt}\large}           
{\titlerule*[.5pc]{.}\contentspage}    

\titlecontents{figure}
[0pt]                                  
{\addvspace{.5cm}}
{\contentsmargin{0pt}                  
    Figure~\thecontentslabel%
    \large}
{\contentsmargin{0pt}\large}           
{\titlerule*[.5pc]{.}\contentspage}    

\title{The Theory of Optical Diffraction Tomography}
\author{Paul Müller, Mirjam Schürmann, Jochen Guck}

\usepackage[hidelinks]{hyperref}
  \hypersetup{%
  	pdftitle={The Theory of Diffraction Tomography},
  	pdfauthor={Paul M{\"u}ller, Mirjam Sch{\"u}rmann, Jochen Guck},
  }

\newcommand{\hyref}[2]{\hyperref[#2]{#1~\ref{#2}}}

\begin{document}
\thispagestyle{empty}
\begin{center}
  \begin{Large}
    \textbf{The Theory of Diffraction Tomography}\\
  \end{Large}
\end{center}
\begin{center}
Paul Müller\footnote{to whom correspondence should be addressed}, Mirjam Schürmann, and Jochen Guck \\
\textit{Biotechnology Center, Technische Universität Dresden, Dresden, Germany} \\
(Dated: \today)
\end{center}

\input{abstract.tex}

\newcommand{\sectionbreak}{\FloatBarrier}

\newpage
\thispagestyle{myfrontbackstyle}
\tableofcontents

\newpage
\addcontentsline{toc}{section}{Introduction}

\counterwithout{figure}{section}
\input{chapter0_introduction.tex}
\counterwithin{figure}{section}
\FloatBarrier
\input{chapter1_tomography_without_diffraction.tex}

\FloatBarrier
\input{chapter2_the_wave_equation.tex}
\FloatBarrier
\input{chapter3_approximations_to_the_scattered_wave.tex}
\FloatBarrier
\input{chapter4_two-dimensional_diffraction_tomography.tex}

\FloatBarrier
\input{chapter5_three-dimensional_diffraction_tomography}

\FloatBarrier
\input{chapter6_implementation.tex}
\FloatBarrier
\addcontentsline{toc}{section}{Conclusions}
\input{conclusion.tex}

\clearpage

\appendix
\input{appendix_notation_in_literature.tex}

\clearpage
\input{appendix_changes.tex}

\clearpage
\pagestyle{myfrontbackstyle}

\addcontentsline{toc}{section}{Bibliography}
\renewcommand{\refname}{Bibliography}
\printbibliography

\clearpage

\section*{List of Figures and Tables}
\addcontentsline{toc}{section}{Lists of Figures, Tables, and Symbols}
\begingroup
\makeatletter
\@starttoc{lof}
\vspace{2em}
\let\clearpage\relax
\@starttoc{lot}
\makeatother
\endgroup

\clearpage

\printglossary[type=symbols]

\clearpage

\section*{Acknowledgements}
We are grateful to Moritz Kreysing, Martin Weigert (Max Planck Institute of Molecular Cell Biology and Genetics, Dresden, Germany), and Kevin Chalut (Cavendish Laboratory, University of Cambridge, Cambridge, UK) for many valuable discussions.

\noindent This project has received funding from the European Union’s Seventh Framework Programme for research, technological development and demonstration under grant agreement no 282060.

\end{document}

%% file: glossary_symbols.tex
\newglossaryentry{a}{
  name		  = {\ensuremath{a}} ,
  sort        = a ,
  description = {amplitude of the scattered wave \gls{u}}
}

\newglossaryentry{c0}{
  name		  = {\ensuremath{c_\mathrm{0}}} ,
  sort        = c0 ,
  description = {speed of light in vacuum}
}

\newglossaryentry{f}{
  name		  = {\ensuremath{f(\mathbf{r})}} ,
  sort        = f ,
  description = {three-dimensional distribution to be reconstructed from two-dimensional projections by inverse methods; in optical tomography \gls{f} is the  inhomogeneity of the wave equation and contains the refractive index distribution $\gls{n}$, $\gls{f} = \gls{km}^2\left[ \left(\frac{\gls{n}}{\gls{nm}}\right)^2 -1 \right]$}
}

\newglossaryentry{F}{
  name		  = {\ensuremath{\widehat{F}(\mathbf{k})}} ,
  sort        = f_fourier ,
  description = {Fourier transform of \gls{f}}
}

\newglossaryentry{G}{
  name		  = {\ensuremath{G(\mathbf{r}-\mathbf{r'})}} ,
  sort        = G ,
  description = {Green's function of the Helmholtz equation}
}

\newglossaryentry{Gf}{
  name		  = {\ensuremath{\widehat{G}(\mathbf{k})}} ,
  sort        = Gf ,
  description = {Fourier transform of the Green's function $G(\mathbf{r})$}
}

\newglossaryentry{k}{
  name		  = {\ensuremath{k(\mathbf{r})}} ,
  sort        = k ,
  description = {spatial distribution of the wave number, $ \gls{k} = \gls{km} \frac{\gls{n}}{\gls{nm}}$}
}

\newglossaryentry{km}{
  name		  = {\ensuremath{k_\mathrm{m}}} ,
  sort        = km ,
  description = {wave number in the medium surrounding a sample, $\gls{km} = \frac{2 \pi \gls{nm}}{\gls{lambda}}$}
}

\newglossaryentry{lambda}{
  name		  = {\ensuremath{\lambda}} ,
  sort        = lambda ,
  description = {vacuum wavelength of the laser light that is used for image acquisition}
}

\newglossaryentry{lD}{
  name		  = {\ensuremath{l_\mathrm{D}}} ,
  sort        = lD ,
  description = {distance between rotational center and detector plane}
}

\newglossaryentry{n}{
  name		  = {\ensuremath{n(\mathbf{r})}} ,
  sort        = n ,
  description = {refractive index distribution inside a sample, $\gls{n} = \gls{nm} \left(1 + \gls{ndelt}\right)$}
}

\newglossaryentry{nm}{
  name		  = {\ensuremath{n_\mathrm{m}}} ,
  sort        = nm ,
  description = {refractive index of the medium surrounding a sample}
}

\newglossaryentry{ndelt}{
  name		  = {\ensuremath{\epsilon_{n}(\mathbf{r})}} ,
  sort        = epsilon ,
  description = {local variation of the refractive index, $\gls{ndelt} = \gls{n} - \gls{nm}$}
}

\newglossaryentry{pc}{
  name		  = {\ensuremath{\varphi(\mathbf{r})}} ,
  sort        = phicomplex ,
  description = {complex phase, $\gls{pc} = i \gls{phase}(\mathrm{r}) + \ln(\gls{a})$, $\gls{u} =  \exp(\gls{pc})$, $\gls{pc} = \gls{pc0} + \gls{pcs}$}
}

\newglossaryentry{pr}{
  name		  = {\ensuremath{p_{\phi_0}(\mathbf{r_D})}} ,
  sort        = pr,
  description = {projection of a sample onto the detector plane}
}

\newglossaryentry{phi0}{
  name		  = {\ensuremath{\phi_0}} ,
  sort        = phi ,
  description = {angle along which a projection of a sample was recorded; in optical tomography the projections are digital holograms}
}

\newglossaryentry{phase}{
  name		  = {\ensuremath{\Phi}} ,
  sort        = phip ,
  description = {phase of an optical wave, $\gls{phase} \in [0,2\pi)$}
}

\newglossaryentry{pc0}{
  name		  = {\ensuremath{\varphi_\mathrm{0}(\mathbf{r})}} ,
  sort        = phicomplex0 ,
  description = {complex phase of a plane wave, $\gls{u0} = \exp(\gls{pc0})$}
}

\newglossaryentry{pcs}{
  name		  = {\ensuremath{\varphi_\mathrm{s}(\mathbf{r})}} ,
  sort        = phicomplexs ,
  description = {scattering component of a complex phase \gls{pc}}
}

\newglossaryentry{pcr}{
  name		  = {\ensuremath{\varphi_\mathrm{R}(\mathbf{r})}} ,
  sort        = phicomplexsr ,
  description = {Rytov approximation to \gls{pcs}}
}

\newglossaryentry{pcrf}{
  name		  = {\ensuremath{\widehat{\varphi}_\mathrm{R}(\mathbf{k})}} ,
  sort        = phicomplexsrf ,
  description = {Fourier transform of \gls{pcr}}
}

\newglossaryentry{psi}{
  name		  = {\ensuremath{\psi}} ,
  sort        = psi ,
  description = {azimuthal angle in the coordinate system of a semi-sphere, $\gls{psi} \in \left[-\pi/2,+\pi/2\right]$}
}

\newglossaryentry{Psi}{
  name		  = {\ensuremath{\Psi(\mathbf{r}, t)}} ,
  sort        = Psi ,
  description = {time-dependent propagator of a scalar wave}
}

\newglossaryentry{R}{
  name		  = {\ensuremath{R_{\phi_0}}} ,
  sort        = R ,
  description = {Radon transform along angle \gls{phi0}}
}

\newglossaryentry{s}{
  name		  = {\ensuremath{\mathbf{s}}} ,
  sort        = s ,
  description = {normal unit vector of an arbitrary plane wave, direction of propagation (3D), $\gls{s} = (p, q, M)$, $p^2 + q^2 + M^2 = 1$ }
}

\newglossaryentry{s0}{
  name		  = {\ensuremath{\mathbf{s}_\mathrm{0}}} ,
  sort        = s0 ,
  description = {normal unit vector of an incident plane wave, direction of propagation (3D), $\gls{s0} = (p_\mathrm{0}, q_\mathrm{0}, M_\mathrm{0})$, $p_\mathrm{0}^2 + q_\mathrm{0}^2 + M_\mathrm{0}^2 = 1$ }
}

\newglossaryentry{theta}{
  name		  = {\ensuremath{\theta}} ,
  sort        = theta ,
  description = {polar angle in the coordinate system of a semi-sphere, $\gls{theta} \in \left[0,\pi\right]$}
}

\newglossaryentry{u}{
  name		  = {\ensuremath{u(\mathbf{r})}} ,
  sort        = u ,
  description = {scattered wave, \gls{u} = \gls{u0}+\gls{us}}
}

\newglossaryentry{u0}{
  name		  = {\ensuremath{u_\mathrm{0}(\mathbf{r})}} ,
  sort        = u0 ,
  description = {incident plane wave, solution to the homogeneous Helmholtz equation}
}

\newglossaryentry{ub}{
  name		  = {\ensuremath{u_\mathrm{B}(\mathbf{r})}} ,
  sort        = usb ,
  description = {Born approximation of \gls{us}}
}

\newglossaryentry{ur}{
  name		  = {\ensuremath{u_\mathrm{R}(\mathbf{r})}} ,
  sort        = usb ,
  description = {Rytov approximation of \gls{us}}
}

\newglossaryentry{us}{
  name		  = {\ensuremath{u_\mathrm{s}(\mathbf{r})}} ,
  sort        = us ,
  description = {scattering component of a scattered wave \gls{u}}
}

\newglossaryentry{U}{
  name		  = {\ensuremath{\widehat{U}(\mathbf{k})}} ,
  sort        = uzfourier ,
  description = {Fourier transform of \gls{u}}
}

\newglossaryentry{Ub}{
  name		  = {\ensuremath{\widehat{U}_\mathrm{B}(\mathbf{k})}} ,
  sort        = uzfourierb ,
  description = {Fourier transform of \gls{ub}}
}

\newglossaryentry{Ubphi0}{
  name		  = {\ensuremath{\widehat{U}_{\mathrm{B}, \phi_0}(\mathbf{k_D})}} ,
  sort        = uzfourierbphi ,
  description = {Fourier transform of $u_\mathrm{B}(\mathbf{r_D})$ at the detector $\mathbf{r_D}$, measured at the angle \gls{phi0}. This notation requires that the direction of the incoming plane wave and the detector normal are parallel.}
}

%% file: abstract.tex
\section*{Abstract}

Tomography is the three-dimensional reconstruction of an object from images taken at different angles. The term classical tomography is used, when the imaging beam travels in straight lines through the object. This assumption is valid for light with short wavelengths, for example in x-ray tomography. For classical tomography, a commonly used reconstruction method is the filtered back-projection algorithm which yields fast and stable object reconstructions. In the context of single-cell imaging, the back-projection algorithm has been used to investigate the cell structure or to quantify the refractive index distribution within single cells using light from the visible spectrum. Nevertheless, these approaches, commonly summarized as optical projection tomography, do not take into account diffraction.
Diffraction tomography with the Rytov approximation resolves this issue. The explicit incorporation of the wave nature of light results in an enhanced reconstruction of the object's refractive index distribution.
Here, we present a full literature review of diffraction tomography. We derive the theory starting from the wave equation and discuss its validity with the focus on applications for refractive index tomography. Furthermore, we derive the back-propagation algorithm, the diffraction-tomographic pendant to the back-projection algorithm, and describe its implementation in three dimensions. Finally, we showcase the application of the back-propagation algorithm to computer-generated scattering data.
This review unifies the different notations in literature and gives a detailed description of the back-propagation algorithm, serving as a reliable basis for future work in the field of diffraction tomography.

%% file: chapter0_introduction.tex
\section*{Introduction\markboth{Introduction}{}}
Computerized tomography (CT) is a common tool to image three-dimensional~(3D) objects like e.g. bone tissue in the human body. The 3D object is reconstructed from transmission images at different angles, i.e. from projections onto a two-dimensional~(2D) detector plane. The assumption that the recorded data are in fact projections is only valid for non-scattering objects, i.e. when the imaging beam travels in straight lines through the sample. A common technique to reconstruct the 3D object from 2D projections is the filtered backprojection algorithm. The term classical tomography is commonly used to identify reconstruction techniques that are based on this straight-line approach, neglecting diffraction.

When the wavelength of the imaging beam is large, i.e. comparable to the size of the imaged object, diffraction can not be neglected anymore and the assumptions of classical tomography are invalidated.
The property that is responsible for diffraction at an object is its complex-valued refractive index distribution. The real part of the refractive index accounts for refraction of light and the imaginary part causes attenuation. Features in the refractive index distribution that are comparable in size to the wavelength of the light cause diffraction, i.e. the wavelike properties of light emerge. For example, organic tissue does not affect the directional propagation of x-ray radiation. Therefore, approximating the propagation of light in straight lines for CT is valid. However, small refractive index changes within single cells lead to diffraction of visible light, requiring a more fundamental model of light propagation.
Diffraction tomography takes wave propagation into account (hence the name backpropagation algorithm) and can thus be applied to tomographic data sets that were imaged with large wavelengths. 
Therefore, diffraction tomography is ideally suited  to resolve the refractive index of sub-cellular structures in single cells using light from the visible spectrum.

Because the refractive index distribution of single cells is mostly real-valued, they do not absorb significant amounts of light. Therefore, the main alteration of  the wave front is due to refraction which is revealed by a measurable phase change. Therefore, diffraction tomography requires imaging techniques that quantify these sample-induced phase changes, such as digital holographic microscopy (DHM).
\begin{figure}[hb]
  \myincludegraphics{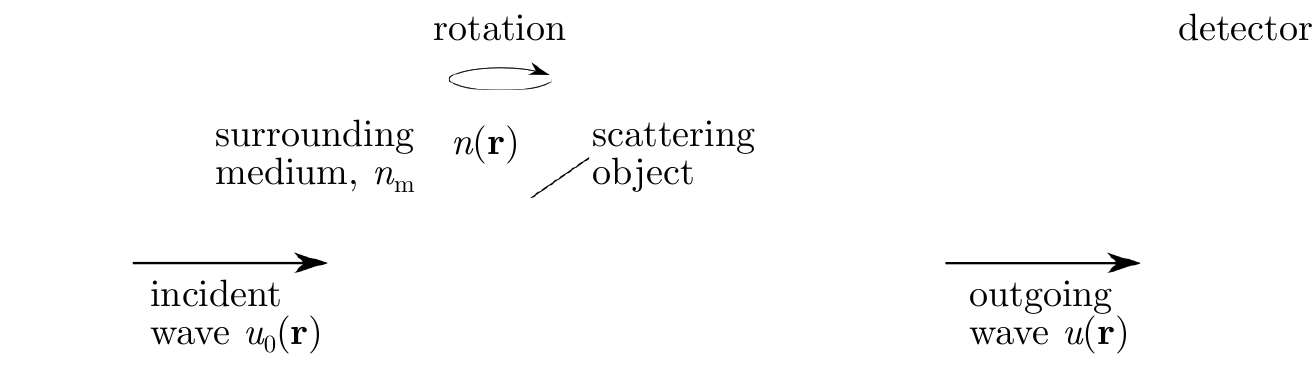}{png}
  \mycaption{Tomographic data acquisition}{An incident plane wave \gls{u0} is scattered by a transparent object with the refractive index distribution \gls{n}. A detector collects the scattered wave \gls{u}. Multi-angular acquisition is facilitated by rotation of the sample.}
  \label{fig:Introduction}
\end{figure}
The general problem is depicted in \hyref{figure}{fig:Introduction}. For different angular positions, images of single-cell sized objects are recorded at the detector.
A plane wave \gls{u0} with a wavelength \gls{lambda} in the visible regime propagates through a biological cell with a certain real refractive index distribution \gls{n}. The recorded set of phase images, measured at different rotational positions of the cell, is called a sinogram. Sinograms are the starting point for the refractive index reconstruction in three dimensions.

The first section of the manuscript is a brief summary of non-diffraction tomographic methods. The two following sections introduce the wave equation and showcase the reconstruction from analytically computed sinograms. The subsequent chapters present the reconstruction algorithms \cite{Wolf1969, Devaney1981, Devaney1982336} and their numerical implementation. The notation that we use here is similar to that used by the relevant literature (e.g.~\cite{Kak2001}). A list of symbols is given at the end of this manuscript.

%% file: chapter1_tomography_without_diffraction.tex
\section{Tomography without Diffraction}
\label{sec:NormalTomo}
The first applications of computerized tomography (CT) used bone tissue as an inherent x-ray-absorbing marker. However, the marker can also be artificially introduced to the specimen. For example, in positron emission tomography (PET) radioactive tracers serve as markers for high metabolic activity.
CT was also applied to biological specimens using wavelengths of the visible spectrum of light. The technique was termed optical projection tomography (OPT) \cite{sharpe2004}. In OPT, instead of measuring the absorption of the sample, the phase change introduced by the sample is measured. Thus, only the real part of the refractive index \gls{n} is reconstructed, whereas in classical  x-ray tomography the imaginary (absorbing) part of the refractive index is measured. In OPT, fluorescent markers can be used to complement the measurement of refractive index, just as PET does for classical x-ray CT.
The algorithms presented in this section do not take into account the wave nature of light. They are only valid in the limit of small wavelengths, i.e. x-ray radiation or for very small refractive index variations $\gls{ndelt} = \gls{n}-\gls{nm}$ of the sample \gls{n} from the surrounding medium \gls{nm}.

\subsection{Radon Transform}
The Radon transform describes the forward tomographic process which is in general the projection of an $n$-dimensional function onto an $(n-1)$-dimensional plane. In the case of computerized tomography (CT), the forward process is the acquisition of two-dimensional~(2D) projections from a three-dimensional (3D) volume\footnote{Keep in mind that in CT, the absorption of e.g. bone tissue is measured, whereas in optical tomography, the phase of the detected wave is measured.}. 
The 3D Radon transform can be described as a series of 2D Radon transforms for adjacent slices of the 3D volume. For the sake of simplicity, we consider only the two-dimensional case in the following derivations.

The value of one point in a projection is computed from the line integral through the detection volume~\cite{Radon1917}. The sample \gls{f} is rotated through~\gls{phi0} along the $y$-axis. For each 2D slice of the sample  $\left. \gls{f} \right|_{y=y_\mathrm{s}}$ at $y=y_\mathrm{s}$, the one-dimensional projection $ \gls{pr} = p_{\phi_0}(x_\mathrm{D},y_\mathrm{s}) $  of this slice onto a detector plane located at $(x_\mathrm{D},y_\mathrm{s})$ is described by the Radon transform operator~\gls{R}.
\begin{align}
p_{\phi_0}(x_\mathrm{D},y_\mathrm{s})  &= 
\gls{R}\lbrace\left. \gls{f} \right|_{y=y_\mathrm{s}} \rbrace(x_\mathrm{D}) \notag \\
&= \int \!\! dv \, f(x(v), \, y_\mathrm{v}, \, z(v)) \notag \\
&= \int \!\! dv \, 
f(x_\mathrm{D} \cos\gls{phi0} - v \sin\gls{phi0}\scalebox{1.4}{,}
~ y_\mathrm{s}\scalebox{1.4}{,}
~ x_\mathrm{D} \sin\gls{phi0} + v \cos\gls{phi0}) \label{eq:Rad}\\
 r_\mathrm{xz}^2 &= x^2 + z^2 =  x_\mathrm{D}^2 + v^2 \label{eq:Rad2} \\
x_\mathrm{D} &= x \cos \gls{phi0} + z \sin\gls{phi0}  \label{eq:Rad3} \\
v &= - x \sin\gls{phi0} + z \cos\gls{phi0} \label{eq:Rad4}
\end{align}
\noindent \hyref{Equation}{eq:Rad2} defines a distance $r_\mathrm{xz}$ from the rotational center of the 2D slice. Equations \ref{eq:Rad3} and \ref{eq:Rad4} describe the dependency of the detector position at~$x=x_\mathrm{D}$ and the parameter~$v$ of the integral on the  rotational position of the sample as defined by~\gls{phi0}.
The equations are illustrated in \hyref{figure}{fig:Radon_projection}. 

Because the 3D Radon transform can be described by multiple 2D Radon transforms, the 3D reconstruction from a sinogram can also be described by multiple 2D reconstructions\footnote{This is not valid for diffraction tomography as discussed in sections \ref{sec:2dODT} and \ref{sec:ReconDiffrObj}.}.

\begin{figure}[hb]
\centering
\subfloat[][3D sketch]
{\myincludegraphics{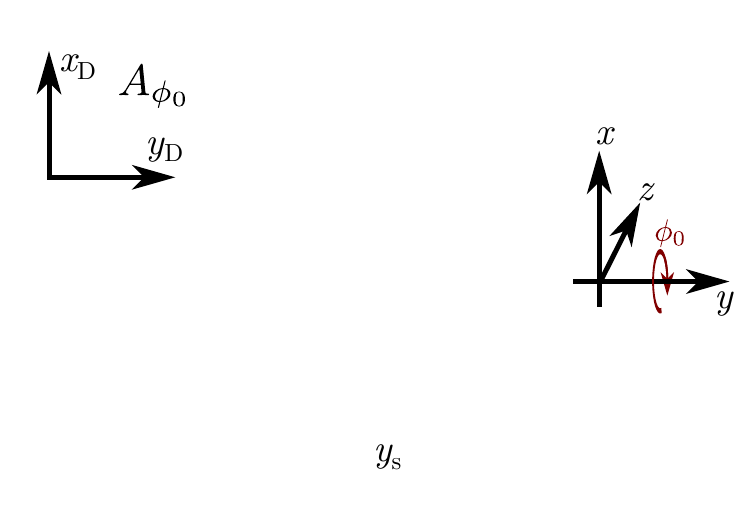}{jpg}\label{fig:Radon_projectWiona}} \qquad
\subfloat[][2D slice integral pathway]
{\includegraphics[]{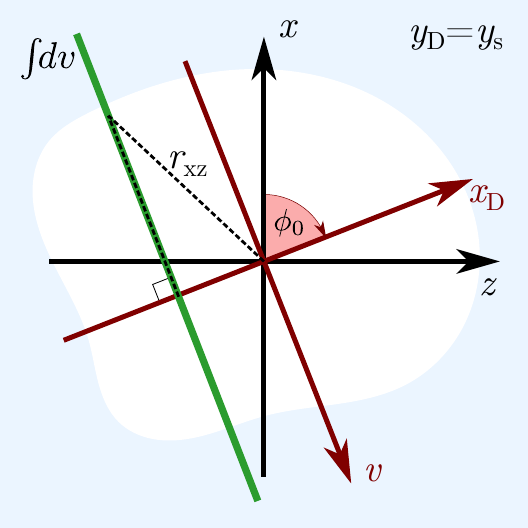}\label{fig:Radon_projectionb}} \qquad
  \mycaption{3D Radon transform}{\textbf{a)} Working principle of the three-dimensional (3D) Radon transform of a 3D object with the rotational axis $y$ and the rotational angle \gls{phi0}. For each slice of the object at $y_\mathrm{s}$ (blue plane), a two-dimensional Radon transform is performed. \textbf{b)} The two-dimensional Radon transform at $y_\mathrm{s}$ is computed by rotation of the object (white) through~\gls{phi0} (red coordinate system) and integration along $v$ (green line) perpendicular to the detector line $x_\mathrm{D}$.
\label{fig:Radon_projection}}
\end{figure}

\subsection{Fourier Slice Theorem}
\label{sec:FSTh}
The Fourier slice theorem is the central theorem in classical tomography. It connects the projection data $p_{\phi_0}(x_\mathrm{D})$ to the original image function $f(x,z)$ in Fourier space, which allows efficient reconstructions by means of the fast Fourier transform.

\noindent The projection of a two-dimensional image \gls{f} onto a detector line at an angle \gls{phi0} can be written as the integral
\begin{align}
p_{\phi_0}(x_\mathrm{D}) = \int \!\! dv \, f(x(v), z(v)) \tag{\ref{eq:Rad}}.
\end{align}
We define the unitary angular frequency Fourier transform of the one-dimensional data at the detector line with
\begin{align}
\widehat{P}_{\phi_0}(k_\mathrm{Dx}) = 
\frac{1}{\sqrt{2\pi}}
\int \!\! dx_\mathrm{D} \,
p_{\phi_0}(x_\mathrm{D})
\exp(-i k_\mathrm{Dx} x_\mathrm{D}).
\end{align}
The actual Fourier transform \gls{F} of the two-dimensional image \gls{f} is given by
\begin{align}
\widehat{F}(k_\mathrm{x}, k_\mathrm{z}) = 
\frac{1}{2\pi}
\iint \!\! dx dz \,
f(x,z)
\exp(-i (k_\mathrm{x}x + k_\mathrm{z}z)).
\end{align}
This formula can be rewritten as (subscript \gls{phi0} denotes rotation; the Jacobian of a rotation is $1$)
\begin{align}
\widehat{F}_{\phi_0}(k_\mathrm{Dx}, k_\mathrm{v}) &= 
\frac{1}{2\pi}
\iint \!\! dx_\mathrm{D} dv \,
f_{\phi_0}(x_\mathrm{D}, v)
\exp(-i (k_\mathrm{Dx}x_\mathrm{D} + k_\mathrm{v}v)) \\
f_{\phi_0}(x_\mathrm{D}, v) &= 
f(x_\mathrm{D} \cos\gls{phi0} - v \sin\gls{phi0}\scalebox{1.4}{,}
~ x_\mathrm{D} \sin\gls{phi0} + v \cos\gls{phi0} ) \\
\widehat{F}_{\phi_0}(k_\mathrm{Dx}, k_\mathrm{v}) &=
F(k_\mathrm{Dx} \cos\gls{phi0} - k_\mathrm{v} \sin\gls{phi0}\scalebox{1.4}{,}
~ k_\mathrm{Dx} \sin\gls{phi0} + k_\mathrm{v} \cos\gls{phi0}).
\end{align}
For the case $k_\mathrm{v} = 0$ (which implies slicing \gls{F} at the angle \gls{phi0}), we arrive at \cite{Bracewell1956, Mersereau1976247, Brooks1976}
\begin{align}
\widehat{F}_{\phi_0}(k_\mathrm{Dx}, 0) = 
\frac{1}{\sqrt{2\pi}} 
\widehat{P}_{\phi_0}(k_\mathrm{Dx}). \label{eq:FSTh}
\end{align}
This formula, known as the Fourier slice theorem, states that the Fourier transform of a projection at an angle \gls{phi0} is distributed along a straight line at the same angle in the Fourier space of the image \gls{f}. This theorem allows us to compute the inverse of the Radon transform operator \gls{R} by interpolating \gls{F} from $\widehat{P}_{\phi_0}(k_\mathrm{Dx})$ in Fourier space and subsequently performing an inverse Fourier transform.

In order to compute the inverse Radon transform in Fourier space, we have to interpolate the data in Fourier space on a rectangular grid. However, this technique is afflicted with interpolation artifacts. Alternatively, one often uses the backprojection algorithm which is introduced in the next section.

\subsection{Backprojection Algorithm}
\label{sec:2DInvRadon}
The backprojection algorithm connects the object function $f(x,z)$ to the Fourier transform of the projection $\widehat{P}_{\phi_0}(k_\mathrm{Dx})$. In order to derive it, we first express the object function $f(x,z)$ as the inverse Fourier transform of \gls{F}.
\begin{align}
f(x,z) = 
\frac{1}{2\pi}
\iint \!\! dk_\mathrm{x} dk_\mathrm{z} \,
\widehat{F}(k_\mathrm{x}, k_\mathrm{z})
\exp(i (k_\mathrm{x}x + k_\mathrm{z}z))
\end{align}
We then perform a coordinate transform from $(k_\mathrm{x}, k_\mathrm{z})$ to $(k_\mathrm{Dx}, \gls{phi0})$. It can be easily shown that the Jacobian computes to
\begin{align}
\left| \det \left( 
    \frac{d(k_\mathrm{x}, k_\mathrm{z})}{d(k_\mathrm{Dx}, \gls{phi0})}
\right) \right| &=
\left| k_\mathrm{Dx} \right|. \\
k_\mathrm{x} &=
k_\mathrm{Dx} \cos\gls{phi0} - k_\mathrm{v} \sin\gls{phi0} \\
k_\mathrm{z} &=
k_\mathrm{Dx} \sin\gls{phi0} + k_\mathrm{v}  \cos\gls{phi0}\\
k_\mathrm{v} &= 0 \label{eq:reqFST3}
\end{align}
Therefore, combining equations \ref{eq:FSTh} to \ref{eq:reqFST3}, we get  \cite{Bracewell1956, Mersereau1976247, Brooks1976, Crowther23061970, ramachandran1971}
\begin{align}
f(x,z) = 
\frac{1}{2\pi}
\int \!\! dk_\mathrm{Dx}
\int_{0}^{\pi} \! \! d\gls{phi0} \,
\left| k_\mathrm{Dx} \right|
\frac{\widehat{P}_{\phi_0}(k_\mathrm{Dx})}{\sqrt{2\pi}}
\exp[i k_\mathrm{Dx}(x\cos \gls{phi0} + z \sin \gls{phi0} )].
\label{eq:Backproj}
\end{align}
Note that the integral over \gls{phi0} runs from $0$ to $\pi$. The  integrals of $k_\mathrm{x}$, $k_\mathrm{z}$, and $k_\mathrm{Dx}$ are computed over the entire $k$-space, i.e. over the interval $(-\infty, +\infty)$.
\noindent The term $\left| k_\mathrm{Dx} \right|$ is a ramp filter in Fourier space which lead to the common term \textit{filtered} backprojection algorithm in literature. However, we refer to it as the backprojection algorithm throughout this manuscript. 

\hyref{Figure}{fig:FBP} depicts the sinogram acquisition and image reconstruction of a two-dimensional test target with the backprojection algorithm from 30 and 100 angular projections.
Note that because of our chosen coordinate system, at the angle $\gls{phi0}=0$, $k_\mathrm{Dx}$ coincides with the $k_\mathrm{x}$ axis ($x_\mathrm{D}\overset{\phi_0 = 0}{=}x$).
\begin{figure}[ht]
\centering
\subfloat[][original image,\\ $500\times 500$ pixels]
{\includegraphics[width=\linewidth/4]{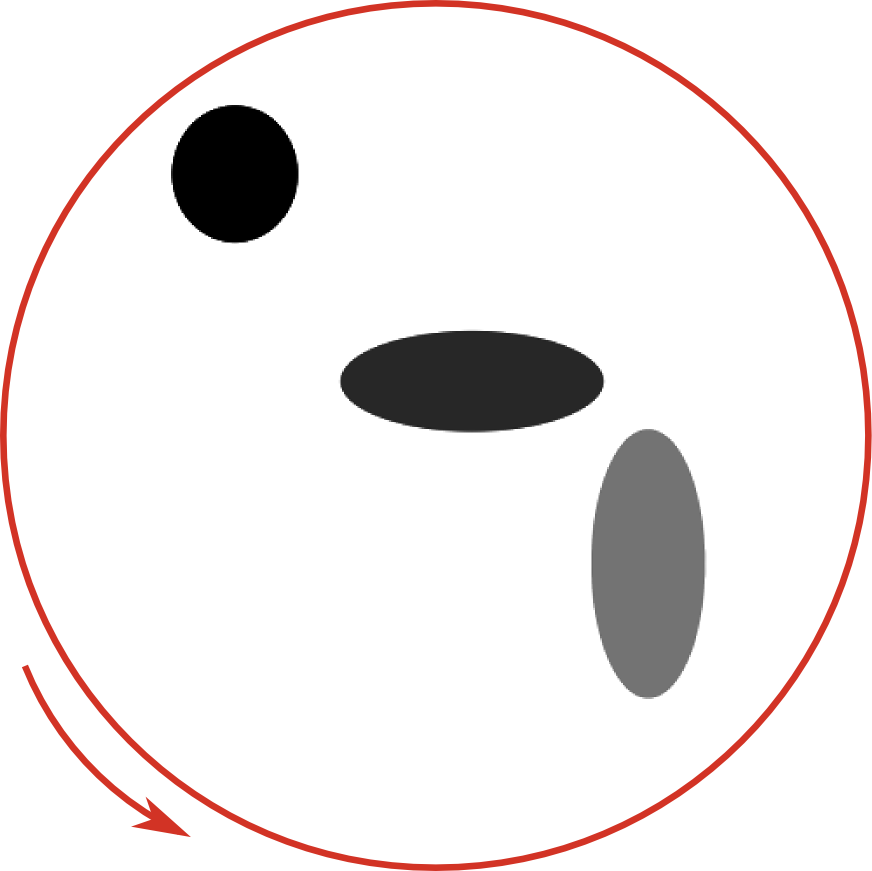}}
\subfloat[][sinogram,\\ 500 projections]  
{\includegraphics[width=\linewidth/4]{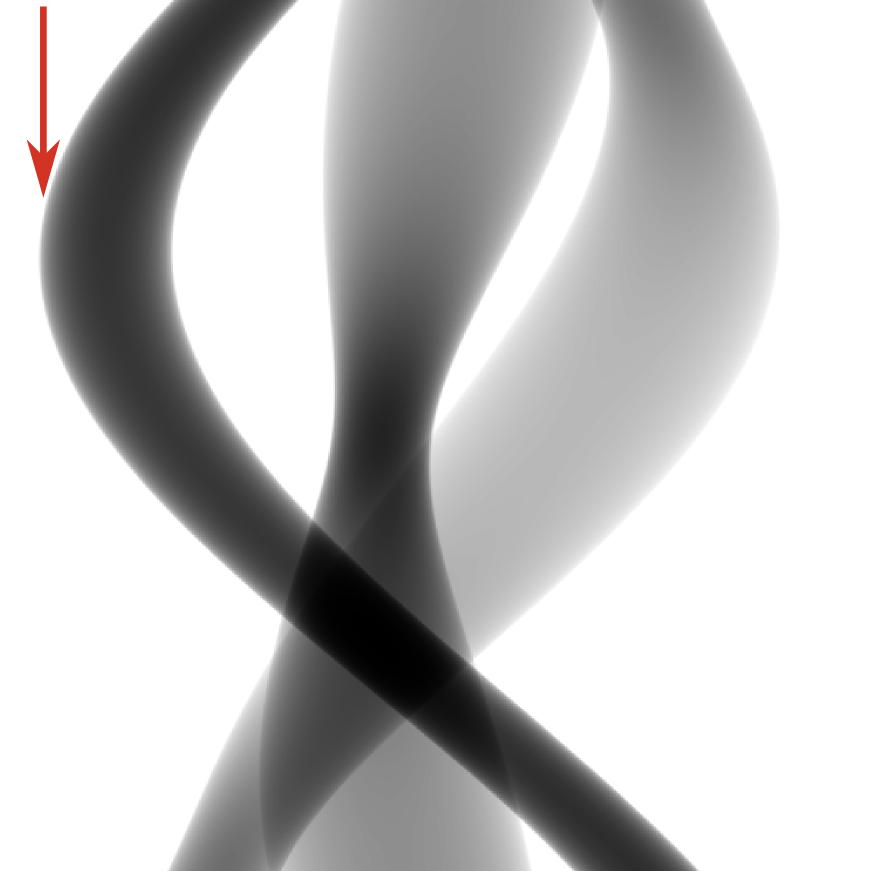}} 
\subfloat[][reconstruction \\ from 30 projections]
{\includegraphics[width=\linewidth/4]{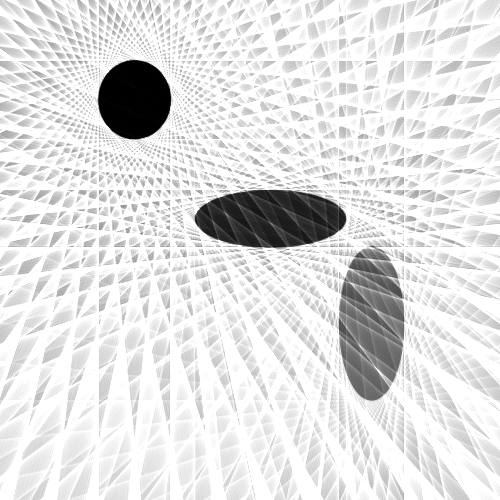}}
\subfloat[][reconstruction \\ from 100 projections]
{\includegraphics[width=\linewidth/4]{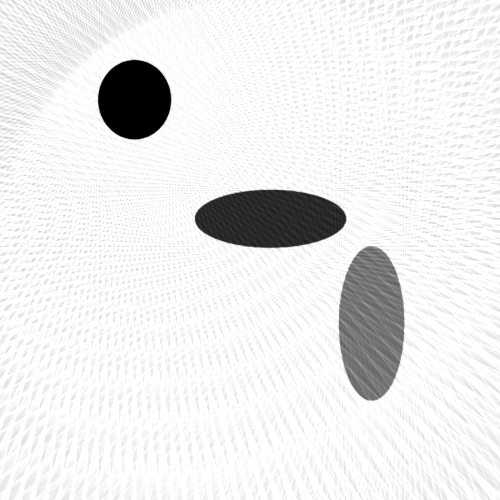}}
  \mycaption{Backprojection}{\textbf{a)} The original two-dimensional image contains ellipses of different gray-scale levels. \textbf{b)} The sinogram shows 500 projection of image (a) from 0$^{\circ}$ to 180$^{\circ}$. For the computation of the sinogram, only the circular region of the original image (red) was used.  \textbf{c)} Reconstruction using 30 equidistant projections. \textbf{d)} Reconstruction with 100 projections. The sinogram and reconstructions were created with the Python library radontea~\cite{radontea}.\label{fig:FBP}}
\end{figure}

As discussed in the introduction, the backprojection algorithm assumes that light propagates along straight lines. It is thus not an optimal method for optical tomography which employs wavelengths that are in the visible spectrum of light. Thus, we need to include the wave nature of light in our reconstruction scheme. The next two chapters introduce the foundation of the Fourier diffraction theorem which yields better reconstruction results for objects that diffract light.

%% file: chapter2_the_wave_equation.tex
\section{The Wave Equation}
\label{sec:waveeq}
The propagation of light through objects in space \vc{r} and time $t$ follows the Maxwell equations. Light propagation through empty space can be described by the wave equation, which simplifies the vectorial description of electromagnetic fields to a scalar description of waves. When light propagates through inhomogeneous media, the wave equation does not anymore provide an exact description because the vectorial components of the electromagnetic field couple at e.g. refractive index boundaries. However, it is known that the wave equation is a good approximation for light propagation in inhomogeneous media \cite{goodman} and we show that this scalar approximation of vectorial fields is valid for the discussed test targets.
The wave equation reads
\begin{equation}
\frac{\partial^2}{\partial t^2} \gls{Psi}  = \left(\frac{\gls{c0}}{\gls{n}}\right)^2 \cdot \nabla^2 \gls{Psi},
\end{equation}
where \gls{Psi} is a time-dependent, scalar wave field in space, \gls{c0} is the speed of light in vacuum, and \gls{n} is the spatial refractive index distribution.
Note that $\frac{\gls{c0}}{\gls{n}}$ is usually a constant coefficient which describes the speed of the propagating wave. Because we are not interested in the temporal information of \gls{Psi}, we may simplify the wave equation by separation of variables. We then obtain the time-independent wave equation, known as the Helmholtz equation~\cite{Cohen-TannoudjiDiuLaloe199211}
\begin{equation}
\left(\nabla^2 + \gls{k}^2\right) \gls{u}  = 0,
\label{eq:notimewave}
\end{equation}
where \gls{k} is the wave number that depends on the local refractive index distribution~\gls{n} and \gls{u} is the scattered field. To simplify the notation, we introduce the wave number inside the medium surrounding a sample \gls{km} and the local variation in refractive index inside a sample \gls{ndelt} as follows,
\begin{eqnarray}
\gls{km} & = & \frac{2 \pi \gls{nm}}{\gls{lambda}} \label{eq:defkm}\\
\gls{n} & = & \gls{nm} +  \gls{ndelt} \label{eq:defn}\\
\gls{k} & = & \gls{km} \frac{\gls{n}}{\gls{nm}} \\
 & = & \gls{km} \left( 1 + \frac{\gls{ndelt}}{\gls{nm}} \right) \notag
\end{eqnarray}
where \gls{lambda} is the vacuum wavelength of the light. 

\subsection{Homogeneous Helmholtz Equation}
For the homogeneous case, i.e. there is no sample ($\gls{ndelt}=0$), the wave equation becomes the  Helmholtz equation for a homogeneous medium
\begin{equation}
\left(\nabla^2 + \gls{km}^2\right) \gls{u0}  = 0.
\label{eq:homwave}
\end{equation}
This second order ordinary differential equation has plane wave solutions of the form
\begin{equation}
\gls{u0} = a_\mathrm{0} \exp \! \left(i \gls{km} \, \gls{s0} \cdot \vc{r} \right),
\label{eq:u0}
\end{equation}
where \gls{s0} is the normal unit vector  and $a_\mathrm{0}$ is the amplitude of the plane wave. Throughout this script, we use the convention that \gls{u0} in \hyref{equation}{eq:u0} defines a wave traveling from left to right, and thus \gls{u0} has a `+'-sign in the exponent. This convention is widely used in the literature dealing with diffraction tomography \cite{Kak2001, Wolf1969, Devaney1982336}.

\subsection{Inhomogeneous Helmholtz Equation}
\label{sec:InhmHelmh}
\hyref{Equation}{eq:notimewave} can be rewritten as the inhomogeneous Helmholtz equation
\begin{align}
\left(\nabla^2 + \gls{km}^2\right) \gls{u}  =  -&\gls{f} \, \gls{u} \label{eq:inhomwave} \\
\text{with~} &\gls{f}  =  \gls{km}^2\left[ \left(\frac{\gls{n}}{\gls{nm}}\right)^2 -1 \right]. \label{eq:fperturbance}
\end{align}
In order to deal with the inhomogeneity \gls{f}, also called scattering potential, we can make use of the Green's function \gls{G} that is defined by \hyref{equation}{eq:greendelta}\footnote{The Green's function is defined as the solution to the inhomogeneous problem: When the operator $(\nabla^2 + \gls{km}^2)$ is applied to \gls{G}, one obtains the Dirac delta distribution $\delta(\vc{r - r'})$.}. The Green's function for the homogeneous Helmholtz equation is shown in \hyref{equation}{eq:greendef}\footnote{Note that our notation implies the ($+$)-sign  and a normalization factor of $4 \pi$ for the Green's function. For a derivation see for example \cite{TaiIEEE199401}, section 2.4 or \cite{MorseFeshback1953}, section 7.2.}.
\begin{align}
\left(\nabla^2 + \gls{km}^2\right) \gls{G} &= - \delta(\vc{r - r'}) \label{eq:greendelta}\\
\gls{G} &=  \frac{\exp \! \left( i \gls{km} \left| \vc{r - r'} \right| \right)}{4 \pi \, \left| \vc{r - r'} \right|}  \label{eq:greendef}
\end{align}
Here, $\delta(\vc{r - r'})$ is the Dirac delta function with the translational property
\begin{equation}
\int \! \! d^3r' \, \delta(\vc{r - r'}) \, g(\mathbf{r'}) = g(\mathbf{r}).
\label{eq:diracdelta}
\end{equation}
By knowledge of the Green's function we can derive an integral representation for the scattered field \gls{u}. Using \hyref{equation}{eq:diracdelta}, the right side of \hyref{equation}{eq:inhomwave} can be integrated to include the Green's function (\hyref{eq.}{eq:greendelta})\footnote{Note that the Green's function depends on the absolute value of $\left|\vc{r-r'}\right|$ and thus \\ $\left(\nabla'^2 + \gls{km}^2\right) \gls{G} = \left(\nabla^2 + \gls{km}^2\right) \gls{G} $}
\begin{align}
\gls{f} \, \gls{u} &= \int \! \! d^3r' \, \delta(\vc{r - r'}) \, f(\mathbf{r'}) \, u(\mathbf{r'}) \\
&= - \int \! \! d^3r' \, \left(\nabla^2 + \gls{km}^2\right) \gls{G} \, f(\mathbf{r'}) \, u(\mathbf{r'}) \\
&= - \left(\nabla^2 + \gls{km}^2\right) \int \! \! d^3r' \, \gls{G} \, f(\mathbf{r'}) \, u(\mathbf{r'}). \label{eq:fuint}
\end{align}
The comparison of equations \ref{eq:inhomwave} and \ref{eq:fuint} suggests that the scattered field \gls{u} can be written as
\begin{align}
\gls{u}  =  \int \! \! d^3r' \, \gls{G} \, f(\mathbf{r'}) \, u(\mathbf{r'}) \label{eq:uint}
\end{align}
which is the integral equation to the inhomogeneous wave equation (eq. \ref{eq:inhomwave}). 

%% file: chapter3_approximations_to_the_scattered_wave.tex
\section{Approximations to the Scattered Wave}
\label{sec:approxwave}
\hyref{Equation}{eq:uint} has no analytical solution. However, under certain conditions approximations can be applied to find a solution. In this section, we derive the Born and Rytov approximations. Both approximations define the scattered field \gls{u} as a superposition of the incident plane wave \gls{u0} and a scattered component \gls{us}.
\begin{equation}
\gls{u} = \gls{u0} + \gls{us}
\end{equation}
\subsection{Born Approximation}
\label{sec:Born}
The Born approximation uses the property of the homogeneous component
\begin{equation}
\left(\nabla^2 + \gls{km}^2\right) \gls{u0}  = 0 \tag{\ref{eq:homwave}}
\end{equation}
to rewrite the inhomogeneous wave equation for the scattered component \gls{us}
\begin{equation}
\left(\nabla^2 + \gls{km}^2\right) \gls{u}  = \left(\nabla^2 + \gls{km}^2\right) \gls{us} = -\gls{f}\gls{u}.
\end{equation}
By once again using the translational property of the delta function (\hyref{eq.}{eq:diracdelta}) we obtain an iterative equation for \gls{us}
\begin{equation}
\gls{us}  =  \int \! \! d^3r' \, \gls{G} \, f(\mathbf{r'}) \, u(\mathbf{r'})
\end{equation}
and we can write the Lippmann-Schwinger equation for the field \gls{u} with the perturbation \gls{f} \cite{Cohen-TannoudjiDiuLaloe199211}
\begin{equation}
\gls{u}  =  \gls{u0} + \int \! \! d^3r' \, \gls{G} \, f(\mathbf{r'}) \, u(\mathbf{r'}).
\end{equation}
By iteratively replacing \gls{u} in the above integral, one obtains the Born series. The Born approximation is actually the first iteration of the Born series \cite{Cohen-TannoudjiDiuLaloe199211}.
\begin{align}
\gls{u}  \overset{\text{Born}}{\approx}  \gls{u0} + &\gls{ub} \\
&\gls{ub} = \int \! \! d^3r' \, \gls{G} \, f(\mathbf{r'}) \, u_\mathrm{0}(\mathbf{r'}). \label{eq:uborn}
\end{align}

\subsubsection*{Validity of the Born Approximation}
In the derivation above, approximating the scattered component \gls{us} as the first term of the Born series \gls{ub} implies that
\begin{align}
\int \! \! d^3r' \, \gls{G} \, f(\mathbf{r'}) \, [u_0(\mathbf{r'}) + u_\mathrm{s}(\mathbf{r'})] &\approx
\int \! \! d^3r' \, \gls{G} \, f(\mathbf{r'}) \, u_0(\mathbf{r'}) \\
\text{or~~~} \gls{u0} + \gls{us} &\approx \gls{u0}.
\end{align}
Thus, the Born approximation holds for the case ``$\gls{us} \ll \gls{u0}$'', i.e. the contributions of amplitude and phase of the scattering component \gls{us} are small when compared to the incident plane wave \gls{u0}. Since both \gls{us} and \gls{u0} are complex-valued functions, the above relation implies that (i) there is only little absorption by the specimen and that (ii) the overall phase change $\Delta \gls{phase}$ must be small. Because we are interested in biological cells whose imaginary part of the refractive index is insignificant, (i) absorption is negligible and we thus focus on (ii) the phase change~$\Delta \gls{phase}$ introduced by the cell. The phase change~$\Delta \gls{phase}$ that a scattered wave \gls{us} experiences when traveling through a sample along a certain path parametrized by $s_\mathrm{path}$ can be approximated with ray optics
\begin{equation}
\Delta \gls{phase}  \approx \frac{2 \pi}{\gls{lambda}} \left( \underset{\text{path}}{\int}\!\!ds_\mathrm{path} \, \gls{n} -  s_\mathrm{tot}\gls{nm} \right),
\end{equation}
where \gls{n} is the refractive index distribution inside the sample, $s$ is the approximate thickness of the sample, and \gls{nm} is the refractive index of the medium surrounding the sample. For a homogeneous sample with the refractive index $n_\mathrm{s}$, the absolute phase change computes to
\begin{equation}
\Delta \gls{phase}  = \frac{2 \pi}{\gls{lambda}} s(n_\mathrm{s}-\gls{nm}) = \frac{2 \pi}{\gls{lambda}} s \epsilon_\mathrm{n}.
\label{eq:phasesimple1}
\end{equation}
This equation can be interpreted as a comparison of the phase change $\Delta \gls{phase}$ over a period of~$2 \pi$ with the change of the optical path length $s(n_\mathrm{s}-\gls{nm})$ over one wavelength~\gls{lambda}. 

We now want to write \hyref{equation}{eq:phasesimple1} in its differential form. We express the total differential of $\Delta \gls{phase}$ with respect to the spatial distance $s$ and the refractive index variation~$\epsilon_\mathrm{n}$ as
\begin{equation}
\frac{d (\Delta \gls{phase})}{2\pi}  = \frac{\epsilon_\mathrm{n}}{\gls{lambda}} ds +
\frac{s}{\gls{lambda}} d\epsilon_n.
\label{eq:phasesimple}
\end{equation}
The phase change $\Delta \gls{phase}$ can have two contributions, namely the thickness of the sample
\begin{equation}
\frac{d (\Delta \gls{phase}_\mathrm{A})}{2\pi}  = \frac{\epsilon_\mathrm{n}}{\gls{lambda}} ds 
\label{eq:phasesimplea}
\end{equation}
 and the refractive index variation inside the sample
\begin{equation}
\frac{d (\Delta \gls{phase}_\mathrm{B})}{2\pi}  = \frac{s}{\gls{lambda}} d\epsilon_n.
\label{eq:phasesimpleb}
\end{equation}
Note that in general $\epsilon_\mathrm{n}$ is dependent on \vc{r} (3D) and that $\Delta \gls{phase}$ is measured at the detector plane \vc{r_D} (2D). If the refractive index of the sample is fixed, then the total phase change solely depends on the thickness of the sample. If the thickness of the sample is fixed, the local variations inside the sample determine the absolute phase change. If we wanted to consider a phase change that is introduced by a refractive index variation over the distance of one wavelength $\gls{lambda}$, then we would have to set $s=\gls{lambda}$.

With these considerations, we can answer the question of the validity of the Born approximation. We interpret the inequality ``$\gls{us} \ll \gls{u0}$'' as a general restriction: The overall phase change must be much smaller than $2 \pi$. According to \hyref{equation}{eq:phasesimple1}, this translates to a restriction for the optical path difference $s(n_\mathrm{s}-\gls{nm})$ which must be much smaller than the wavelength \gls{lambda} of the used light
\begin{align}
\Delta \gls{phase} & \ll 2 \pi \\
s(n_\mathrm{s}-\gls{nm}) & \ll \gls{lambda}.
\end{align}
The fact that the Born approximation only allows to observe optically thin samples is a serious drawback. A second approach to the problem is the Rytov approximation. 

\subsection{Rytov Approximation}
\label{sec:Rytov}
In order to derive the Rytov approximation we assume that the scattered wave~\gls{u} and the incident wave \gls{u0} have the form
\begin{align}
\gls{u} &= \exp(\gls{pc}) = \gls{u0} + \gls{us} \\
\gls{u0} &= \exp(\gls{pc0}) \\
\gls{pc} &= \gls{pc0} + \gls{pcs} \label{eq:rytovphasecomp}
\end{align}
where we use the complex phases
\begin{align}
\gls{pc} &= i \gls{phase}(\mathbf{r}) + \ln(a(\vc{r})) \\
\gls{pc0} &= i \gls{phase}_0 (\mathbf{r}) + \ln(a_\mathrm{0}(\vc{r}))
\end{align}
to denote the complex exponent containing phase and amplitude of the wave function. Note that \gls{us} is now computed from \gls{pcs} with
\begin{align}
\gls{us}  &= \gls{u} - \gls{u0} \\
&= \exp(\gls{pc0}) \left[ \exp(\gls{pcs}) - 1 \right].
\end{align}
Using the equations above, the inhomogeneous wave equation becomes
\begin{align}
(\nabla^2 + \gls{km}^2) \gls{u} &= - \gls{f}  \gls{u} \\
(\nabla^2 + \gls{km}^2) \exp(\gls{pc}) &= - \gls{f} \exp(\gls{pc}).
\end{align}
We can then compute the term $ \nabla^2 \exp(\gls{pc}) $
\begin{align}
\nabla^2 \exp(\gls{pc}) &= \nabla \left[ \exp(\gls{pc})\cdot \nabla\gls{pc} \right] \\
\nabla^2 \exp(\gls{pc}) &= \exp(\gls{pc}) \left[\nabla^2 \gls{pc} + \left(\nabla \gls{pc}\right)^2 \right]
\end{align}
to obtain a differential equation for \gls{pc}.
\begin{align}
\exp(\gls{pc}) \left[\nabla^2 \gls{pc} + \left(\nabla \gls{pc}\right)^2 + \gls{km}^2 \right] &= - \gls{f} \exp(\gls{pc})\\
\nabla^2 \gls{pc} + \left(\nabla \gls{pc}\right)^2 + \gls{km}^2 &= - \gls{f} \label{eq:rytovwavediff}
\end{align}
\hyref{Equation}{eq:rytovwavediff} is a non-linear differential equation for the complex phase \gls{pc}. In the same manner, a differential equation for \gls{pc0} can be derived
\begin{align}
\nabla^2 \gls{pc0} + \left(\nabla \gls{pc0}\right)^2 + \gls{km}^2 = 0.  \label{eq:rytovwavediff0}
\end{align}
The next step is to insert \hyref{equation}{eq:rytovphasecomp} into \hyref{equation}{eq:rytovwavediff} to find a differential equation for \gls{pcs}.
\begin{align}
\nabla^2 [\underline{\gls{pc0}} + \gls{pcs}] + \underbrace{\left(\nabla [\gls{pc0} + \gls{pcs}]\right)^2}_{\underline{(\nabla \gls{pc0})^2} + 2\nabla \gls{pc0} \cdot \nabla \gls{pcs} + (\nabla \gls{pcs})^2} + \underline{\gls{km}^2} &= - \gls{f}
\end{align}
The terms marked with an underline compute to zero (\hyref{eq.}{eq:rytovwavediff0}) and the  equation above becomes
\begin{align}
\nabla^2 \gls{pcs} + 2\nabla \gls{pcs} \cdot \nabla \gls{pc0} + (\nabla \gls{pcs})^2 = - \gls{f}. \label{eq:pcs1}
\end{align}
To simplify this expression we need to consider
\begin{align}
\nabla^2 \gls{u0} \gls{pcs} = \underbrace{\nabla^2 \gls{u0}}_{-\gls{km}^2 \gls{u0}} \cdot \gls{pcs} & + 2 \underbrace{\nabla\gls{u0}}_{\gls{u0} \nabla \gls{pc0}} \cdot \nabla \gls{pcs} + \gls{u0} \nabla^2 \gls{pcs}. \\
& \downarrow \notag \\
(\nabla^2+\gls{km}^2)\gls{u0} \gls{pcs} & = 
  2\gls{u0} \nabla \gls{pc0} \cdot \nabla \gls{pcs} + \gls{u0} \nabla^2 \gls{pcs} \label{eq:pcs2}
\end{align}
If we multiply \hyref{equation}{eq:pcs1} by \gls{u0} then we can substitute with \hyref{equation}{eq:pcs2} to obtain
\begin{equation}
(\nabla^2+\gls{km}^2)\gls{u0} \underbrace{\gls{pcs}}_{\overset{\text{Rytov}}{\approx} \gls{pcr}} = -\gls{u0} \underbrace{[(\nabla \gls{pcs})^2 + \gls{f}]}_{\overset{\text{Rytov}}{\approx} \gls{f}}.
\label{eq:rytapprxeq}
\end{equation}
The Rytov approximation assumes that the phase gradient $ \nabla \gls{pcs} $ is small compared to the perturbation \gls{f}. We can now make use of the Green's function again (eqns. \ref{eq:greendelta} to \ref{eq:uint}) and arrive at the formula for the Rytov Phase \gls{pcr} \cite{Kak2001}.
\begin{align}
\gls{u0} \gls{pcr} & =  \int \! \! d^3r' \, \gls{G} \, f(\mathbf{r'}) \, u_\mathrm{0}(\mathbf{r'}) \\
\gls{pcr} & = \frac{\int \! \! d^3r' \, \gls{G} \, f(\mathbf{r'}) \, u_\mathrm{0}(\mathbf{r'})}{\gls{u0}}
\end{align}
By comparing this expression to the first Born approximation (\hyref{eq.}{eq:uborn}) we find that we can compute the Rytov approximation \gls{ur} from the Born approximation \gls{ub} and vice versa.
\begin{align}
\gls{pcr} & = \frac{\gls{ub}}{\gls{u0}} \\
\gls{ur}  & = \gls{u0} \left[ \exp \!\left(\frac{\gls{ub}}{\gls{u0}}\right) -1  \right] \\
\gls{ub} & = \gls{u0} \ln\!\left(  \frac{\gls{ur}}{\gls{u0}} +1 \right) \label{eq:ubornfromrytov} \\
& = \gls{u0} \gls{pcr} \label{eq:ubornfromrytovphase} \\
\gls{u} & \overset{\text{Born}}{\approx} \gls{u0} + \gls{ub} \notag \\
\gls{u} & \overset{\text{Rytov}}{\approx} \gls{u0} + \gls{ur} \notag
\end{align}
Note that the complex Rytov phase \gls{pcr} also contains the amplitude information of the scattered wave.
In \hyref{section}{chap:backprop} we introduce a backpropagation algorithm for the Born approximation. This algorithm may also be used in combination with the Rytov approximation by using  \hyref{equation}{eq:ubornfromrytov}. In practice, calculating the logarithm of ${\exp(\gls{pc}-\gls{pc0})}$ involves calculating the logarithm of the amplitudes and the phases.
\begin{align}
\ln(\exp(\gls{pc}-\gls{pc0})) = \ln \left( \frac{a(\vc{r})}{a_0(\vc{r})}\right) + i \left(\Phi(\vc{r}) - \Phi_0(\vc{r}) \right)
\end{align}
Because the phase $\Phi(\vc{r}) - \Phi_0(\vc{r})$ is modulo $2\pi$, a computational implementation of the Rytov approximation must contain a phase-unwrapping algorithm \cite{Chen:98}.

\subsubsection*{Validity of the Rytov Approximation}
From our approximation in \hyref{equation}{eq:rytapprxeq}, we can tell that the Rytov approximation is valid for the case
\begin{align}
(\nabla \gls{pcs})^2 & \ll \gls{f} \notag \\
& \overset{\text{\hyref{eq.}{eq:fperturbance}}}{\ll} 
\gls{km}^2\left[ \left(\frac{\gls{n}}{\gls{nm}}\right)^2 -1 \right] \\
\gls{n}^2 & \gg \gls{nm}^2 \left[ \frac{(\nabla \gls{pcs})^2}{\gls{km}^2} + 1 \right].
\end{align}
We are interested in a validity condition that connects the refractive index with its gradient.
We insert the definitions of the wave vector \gls{km} and the refractive index distribution \gls{n} (eq. \ref{eq:defkm}, \ref{eq:defn}) to retrieve a condition for the variation in refractive index \gls{ndelt}.
\begin{align}
\gls{n}^2 & \gg \gls{nm}^2 \left( \frac{\nabla \gls{pcs} \gls{lambda}}{2 \pi \gls{nm}}\right)^2 + \gls{nm}^2 \\
\gls{n}^2 - \gls{nm}^2 & \gg  \left( \frac{\nabla \gls{pcs} \gls{lambda}}{2 \pi}\right)^2 \\
\underbrace{\gls{ndelt}^2}_{\approx 0} + 2 \gls{nm} \gls{ndelt} & \gg  \left( \frac{\nabla \gls{pcs} \gls{lambda}}{2 \pi}\right)^2 \label{eq:ndeltquadreq}
\end{align}
Because the local variation \gls{ndelt} is small, we may neglect\footnote{This can be shown by solving the quadratic \hyref{equation}{eq:ndeltquadreq} for \gls{ndelt} and Taylor-expanding for small $\nabla \gls{pcs}$ to the second order.} $\gls{ndelt}^2$. The resulting constraint for the phase gradient is \cite{Kak2001}
\begin{align}
 \frac{ \left| \nabla \gls{pcs}  \right|}{2 \pi}  & \ll \frac{\sqrt{2 \gls{nm}  \left| \gls{ndelt} \right|}}{\gls{lambda}}  \\
\frac{ \left| \operatorname{d}\!\gls{pcs}  \right|}{2 \pi} & \ll \frac{ \sqrt{2 \gls{nm} \left| \gls{ndelt} \right|} \cdot \left|\operatorname{d}\! \vc{r} \right|}{\gls{lambda}}. \label{eq:rytovboundary}
\end{align}
For any position $\vc{r}$ inside a sample, \hyref{equation}{eq:rytovboundary} reads: \\
\textit{The sample induces a phase change over a period of $2 \pi$ radians. This number must be smaller than the variation in refractive index \gls{ndelt} along the corresponding optical path scaled by the used wavelength \gls{lambda}.}\\
Note that the total phase gradient is computed from
\begin{align}
|\nabla \gls{pc}| & = |\nabla \gls{pc0} + \nabla \gls{pcs}| \\
|\nabla \gls{pc}| & = |i \vc{\gls{km}} + \nabla \gls{pcs}|.
\end{align}
Thus, compared to the Born approximation, where the overall phase change must be smaller than $2 \pi$, the Rytov approximation is also valid for optically thicker samples.

The Rytov approximation is accurate for small wavelengths and breaks down for large variations in the refractive index \cite{Slaney1984}. The following calculations attempt to derive a statement of validity for the Rytov approximation that only depends on the refractive index, which is difficult considering its non-linear, but exponential description of wave propagation.
When we insert \hyref{equation}{eq:phasesimpleb} into \hyref{equation}{eq:rytovboundary} ($\Delta \gls{phase}_\mathrm{B}(\vc{r})=\gls{pcs}$), we obtain the restriction for the Rytov approximation.
\begin{equation}
  \frac{d\gls{ndelt}}{|d\vc{r}|} \ll \frac{\sqrt{2 \gls{nm} |\gls{ndelt}|}}{s}
\end{equation}
Here, we assumed that we can replace a change in the local complex phase $d\varphi (\vc{r})$ with a variation in the local refractive index $d\gls{ndelt}$. This estimate is valid when light propagates approximately along straight lines, as described by \hyref{equation}{eq:phasesimpleb}. Furthermore, this estimate does not cover scattering at small objects and thus, $s>\lambda$. The quantity $s$ can be interpreted as a characteristic length scale within the imaged object below which light propagation can be approximated along straight lines (\hyref{eq.}{eq:phasesimpleb}).
\begin{equation}
  | \nabla \gls{n} | \ll \frac{\sqrt{2 \gls{nm} |\gls{n}-\gls{nm}|}}{s}, \hspace{2em} s > \lambda
\end{equation}
The validity of the Rytov approximation is not dependent on the absolute phase change introduced by the sample, but on the gradient of the refractive index within the sample. This makes the Rytov approximation applicable to biological cells. The application of the Rytov approximation to the three-dimensional backpropagation algorithm is described in \hyref{secion}{sec:FDTandRytov}.

%% file: chapter4_two-dimensional_diffraction_tomography.tex
\section{Two-Dimensional Diffraction Tomography}
\label{sec:2dODT}
The name \textit{Fourier diffraction theorem} was introduced by Slaney and Kak \cite{Kak2001} in the 1980's. In the limit of small wavelengths, the Fourier diffraction theorem in the Rytov approximation converges to the Fourier slice theorem (\hyref{section}{sec:FDTandRytov}, \cite{Devaney1981}). 
In this section, we derive the two-dimensional backpropagation algorithm, as described by Kak and Slaney~\cite{Kak2001}, which was implemented in the C~programming language in 1988\footnote{implemented by Malcolm Slaney, available at \url{http://slaney.org}; see also \cite{Slaney1985}}. At the end of the section, we test the backpropagation algorithm with artificial data that were computed using Mie theory. Mie theory provides solutions to the Maxwell equations in the form of infinite series for scattering objects consisting of simple geometric shapes. It is thus ideally suited to test the approximations made in sections \ref{sec:waveeq} and \ref{sec:approxwave}.
The 3D theory can be derived analogous to the 2D theory. It is described and discussed in greater detail in \hyref{section}{sec:ReconDiffrObj}. 
Throughout this section we use two-dimensional vectors, i.e.~{$ \vc{r} = (x,\, z)$}.
\subsection{Fourier Diffraction Theorem}
We start from the inhomogeneous wave equation as discussed in \hyref{section}{sec:InhmHelmh}.
\begin{equation}
\left(\nabla^2 + \gls{km}^2\right) \gls{u} = - \gls{f} \gls{u} 
\label{eq:2Dnotimewave}
\end{equation}
The Green's function in the two-dimensional case is the zero-order Hankel function of the first kind.
\begin{align}
\left(\nabla^2 + \gls{km}^2\right) \gls{G} &= - \delta(\vc{r - r'}) \\
\gls{G} &= \frac{i}{4} H_0^\text{(1)}(\gls{km}\left|\vc{r-r'}\right|) \\
H_0^\text{(1)}(\gls{km}\left|\vc{r-r'}\right|) &= \frac{1}{\pi} \int \!\! dk_\mathrm{x} \frac{1}{\gamma} \exp\! \left\lbrace i \left[ k_\mathrm{x}(x-x') + \gamma(z-z') \right] \right\rbrace \\
\gamma &= \sqrt{\gls{km}^2 - k_\mathrm{x}^2}
\end{align}
We observe restrictions for the Cartesian coordinates for the vector \gls{s0} describing the incoming plane wave and the vector \gls{s} describing an arbitrarily scattered wave. The following substitutions are defined by our notation:
\begin{align}
k_\mathrm{x} = \gls{km} p\, &, \hspace{2em} \gamma = \gls{km}M \\
M = \sqrt{1-p^2} \, &, \hspace{2em} M_0 = \sqrt{1-p_0^2} \label{eq:2DM} \\
 \gls{s} = (p, \, M) \, &, \hspace{2em}  \gls{s0} = (p_0, \, M_0)
\end{align}
The parameters $p$ and $p_0$ describe the $k_\mathrm{x}$-component of the vectors \gls{s} and \gls{s0}. They must fulfill the relation $0\geq p, p_0 \geq 1$.
In order to have the data consistent with \hyref{section}{sec:NormalTomo} (Fourier slice theorem), \gls{s0} points into the $z$-direction when $\gls{phi0}=0$. This means that $\gls{s0} = (-\sin\gls{phi0}, \, \cos\gls{phi0})$.
We rewrite the Green's function accordingly.
\begin{align}
\label{eq:2Dgreen}
\gls{G} &= \frac{i}{4\pi} \int \!\! dp \frac{1}{M} 
\exp\! \left\lbrace i \gls{km} \left[ p(x-x') + M(z-z') \right] \right\rbrace
\end{align}
In our notation, the incoming plane wave \gls{u0} with amplitude $a_0$, propagation direction \gls{s0}, and wavenumber \gls{km} can be written as 
\begin{align}
\gls{u0} = a_0 \exp(i \gls{km} \gls{s0} &\vc{r}).
\end{align}
The first Born approximation in two dimensions then reads (see \hyref{section}{sec:Born}).
\begin{align}
\label{eq:2Dborn}
\gls{ub} = \iint \!\! d^2r' 
\gls{G} f(\vc{r'}) u_0(\vc{r'})
\end{align}
We define the two-dimensional Fourier transform~\gls{F} of the scattering potential~\gls{f}
\begin{align}
\gls{F} &= \frac{1}{2 \pi} \iint \!\! d^2r \, \gls{f} \exp(-i\vc{kr}) \\
\gls{f} &= \frac{1}{2 \pi} \iint \!\! d^2k \, \gls{F} \exp(i\vc{kr})
\end{align}
and we keep in mind the identity of the Dirac delta distribution for any given $p$ and $a$
\begin{align}
\delta(p-a) =
\frac{1}{2 \pi} \int \!\! dx \,
\exp(i(p-a)x).
\end{align}
We now insert \hyref{equation}{eq:2Dgreen} into \hyref{equation}{eq:2Dborn}.
\begin{align}
\gls{ub} =  \frac{i}{4\pi} \iint \!\! d^2r'  \int \!\! dp \,
\frac{1}{M} \exp \! \left[ 
i \gls{km} p(x-x') + i \gls{km} M(z-z') \right] \times \notag \\
 \times f(\vc{r'}) a_0
\exp(i \gls{km}(p_0x' + M_0z'))
\end{align}
The integral over $r'$ can be replaced with the Fourier transform of $f(\vc{r'})$.
\begin{align}
\widehat{F}(\gls{km}(\vc{s} - \gls{s0})) = 
\frac{1}{2\pi}  \iint d^2r' f(\vc{r'})
\exp( -i\gls{km}(\vc{s} - \gls{s0}) \vc{r'} )
\end{align}
The equation for the Born approximation now reads
\begin{align}
\gls{ub} =  \frac{ia_0}{4\pi} \int \!\! dp \,
\frac{2 \pi}{M} 
\widehat{F}(\gls{km}(\vc{s} - \gls{s0}))
\exp(i \gls{km}\vc{sr}).
\end{align}
We measure the field at the detector line  $\vc{r} \rightarrow \vc{r_D} = (x_\mathrm{D}, \gls{lD})$ and at the angle \gls{phi0}. Here, \gls{lD} is the distance of the detector plane from the rotational center of the sample.
\begin{align}
\label{eq:2Dborn2}
u_{\mathrm{B},\phi_0}(\vc{r_D}) =  \frac{ia_0}{2} \int \!\! dp \,
\frac{1}{M} 
\widehat{F}(\gls{km}(\vc{s} - \gls{s0}))
\exp(i \gls{km}\vc{sr_D})
\end{align}
The subscript \gls{phi0} denotes the angular position of the sample and the direction of the incoming plane wave with respect to the sample.
In order to arrive at the Fourier diffraction theorem in two dimensions, we perform a one-dimensional Fourier transform of $u_\mathrm{B}(\vc{r_D})$ along $x_\mathrm{D}$.
\begin{align}
\widehat{U}_{\mathrm{B},\phi_0}(k_\mathrm{Dx}) =  \frac{ia_0}{2 \sqrt{2\pi}} 
\int \!\! dx_\mathrm{D} \,
\int \!\! dp \,
\frac{1}{M} 
\widehat{F}(\gls{km}(\vc{s} - \gls{s0}))
\exp(i \gls{km}(p x_\mathrm{D} + M \gls{lD})) \times \notag \\
\times \exp(- i k_\mathrm{Dx} x_\mathrm{D})
\end{align}
We now identify the delta distribution
\begin{align}
\delta(\gls{km}p - k_\mathrm{Dx}) = 
\frac{1}{2 \pi} \int \!\! dx_\mathrm{D} \,
\exp(i (\gls{km} p - k_\mathrm{Dx})x_\mathrm{D})
\end{align}
which in turn we can use to solve the integral over $dp$. \newline
\noindent Note that ${\delta(\gls{km}p - k_\mathrm{Dx}) = \frac{1}{\left| \gls{km} \right|} \delta(p - k_\mathrm{Dx}/\gls{km})}$. 
\begin{align}
\widehat{U}_{\mathrm{B},\phi_0}(k_\mathrm{Dx}) =  \frac{ia_0 2\pi}{2 \sqrt{2\pi}} 
\int \!\! dp \,
\frac{1}{M} 
\widehat{F}(\gls{km}(\vc{s} - \gls{s0}))
\exp(i \gls{km} M \gls{lD})
\delta(\gls{km}p - k_\mathrm{Dx})
\end{align}
Solving the integral implies replacing all occurrences of $p$ with $\frac{k_\mathrm{Dx}}{\gls{km}}$.
\begin{align}
\widehat{U}_{\mathrm{B},\phi_0}(k_\mathrm{Dx}) =
  \frac{ia_0 \pi}{ \sqrt{2\pi} \gls{km}} 
\frac{1}{\sqrt{1-\left(\frac{k_\mathrm{Dx}}{\gls{km}}\right)^2}} 
\widehat{F}(\gls{km}(\vc{s} - \gls{s0}))
\exp \! \left(i \gls{km} \sqrt{1-\left(\frac{k_\mathrm{Dx}}{\gls{km}}\right)^2} \gls{lD}\right)
\end{align}
Finally, we arrive at the 2D Fourier diffraction theorem
\begin{align}
\label{eq:2DFDTh}
\widehat{U}_{\mathrm{B},\phi_0}(k_\mathrm{Dx}) = &
\frac{ia_0}{\gls{km}}
\sqrt{\frac{\pi}{2}} 
\frac{1}{M} 
\widehat{F}(\gls{km}(\vc{s} - \gls{s0}))
\exp \! \left(i \gls{km} M \gls{lD}\right).
\end{align}
Note that we have used the substitution $M=\sqrt{1-\left(\frac{k_\mathrm{Dx}}{\gls{km}}\right)^2}$ to simplify the expression.
Solving for the Fourier transformed object $\widehat{F}$ yields
\begin{align}
\widehat{F}(\gls{km}(\vc{s} - \gls{s0})) = &
- \sqrt{\frac{2}{\pi}} 
\frac{i \gls{km}}{a_0} M
\widehat{U}_{\mathrm{B},\phi_0}(k_\mathrm{Dx})
\exp \! \left(-i \gls{km} M \gls{lD}\right). \label{eq:2DFDTF}
\end{align}
The restriction in \hyref{equation}{eq:2DM}, rewritten in \hyref{equation}{eq:2DMkm}, forces the one-dimensional Fourier transform of the scattered wave $\widehat{U}_{\mathrm{B},\phi_0}(k_\mathrm{Dx})$ to be placed on circular arcs in Fourier space:
\begin{align}
\gls{km} \vc{s} =  
 (k_\mathrm{Dx} \cos\gls{phi0} - \gls{km}&M \sin\gls{phi0}\scalebox{1.4}{,}\,
  k_\mathrm{Dx} \sin\gls{phi0} +  \gls{km}M \cos\gls{phi0})\\
\gls{km} &M  =  \sqrt{\gls{km}^2 - k_\mathrm{Dx}^2} \label{eq:2DMkm}
\end{align}
The argument $\gls{km}(\vc{s} - \gls{s0})$ shifts the circular arcs in Fourier space such that $\widehat{U}_{\mathrm{B},\phi_0}(0)$ is centered at $\widehat{F}(0,0)$.

\subsubsection*{Comparison to the Fourier Slice Theorem}
\hyref{Equation}{eq:2DFDTF} describes the Fourier diffraction theorem in 2D. When compared to the Fourier slice theorem from \hyref{equation}{eq:FSTh} in \hyref{section}{sec:FSTh}, a few differences become apparent.
We can write \hyref{equation}{eq:2DFDTF} in the same manner as \hyref{equation}{eq:FSTh}, with the subscript \gls{phi0} denoting the rotation of the object \gls{f}. The main differences are a complex factor and a different distribution of the data in Fourier space, as illustrated in \hyref{table}{tab:2DSliceDiff}.

\begin{table}[ht]
\centering
\begin{tabular}{l|c|c|}
 & \multicolumn{2}{c}{\parbox{0.65\linewidth}{
    \begin{align*}         
       \widehat{F}_{\phi_0}(k_\mathrm{x}, k_\mathrm{z}) = A \cdot 
       \sqrt{\frac{1}{2\pi}} \widehat{P}_{\phi_0}(k_\mathrm{Dx})
	\end{align*}
	}} \\
 & 
 \parbox[][3em][c]{0.41\linewidth}{\centering
 	\textbf{Fourier Slice Theorem} \newline (\hyref{equation}{eq:FSTh})}  &
 \parbox[][3em][c]{0.41\linewidth}{\centering
 	\textbf{Fourier Diffraction Theorem} \newline (\hyref{equation}{eq:2DFDTF})} \\ 
\hline 
\parbox{0.15\linewidth}{
    Sinogram \\ $\widehat{P}_{\phi_0}  (k_\mathrm{Dx})$} &
\parbox[][5em][c]{0.31\linewidth}{\hspace*{.5em}\centering
    Fourier transform of~projections
    $\widehat{P}_{\phi_0}  (k_\mathrm{Dx})$} &
\parbox{0.39\linewidth}{\centering
    Fourier transform of~complex scattered waves
    $\widehat{U}_{\mathrm{B},\phi_0}(k_\mathrm{Dx})$} \\ 
\hline 
\parbox{0.15\linewidth}{
	Factor \\ $A$} &
$A=1$ & 
\parbox{0.30\linewidth}{
	\begin{align*}
	A=
    -\frac{2 i \gls{km} M }{a_0}
    \exp \! \left(-i \gls{km} M \gls{lD}\right)
    \end{align*}
    } \\
\hline
\parbox{0.15\linewidth}{
	Coordinates \\ $(k_\mathrm{x}, k_\mathrm{z})$ \\
	sliced at $\gls{phi0}$} &
\parbox{0.30\linewidth}{ 
	\begin{align*}
	  k_\mathrm{x} &= k_\mathrm{Dx} \\
      k_\mathrm{z} &= k_\mathrm{t} = 0 \\
      &\text{(straight line)}
	\end{align*}
	} &
\parbox{0.35\linewidth}{ 
	\begin{align*}
	  k_\mathrm{x} &= k_\mathrm{Dx} \\
      k_\mathrm{z} &= \sqrt{\gls{km}^2 - k_\mathrm{Dx}^2} - \gls{km} \\
      &\text{(semicircular arc)}
	\end{align*}
	} \\
\hline
\parbox{0.13\linewidth}{
	Fourier space~\gls{F} coverage (180$^\circ$)} &
\parbox{0.38\linewidth}{ 
	\includegraphics[width=\linewidth]{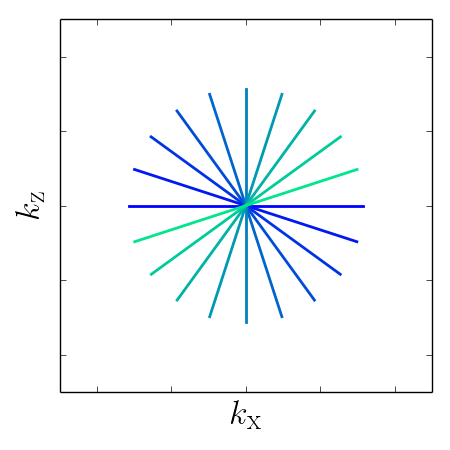}
	} &
\parbox{0.38\linewidth}{
	\includegraphics[width=\linewidth]{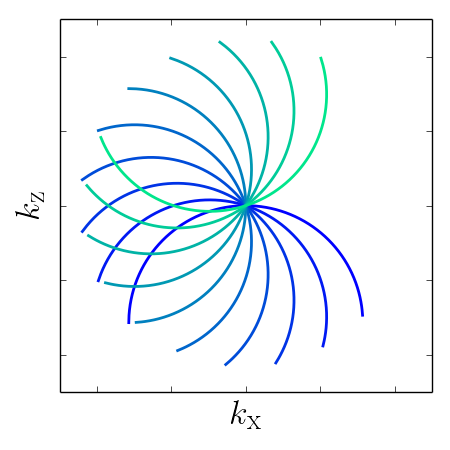}
	} \\
\hline
\end{tabular} 
\mycaption{Classical versus diffraction tomography}{The equation above the table combines the Fourier slice theorem and the Fourier diffraction theorem, pointing out their common structure. However, there are two major differences between the theorems.  \textbf{(1)}~Diffraction tomography data are multiplied by a complex factor $A \neq 1$. \textbf{(2)}~For diffraction tomography, the Fourier transform of the obtained data is distributed along circular arcs (not straight lines) in Fourier space. The plots show a visualization in Fourier space for data that are sampled at a frequency of $\nicefrac{1}{\gls{km}}$ for angles between 0$^\circ$ and 180$^\circ$. \label{tab:2DSliceDiff}}
\end{table}

\FloatBarrier

\subsection{Backpropagation Algorithm}
\label{sec:2DBackprop}
The term \textit{backpropagation} comes from an interpretation of the mathematical formalism which is similar to the backprojection formula derived in \hyref{section}{sec:2DInvRadon}.
The backpropagation algorithm \cite{Kak2001} is a solution to the inverse scattering problem. The reconstructed scattering potential \gls{f} is directly computed from the input data \gls{Ub}. For this, we need to perform a coordinate transform from $(k_\mathrm{x},k_\mathrm{z})$ to $(k_\mathrm{Dx}, \gls{phi0})$. We start by computing the inverse two-dimensional Fourier transform of \hyref{equation}{eq:2DFDTF}.
\begin{align}
\widehat{F}(\gls{km}(\vc{s} - \gls{s0})) = &
- \sqrt{\frac{2}{\pi}} 
\frac{i \gls{km}}{a_0} M
\widehat{U}_{\mathrm{B},\phi_0}(k_\mathrm{Dx})
\exp \! \left(-i \gls{km} M \gls{lD}\right) \tag{\ref{eq:2DFDTF}} \\
\gls{f} = & 
- \sqrt{\frac{2}{\pi}} 
\frac{i \gls{km}}{2 a_0 \pi}
\iint \!\! dk_\mathrm{x} dk_\mathrm{z}  \,
M
\widehat{U}_{\mathrm{B},\phi_0}(k_\mathrm{Dx})
\exp \! \left(-i \gls{km} M \gls{lD}\right) \times \notag \\
      & \hphantom{{}- \sqrt{\frac{2}{\pi}} 
        \frac{i \gls{km}}{2 a_0 \pi^2}
        \iint \!\! dk_\mathrm{x} dk_\mathrm{z}  \,}
 \times \exp \! \left( i \gls{km}(\vc{s} - \gls{s0}) \vc{r} \right) \label{eq:2DFDSpace} \\
 (k_\mathrm{x}, k_\mathrm{z}) = & \gls{km} (\vc{s} - \gls{s0})
\end{align}
As described in \hyref{table}{tab:2DSliceDiff}, the input data are distributed along circular arcs in Fourier space. The orientation of these arcs is defined by the acquisition angle \gls{phi0} and is described by~$D_{\phi_0}$.
\begin{align}
D_{\phi_0} &= 
\begin{pmatrix}
\cos\gls{phi0} & -\sin\gls{phi0} \\
\sin\gls{phi0} &  \cos\gls{phi0} \\
\end{pmatrix} \\
\vc{k} &= D_{\phi_0} \vc{k'}
\end{align}
Here, $\vc{k}$ denotes the non-rotated Fourier space, whereas $\vc{k'}$ denotes the positions of acquisition at a certain angle \gls{phi0}. At \gls{phi0} = 0, \vc{k} and \vc{k'} coincide. We have defined the angle~\gls{phi0} such that, $k'_\mathrm{x} =k_\mathrm{Dx}$ and therefore $k'_\mathrm{z} = \sqrt{\gls{km}^2 - k_\mathrm{Dx}^2} - \gls{km}$. The coordinate transform from $(k_\mathrm{x},k_\mathrm{z})$ to $(k_\mathrm{Dx}, \gls{phi0})$ is now fully described.
\begin{align}
k_\mathrm{x} &=  k_\mathrm{Dx} \cos\gls{phi0}
			   -  \left[\sqrt{\gls{km}^2 - k_\mathrm{Dx}^2} - \gls{km} \right] \sin\gls{phi0}  \label{eq:2Dsoperp1} \\
k_\mathrm{z} &= k_\mathrm{Dx} \sin\gls{phi0} 
			   +  \left[\sqrt{\gls{km}^2 - k_\mathrm{Dx}^2} - \gls{km} \right] \cos\gls{phi0} \label{eq:2Dsoperp2}
\end{align}
In order to replace the integrals over $k_\mathrm{x}$ and $k_\mathrm{z}$ with the integrals over $k_\mathrm{Dx}$ and \gls{phi0}, we compute the Jacobian matrix $J$ and its determinant.
\begin{align}
J &= \frac{\partial k_\mathrm{x} \partial k_\mathrm{z}}{\partial k_\mathrm{Dx} \partial \gls{phi0}} \\
  &=
\begin{pmatrix}
   \cos\gls{phi0}  + \frac{k_\mathrm{Dx}}{\sqrt{\gls{km}^2 - k_\mathrm{Dx}^2}} \sin\gls{phi0} & 
   			   -	 k_\mathrm{Dx} \sin\gls{phi0}
			   - \left[\sqrt{\gls{km}^2 - k_\mathrm{Dx}^2} - \gls{km} \right] \cos\gls{phi0} \\
   \sin\gls{phi0}  - \frac{k_\mathrm{Dx}}{\sqrt{\gls{km}^2 - k_\mathrm{Dx}^2}} \cos\gls{phi0} & 
				k_\mathrm{Dx} \cos\gls{phi0} 
			   - \left[\sqrt{\gls{km}^2 - k_\mathrm{Dx}^2} - \gls{km} \right] \sin\gls{phi0}
\end{pmatrix}
\end{align}
The determinant of the Jacobian $J$ computes to 
\begin{align}
\det(J) &= k_\mathrm{Dx} - \left(
 				k_\mathrm{Dx} - \frac{\gls{km} k_\mathrm{Dx}}{\sqrt{\gls{km}^2 - k_\mathrm{Dx}^2}}
 					\right) \\
 	&= \frac{\gls{km} k_\mathrm{Dx}}{\sqrt{\gls{km}^2 - k_\mathrm{Dx}^2}}.
\end{align}
We insert the coordinate transform into \hyref{equation}{eq:2DFDSpace} and obtain the backpropagation formula.
\begin{align}
\gls{f} = 
- \sqrt{\frac{2}{\pi}} 
\frac{i \gls{km}}{2 a_0 \pi}
\int \!\! dk_\mathrm{Dx} \, \frac{1}{2} \int_0^{2 \pi} \!\!  d\gls{phi0} \,
\left| \frac{\gls{km} k_\mathrm{Dx}}{\sqrt{\gls{km}^2 - k_\mathrm{Dx}^2}} \right |
M
\widehat{U}_{\mathrm{B},\phi_0}(k_\mathrm{Dx})
\exp \! \left(-i \gls{km} M \gls{lD}\right) \times \notag \\
\times \exp \! \left( i \gls{km}(\vc{s} - \gls{s0}) \vc{r} \right)
\end{align}
Note that the integration over \gls{phi0} goes from $0$ to $2\pi$. This is necessary because the Fourier space needs to be covered homogeneously by the projection data. A coverage of only 180$^\circ$ leads to an incomplete coverage in Fourier space as depicted in \hyref{table}{tab:2DSliceDiff}. This aspect is important to keep in mind for later experimental realization. The necessary double-coverage in Fourier space leads to the additional factor $\nicefrac{1}{2}$. Furthermore, we express $(\vc{s-s_0})$ in terms of a lateral (\vc{t_\perp}) and an axial (\gls{s0}) component (eq. \ref{eq:2Dsoperp1} and \ref{eq:2Dsoperp2}).
\begin{align}
\gls{km} (\vc{s-s_0}) &= k_\mathrm{Dx} \, \vc{t_\perp} + \gls{km}(M - 1) \, \gls{s0} \\
\vc{s_0} &= \left(p_0 , \, M_0 \right) = (-\sin\gls{phi0}, \, \cos\gls{phi0})\\
\vc{t_\perp} &= \left(- M_0 , \, p_0 \right) =  (\cos\gls{phi0}, \, \sin\gls{phi0})
\end{align}
By assuming that $(\gls{km}M)^2 = \gls{km}^2 - k_\mathrm{Dx}^2 \overset{!}{>} 0$, we can rewrite the backpropagation formula as
\begin{align}
\gls{f} = 
- \frac{i \gls{km}}{a_0 (2\pi)^{3/2}}
\int \!\! dk_\mathrm{Dx} \int_0^{2 \pi} \!\!  d\gls{phi0} \,
\left| k_\mathrm{Dx} \right |  
\widehat{U}_{\mathrm{B},\phi_0}(k_\mathrm{Dx})
\exp( -i \gls{km} M \gls{lD} ) \times \notag \\
\times \exp \! \left[i (k_\mathrm{Dx} \, \vc{t_\perp} + \gls{km}(M - 1) \, \gls{s0})\vc{r}\right].
\label{eq:2DBackprop}
\end{align}
\hyref{Figure}{fig:FDT2D} illustrates the acquisition and reconstruction process for a non-centered cylinder. The data were computed according to Mie theory. Note that the reconstruction with the Born approximation is not resembling the original structure of the cylinder and that the reconstruction quality with the Rytov approximation is dependent on the total number of projections that were acquired for the sinogram. Also note that the main contribution to the reconstructed refractive index distribution comes from the measured phase. The measured amplitude only slightly influences the reconstruction. Details of the simulation and reconstruction process can be found in \hyref{figure}{fig:FDT2D_cross-section}.

\begin{figure}
\centering
  \subfloat[][phantom, \gls{ndelt}]
  {\includegraphics[width=\linewidth/3]{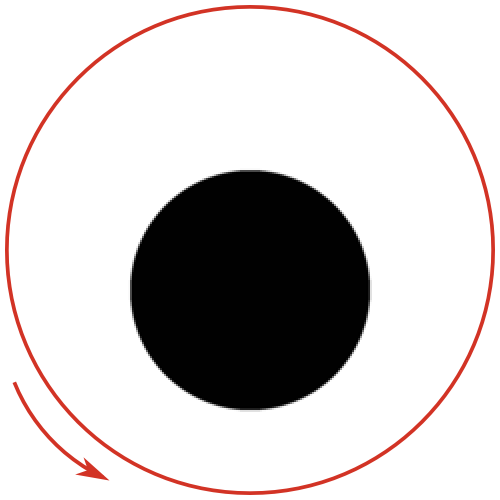}}
  \subfloat[][sinogram amplitude] 
  {\includegraphics[width=\linewidth/3]{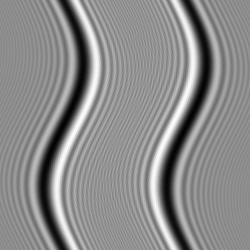}}
  \subfloat[][sinogram phase]  
  {\includegraphics[width=\linewidth/3]{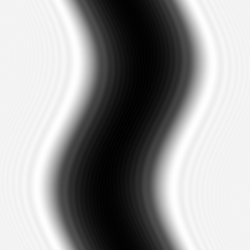}} \\
  \subfloat[][Born (250 projections)]
  {\includegraphics[width=\linewidth/3]{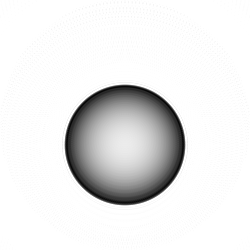}} 
  \subfloat[][Rytov (50 projections)]
  {\includegraphics[width=\linewidth/3]{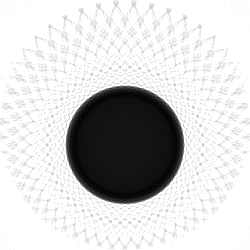}}
  \subfloat[][Rytov (250 projections)]
  {\includegraphics[width=\linewidth/3]{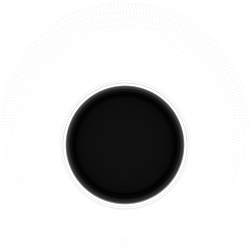}}\\

\mycaption{2D Backpropagation}{Image (a) shows the refractive index distribution of a dielectric cylinder. White indicates the refractive index of the medium and black the higher refractive index of the cylinder. The forward-scattering process was computed according to Mie theory. Because the cylinder is not centered, the characteristic shape of the sinogram is visible in amplitude (b) and phase (c). 
The lower images show the reconstruction of the refractive index with (d) the Born approximation using 250 projections, (e) the Rytov approximation using 50 projections, and (f) the Rytov approximation with a total of 250 projections. See \hyref{figure}{fig:FDT2D_cross-section} for details.
\label{fig:FDT2D}}
\end{figure}

\begin{figure}
\centering
  \includegraphics[width=\linewidth]{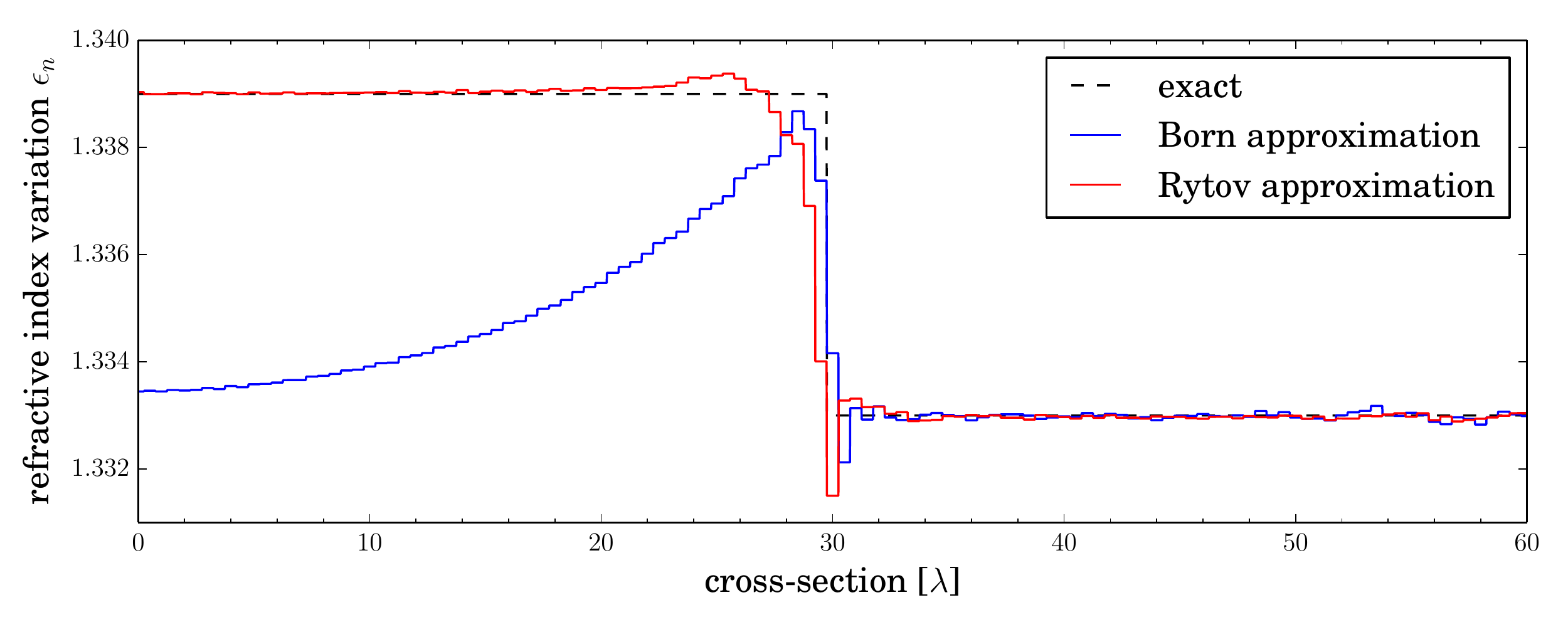}
\mycaption{2D Backpropagation cross-sections}{The refractive index of the medium is $\gls{nm} = 1.333$ and the local variation inside the cylinder is $\gls{ndelt} = \gls{n} - \gls{nm} = 0.006$.
The radius of the cylinder is $30 \gls{lambda}$ (vacuum wavelength \gls{lambda}). 
The scattered wave is computed according to Mie theory at an optical distance of $l_\mathrm {D} = 60 \gls{lambda}$ from the center of the cylinder and sampled at $\gls{lambda}/2$ over 250 pixels. The Mie theory computations are based on \cite{Zhu11}.
The refractive index map is reconstructed on a grid of $250\times 250$ pixels from 250 projections over $2 \pi$. The reconstruction was performed with the Python library ODTbrain \cite{Mueller15}.
\label{fig:FDT2D_cross-section}}
\end{figure}

\FloatBarrier
\clearpage
\subsubsection*{Comparison to Backprojection}
The main differences between the backprojection and backpropagation algorithms (dependencies on \gls{lD} and~\gls{s0}) are summarized in \hyref{table}{tab:2DProjProp}.
\begin{table}[ht]
\centering
\begin{tabular}{l|c|c|}
 & \multicolumn{2}{c}{\parbox{0.80\linewidth}{
    \begin{align*}         
       f(x,y) = \frac{A}{(2\pi)^{3/2}} 
       \int \!\! dk_\mathrm{Dx} \int_0^{2 \pi} \!\!  d\gls{phi0} \,
		\exp(iB)
       \left| k_\mathrm{Dx} \right |  
       \widehat{P}_{\phi_0}(k_\mathrm{Dx})
	\end{align*}
	}} \\
 & \parbox[][3em][c]{0.38\linewidth}{\centering
    \textbf{Backprojection} \\(\hyref{equation}{eq:Backproj})}
 & \parbox{0.38\linewidth}{\centering
 \textbf{Backpropagation}\\(\hyref{equation}{eq:2DBackprop})} \\ 
\hline 
\parbox{0.12\linewidth}{
    Sinogram \\ $\widehat{P}_{\phi_0}  (k_\mathrm{Dx})$} &
\parbox[][5em][c]{0.38\linewidth}{\hspace*{.5em}\centering
    Fourier transform of~projections
    $\widehat{P}_{\phi_0}  (k_\mathrm{Dx})$} &
\parbox{0.38\linewidth}{\centering
    Fourier transform of~complex~scattered~waves \\
    $\widehat{U}_{\mathrm{B},\phi_0}(k_\mathrm{Dx})$} \\ 
\hline 
\parbox{0.12\linewidth}{
	Factor \\ $A$} &
\parbox{0.38\linewidth}{ \centering
$A=\frac{1}{2}$ \\ (double coverage)} & 
\parbox{0.38\linewidth}{
	\begin{align*}
	A=
    -\frac{i \gls{km} }{a_0}
    \end{align*}
    } \\
\hline
\parbox{0.12\linewidth}{
	Exponent \\ $B$} &
\parbox{0.38\linewidth}{ 
	\begin{align*}
	  B &= k_\mathrm{Dx}(\vc{t_\perp r} ) \\
	  \vc{t_\perp} &= (\cos \gls{phi0}, \, \sin \gls{phi0}) \\
	\end{align*}
	} &
\parbox{0.38\linewidth}{ 
	\begin{align*}
	  B &= - \gls{km} M \gls{lD} + k_\mathrm{Dx}\vc{t_\perp r}\, + \\
	  	& \hspace{3em} + \gls{km}(M - 1)\gls{s0}\vc{r} \\
	  \vc{t_\perp} &= (\cos \gls{phi0}, \, \sin \gls{phi0}) \\
	  \gls{s0} &= (-\sin \gls{phi0}, \, \cos \gls{phi0}) 
	\end{align*}
	} \\
\hline
\end{tabular} 
\mycaption{Backprojection versus backpropagation}{The equation above the table illustrates the similar structure of the backprojection and the backpropagation formula. Nevertheless, the backpropagation formula contains an exponent $B$, resulting from the first Born approximation, that depends on \gls{lD} and \gls{s0} which increases the complexity (See also \hyref{table}{tab:2DSliceDiff}). Note that the backprojection formula has a factor of $\nicefrac{1}{2}$ due to the double coverage over \gls{phi0}. The necessity of the $2\pi$-coverage (360$^\circ$) for the backpropagation algorithm is illustrated by the visualization of the Fourier slice and diffraction theorems in the figures of \hyref{table}{tab:2DSliceDiff}. \label{tab:2DProjProp}}
\end{table}

%% file: chapter5_three-dimensional_diffraction_tomography.tex
\section{Three-Dimensional Diffraction Tomography}
\label{sec:ReconDiffrObj}
Two-dimensional (2D) diffraction tomography is valid in three dimensions (3D) only for infinitely elongated objects (i.e. cylinders). For objects that exhibit inhomogeneities along the third dimension, a 3D theory needs to be applied. The 2D and the 3D theory are very similar. There are only a few dimension-specific differences like the position of the Fourier transformed fields along spherical surfaces (3D) instead of circular arcs~(2D). This section uses the previous sections as a basis to introduce and discuss the 3D version of diffraction tomography. The notation stays the same, except that a third axis $y$ is introduced in every vector. As in \hyref{section}{sec:2dODT}, we conclude by testing the 3D backpropagation algorithm with data computed according to Mie theory. 

\subsection{Fourier Diffraction Theorem}
There are at least two ways to derive the 3D Fourier diffraction theorem. The first uses a double integral representation of the Green's function. The second makes use of the convolution theorem that connects the convolution of two functions with their product in the Fourier domain. We will thoroughly investigate the first and only briefly discuss the second method.

The Green's function (\hyref{equation}{eq:greendef}) of the Helmholtz equation can be rewritten using the double integral representation as shown by Banos et. al\footnote{Banos' notation does not include the prefactor $1/(4\pi)$ in the Green's function (see \hyref{eq}{eq:greendef}). Therefore, the prefactor in \hyref{equation}{eq:GreenBanos} is $1/(8\pi^2)$ and not $1/(2\pi)$.} (\cite{Banos1966}, section 2.11). 
\begin{align}
\gls{G} = \frac{i\gls{km}}{8\pi^2} \iint \!\! dpdq \frac{1}{M}& \exp\! \left\lbrace i \gls{km} \left[ p(x-x') + q(y-y') + M(z-z') \right] \right\rbrace
\label{eq:GreenBanos} \\
M& = \sqrt{1-p^2-q^2} \\
\gls{s}& = (p, q, M)
\end{align}
Here, we define $\gls{km}\gls{s}$ as the wave vector of a plane wave with the wave number \gls{km}. Note that we have introduced the coordinate $q$ in addition to $p$ from the 2D case. In order to keep $M$ real, we now have the restriction $p^2 + q^2 \le 1$. If this restriction was violated, we would allow evanescent waves which would complicate the inversion process \cite{Kak2001}.
An incoming plane wave has the normal vector \gls{s0} that is defined by
\begin{align}
\gls{u0} =  a_\mathrm{0} \exp(i \gls{km} \gls{s0}& \vc{r}) \\
\text{with~~~} \gls{s0}& = (p_\mathrm{0}, q_\mathrm{0}, M_\mathrm{0}).
\end{align}
In \hyref{section}{sec:Born} we showed that the Born approximation of \gls{us} reads
\begin{align}
\gls{ub} =  \iiint \!\!d^3r' \, \gls{G} f(\vc{r'}) u_\mathrm{0}(\vc{r'}).
\end{align}
We define the unitary Fourier transform\footnote{Please note that other authors, e.g. Devaney \cite{Devaney1981} may use the non-unitary Fourier transform. The prefactors ($1/(2\pi)^{3/2}$) may differ from our notation.} \gls{F} of \gls{f} in 3D as
\begin{align}
\gls{F} & = \frac{1}{(2\pi)^{3/2}} \iiint \!\!d^3r \, \gls{f} \exp(- i \, \vc{kr}) \\
\gls{f} & = \frac{1}{(2\pi)^{3/2}} \iiint \!\!d^3k \, \gls{F} \exp(i \, \vc{kr}).
\end{align}
Furthermore, we define the position of the detector such that its normal vector $\vc{n_D}$ is parallel to the incident plane wave vector $\gls{km}\gls{s0}$ ($\vc{n_D}=\gls{s0}$). This is a valid assumption, because we only rotate the cell and thus the spatial arrangement of incoming plane wave~\gls{u0} and  detector do not change.

\begin{landscape}
\noindent In the derivations that follow, we make use of the following definition of the Dirac delta function.
\begin{align}
\delta(x-a) = \frac{1}{2\pi} \int \!\!dp \, \exp(ip(x-a))
\end{align}

\noindent We insert the definitions of the Green's function and the incident plane wave into the formula for the Born approximation
\begin{align}
\gls{ub} = \frac{i a_\mathrm{0} \gls{km}}{8 \pi^2} 	
\iiint \!\! d^3r' \iint \!\! dpdq \,		
 \frac{1}{M}  f(\vc{r'})					
\exp\!\left\lbrace						
-i \gls{km} \left[ (p-p_\mathrm{0})x' + 
				   (q-q_\mathrm{0})y' +
				   (M-M_\mathrm{0})z'   \right]
	  \right\rbrace 
\cdot \exp( i \gls{km} \underbrace{(px +qy + Mz)}_{\gls{s}\vc{r}}).
\end{align}
We may interpret the integral over $\vc{r'}$ as the Fourier transform of \gls{f} that is shifted by $-\gls{km}\gls{s0}$ in Fourier space
\begin{align}
\gls{ub} = 
\frac{i a_\mathrm{0} \gls{km}}{8 \pi^2} (2\pi)^{3/2} 		
\iint \!\! dpdq	\,					        
\frac{1}{M}      							
\widehat{F}(	\underbrace{						
			\gls{km}(p-p_\mathrm{0}),
			\gls{km}(q-q_\mathrm{0}),
			\gls{km}(M-M_\mathrm{0}) 
			}_{\gls{km}(\gls{s} -\gls{s0})})
\cdot \exp( i \gls{km} (px +qy + Mz)).
\end{align}
Any vector $\vc{r_D} = (x_\mathrm{D}, y_\mathrm{D}, z_\mathrm{D})$ in the detector plane  is defined by  $\vc{n_D}(\vc{r} - \vc{r_D}) = 0$, where $\vc{n_D}$ is the normal vector on the detector plane.  
We previously placed our detector according to $\vc{n_D}=\gls{s0}$, which implies a rotation of the sample perpendicular to the propagation direction \gls{s0} of the incident plane wave \gls{u0}. Thus, $z_\mathrm{D}=\gls{lD}$ is a constant and describes the distance of the detector from the center of the rotation axis. The detected field at the detector plane is thus
\begin{align}
\left. \gls{ub} \right|_D =  u_{\mathrm{B},\phi_0}(\vc{r_D}) = 
\frac{i \pi a_\mathrm{0} \gls{km}}{(2\pi)^{3/2}}  		
\iint \!\! dpdq	\,					        			
\frac{1}{M}      										
\widehat{F}(	\gls{km}(\gls{s} -\gls{s0}))     		
\cdot \exp( i \gls{km} (px_\mathrm{D} +qy_\mathrm{D} + M\gls{lD})).
\end{align}
Now we compute the two-dimensional Fourier transform of the recorded data $u_{\mathrm{B},\phi_0}(\vc{r_D})$ in the detector plane ($x_\mathrm{D}$,$y_\mathrm{D}$). 
\begin{align}
\widehat{U}_{\mathrm{B},\phi_0}(\vc{k_D}) = 
\frac{i \pi a_\mathrm{0} \gls{km}}{(2\pi)^{5/2}} 		
\iint \!\! dx_\mathrm{D}dy_\mathrm{D}					
 \iint \!\! dpdq	\, 				 	         		
\frac{1}{M}      										
\widehat{F}(	\gls{km}(\gls{s} -\gls{s0}))     		
\cdot \exp( i \gls{km} (px_\mathrm{D} +qy_\mathrm{D} + M\gls{lD})) \cdot \exp(-i(k_\mathrm{Dx}x_\mathrm{D} + k_\mathrm{Dy}y_\mathrm{D})
\end{align}
Note that since ${k_\mathrm{Dx}^2 + k_\mathrm{Dy}^2 + k_\mathrm{Dz}^2 = \gls{km}^2}$, i.e. there is no inelastic scattering, $\widehat{U}_{\mathrm{B},\phi_0}(\vc{k_D})$ must be on a  surface of a semi-sphere in Fourier space. When we combine the exponential functions that contain the components of $\vc{r_D}$, we can identify the two-dimensional Dirac delta function
\begin{align}
\iint \!\! dx_\mathrm{D} dy_\mathrm{D} \,
\exp(ix_\mathrm{D}(\gls{km}p-k_\mathrm{Dx}) +
     iy_\mathrm{D}(\gls{km}q-k_\mathrm{Dy})) =
(2\pi)^2 \delta(\gls{km}p-k_\mathrm{Dx}, \gls{km}q-k_\mathrm{Dy})).
\end{align}
We can now evaluate the integral over $p$ and $q$.
\begin{align}
\widehat{U}_{\mathrm{B},\phi_0}(\vc{k_D}) &= 
\frac{i \pi a_\mathrm{0} \gls{km}}{(2\pi)^{1/2}}	
\iint \!\! dpdq	\, 				 	            	
\frac{1}{M}      							    	
\widehat{F}(	\gls{km}(\gls{s} -\gls{s0}))    	
\cdot \exp(i\gls{km}M\gls{lD})
\cdot \delta(\gls{km}p-k_\mathrm{Dx}, \gls{km}q-k_\mathrm{Dy})) \\
\widehat{U}_{\mathrm{B},\phi_0}(\vc{k_D}) &= 
\frac{i \pi a_\mathrm{0} }{(2\pi)^{1/2} \gls{km}}	
\iint \!\! dpdq	\, 				 	            	
\frac{1}{M\!(p,q)}      							
\widehat{F}(	\gls{km}\left[\gls{s}(p,q) - \gls{s0}\right])  
\cdot \exp(i\gls{km}M\!(p,q)\gls{lD})
\cdot \delta\!\left(p-\frac{k_\mathrm{Dx}}{\gls{km}}, q-\frac{k_\mathrm{Dy}}{\gls{km}}\right) \\
\widehat{U}_{\mathrm{B},\phi_0}(\vc{k_D}) &= 
\frac{i  \pi a_\mathrm{0} }{(2\pi)^{1/2}}			
\frac{												
 \exp\!\left(i \gls{lD} \sqrt{\gls{km}^2-k_\mathrm{Dx}^2-k_\mathrm{Dy}^2}\right)
 }{													
 \sqrt{\gls{km}^2-k_\mathrm{Dx}^2-k_\mathrm{Dy}^2}
 }    						
\widehat{F}\left(k_\mathrm{Dx}-\gls{km}p_\mathrm{0}\scalebox{1.5}{,}
			k_\mathrm{Dy}-\gls{km}q_\mathrm{0}\scalebox{1.5}{,}
			\sqrt{\gls{km}^2-k_\mathrm{Dx}^2-k_\mathrm{Dy}^2} - 
			\gls{km} \sqrt{1-p_\mathrm{0}^2 -q_\mathrm{0}^2}\right)
\label{eq:FourierDifflong}
\end{align}
We used the identity of the delta function ${\delta(Ap) = \frac{1}{\left|A \right| } \delta(p)}$, and inserted the definition of $M=\sqrt{1-p^2-q^2}$ and ${\gls{s}=(p,q,M)}$.
\end{landscape}
\noindent We may write the short form of \hyref{equation}{eq:FourierDifflong} by setting 
\begin{equation}
k_\mathrm{Dx} = \gls{km}p  \text{~~and~~} k_\mathrm{Dy} = \gls{km}q \notag
\end{equation}
\begin{align}
\widehat{U}_{\mathrm{B},\phi_0}(\vc{k_D}) = 
\frac{i \pi a_\mathrm{0} }{(2\pi)^{1/2}}	
\frac{											
 \exp(i\gls{km}M \gls{lD})
 }{												
 \gls{km}M
 }    						
\widehat{F}\left(\vc{k_D} -\gls{km}\gls{s0}\right)
\label{eq:FfandUf}
\end{align}
When we solve this equation for $\widehat{F}$, we get \cite{Kak2001, Devaney1982336}\footnote{If the non-unitary Fourier transform and the sign of the Green's function is taken into account, the term computed by Devaney et al. \cite{Devaney1982336} (equation 48) differs by a factor of $1/(\gls{km}^2)$. This factor $\gls{km}^2$ comes from a different definition of the data: $\gls{f} = -\gls{km}^2 O(\vc{r})$ in that paper. }
\begin{align}
\widehat{F}\left(\vc{k_D} -\gls{km}\gls{s0}\right) = 
\underbrace{
- \sqrt{\frac{2}{\pi}} \frac{i M \gls{km}}{ a_\mathrm{0} }  	
 \exp(-i\gls{km}M \gls{lD})					
}_{\text{complex factor}}
\widehat{U}_{\mathrm{B},\phi_0}(\vc{k_D}).
\label{eq:UfandFf}
\end{align}

\subsection{Interpretation of the Fourier Diffraction Theorem}
\hyref{Equation}{eq:UfandFf} shows a direct relation between the Fourier transform of the measured wave (Born approximation) $\widehat{U}_{\mathrm{B},\phi_0}(\vc{k_D})$ and the Fourier transform of the inhomogeneity of the sample \gls{F}.

\subsubsection{Surface of a Semi-Sphere in Fourier Space}
\noindent A closer look at the arguments reveals that the information defined by $\widehat{U}_{\mathrm{B},\phi_0}(\vc{k_D})$ is distributed along a semi-spherical surface with radius $\gls{km}$ in Fourier space that is shifted by~$-\gls{km}\gls{s0}$, as is explained in the following.

The spherical surface is defined by the wave vector with the magnitude \gls{km}.
\begin{equation*}
\gls{km}^2 = k_\mathrm{Dx}^2+k_\mathrm{Dy}^2+k_\mathrm{Dz}^2
\end{equation*}
Because $k_\mathrm{Dx}$ and $k_\mathrm{Dy}$ are given by the size of the detector image, we are left with a restriction for $k_\mathrm{Dz}$. This restriction forces the data $\widehat{U}_{\mathrm{B},\phi_0}(\vc{k_D})$ to be placed on a spherical surface with radius \gls{km} in Fourier space. Thus, the two-dimensional Fourier transform is projected onto a semi-spherical surface as depicted in \hyref{figure}{fig:FourierDiffract}.

The shift in Fourier space is defined by the argument of $\widehat{F}\left(\vc{k_D} -\gls{km}\gls{s0}\right)$.
The information of the Fourier transform of \gls{ub} at the detector $\widehat{U}_\mathrm{B}(\vc{k_D})$ is shifted in Fourier space in the direction $-\gls{s0}$. The vector \gls{s0} is constant for a fixed value of \gls{phi0} and describes the propagation direction of the incident wave onto the sample. The magnitude of the shift is equal to \gls{km} which is the radius of the spherical surface described above.

\begin{figure}
\centering
  \myincludegraphics{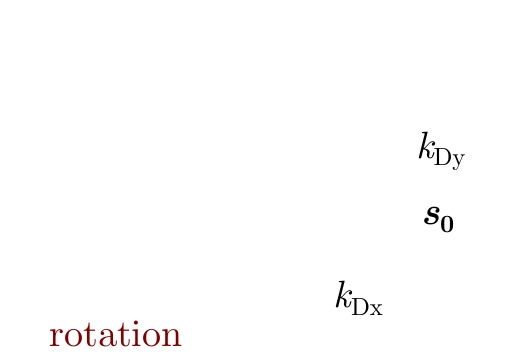}{jpg}
  \mycaption{Fourier diffraction theorem}{The data $\widehat{U}_{\mathrm{B},\phi_0}(\vc{k_D})$ (green) are projected onto a semi-sphere in Fourier space according to $\gls{km}^2 = k_\mathrm{Dx}^2+k_\mathrm{Dy}^2+k_\mathrm{Dz}^2$. The radius of the sphere is~\gls{km}. The surface of the sphere is oriented along the \gls{s0} axis in the case of coaxial illumination (detector aligned with incoming wave \gls{u0}). The axes of the two-dimensional Fourier transformed detector image are labeled $k_\mathrm{Dx}$ and $k_\mathrm{Dy}$. If the sample is rotated about one axis (red, here it is the \mbox{$y$-axis}) and the resolution of the setup is high enough (see sampling considerations below), then one obtains a horn torus-like shape in Fourier space (\hyref{fig.}{fig:HornTorus}).}
  \label{fig:FourierDiffract}
\end{figure}

\subsubsection{Sampling Considerations}
Diffraction tomography inherently increases and limits the resolution at the same time.
As a result of the data distribution on semi-spherical surfaces in Fourier space, the maximum value of $|\vc{k}| = \left|\vc{k_D} -\gls{km}\gls{s0}\right|$ is $\sqrt{2}\gls{km}$. Apparently, the resolution is increased by a factor of $\sqrt{2}$ when compared to the data measured. However, at the same time a maximum possible resolution is enforced by the restriction $|\vc{k}| \le \sqrt{2}\gls{km}$ in Fourier space. For the 2D case, the difference in resolution between projection tomography and diffraction tomography (maximum frequency in Fourier space) is visualized in \hyref{table}{tab:2DSliceDiff}.

It follows that all data distributed in Fourier space are located within a sphere of radius $\sqrt{2}\gls{km}$ according to
\begin{align}
k_\mathrm{Dx}^2 + k_\mathrm{Dy}^2 \leq \gls{km}^2.
\end{align}
This frequency-limit in Fourier space infers a resolution limit in real space. The maximum frequency in Fourier space computes to
\begin{align}
f_\mathrm{max}= \sqrt{2} \cdot \frac{\gls{km}}{2\pi} = \frac{\sqrt{2} \gls{nm}}{\gls{lambda}}.
\end{align}
In other words, the optical resolution is limited to $\nicefrac{\gls{lambda}}{\sqrt{2}\,\gls{nm}}$ and depends on the refractive index of the surrounding medium \gls{nm}.

\subsubsection{Comparison to two-dimensional diffraction tomography}
When comparing equations \ref{eq:2DFDTF} and \ref{eq:UfandFf}, it turns out that the Fourier diffraction theorem in two dimensions is similar to the Fourier diffraction theorem in three dimensions. \hyref{Table}{tab:2D3DDiffract} compares these two equations with respect to their dimensionality.
\begin{table}[H]
\centering
\begin{tabular}{l|c|c|}
 & \multicolumn{2}{c}{\parbox{0.65\linewidth}{
    \begin{align*}
       \widehat{F}(\vc{k}) = A \cdot 
       \sqrt{\frac{1}{2\pi}} \widehat{U}_{\mathrm{B}, \phi_0}(\vc{k_D})
	\end{align*}
	}} \\
 & \textbf{2D} (\hyref{equation}{eq:2DFDTF})&
 \parbox[][2em][c]{0.41\linewidth}{\centering
 	\textbf{3D} (\hyref{equation}{eq:UfandFf})} \\ 
\hline 
\parbox{0.15\linewidth}{
    Sinogram \\ $\widehat{U}_{\mathrm{B}, \phi_0}  (\vc{k_D})$} &
\parbox[][5em][c]{0.31\linewidth}{\hspace*{.5em}\centering
    Fourier transform of~complex scattered waves
    $\widehat{U}_{\mathrm{B},\phi_0}(k_\mathrm{Dx})$} &
\parbox{0.39\linewidth}{\centering
    Fourier transform of~complex scattered waves
    $\widehat{U}_{\mathrm{B},\phi_0}(k_\mathrm{Dx},k_\mathrm{Dy})$} \\ 
\hline 
\parbox{0.15\linewidth}{
	Factor \\ $A$} &
	\multicolumn{2}{c|}{\parbox{0.65\linewidth}{
	\begin{align*}
	A=
    -\frac{2 i \gls{km} M }{a_0}
    \exp \! \left(-i \gls{km} M \gls{lD}\right)
    \end{align*}
    }} \\
\parbox{0.15\linewidth}{
	} &
\parbox{0.30\linewidth}{ 
	\begin{align*}
	  M = \frac{1}{\gls{km}} \sqrt{\gls{km}^2 - k_\mathrm{Dx}^2}
	\end{align*}
	} &
\parbox{0.35\linewidth}{ 
	\begin{align*}
	  M = \frac{1}{\gls{km}} \sqrt{\gls{km}^2 - k_\mathrm{Dx}^2 - k_\mathrm{Dy}^2}		
	\end{align*}
	} \\
\hline
\parbox{0.15\linewidth}{
	Coordinates \\ \vc{k} at $\gls{phi0} = 0$} &
\parbox{0.30\linewidth}{ 
	\begin{align*}
	  \vc{k} &= (k_\mathrm{x}, k_\mathrm{z}) \\
	  k_\mathrm{x} &= k_\mathrm{Dx} \\
      k_\mathrm{z} &= \sqrt{\gls{km}^2 - k_\mathrm{Dx}^2} - \gls{km} \\
      &\text{(semicircular arc)}
	\end{align*}
	} &
\parbox{0.35\linewidth}{ 
	\begin{align*}
	  \vc{k} &= (k_\mathrm{x}, k_\mathrm{y}, k_\mathrm{z}) \\
	  k_\mathrm{x} &= k_\mathrm{Dx}, \hspace{1em} 	  k_\mathrm{y} = k_\mathrm{Dy} \\
      k_\mathrm{z} &= \sqrt{\gls{km}^2 - k_\mathrm{Dx}^2 - k_\mathrm{Dy}^2} - \gls{km} \\
      &\text{(semispherical surface)}
	\end{align*}
	} \\
\hline
\end{tabular} 
\mycaption{Comparison of 2D and 3D diffraction tomography}{Above the table is the generalized  Fourier diffraction theorem for 2D and 3D. The object \gls{f} is rotated about the $y$-axis, which is the axis pointing away from the 2D $(x,z)$-plane. The shape of the diffraction tomography formula is identical in 2D and in 3D. The only differences come from the different numbers of dimensions. For a comparison to the Fourier slice theorem, see \hyref{table}{tab:2DSliceDiff}. \label{tab:2D3DDiffract}}
\end{table}

\FloatBarrier

\subsubsection{Artifacts from Uniaxial Rotation}
\label{sec:FourierDiffrRelevance1axis}
Rotating the sample only about one axis results in missing-angle artifacts for 3D diffraction tomography. According to the Fourier diffraction theorem, the resulting reciprocal volume in Fourier space has a shape of a missing apple core as depicted in \hyref{figure}{fig:HornTorus} \cite{Vertu09}. This results in directional blurring along the axis of rotation ($y$).
In order to address directional blurring, regularization techniques need to be applied, as is described in e.g. \cite{Sung2009, Aganj2007, Ma14, LaRoque:08}.

\begin{figure}[ht]
\centering
  \myincludegraphics{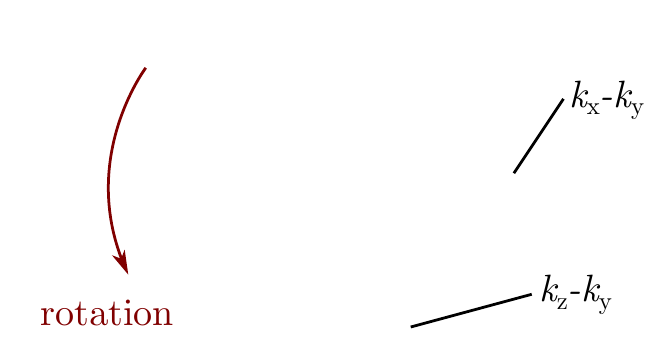}{jpg}
  \mycaption{Horn torus}{As a result of the Fourier diffraction theorem, images acquired perpendicularly to the sample-rotating axis ($y$) fill a horn torus-like shape in Fourier space. The resulting reconstruction exhibits missing angle artifacts, as shown in figures \ref{fig:FDT3D} and \ref{fig:FDT3D_cross-section}}.
  \label{fig:HornTorus}
 \end{figure}

\subsection{The Limes of the Rytov Approximation}
\label{sec:FDTandRytov}
In this section we show that diffraction tomography with the Rytov approximation converges to classical tomography when the wavelength \gls{lambda} becomes small \cite{Devaney1981}.
We start with the Fourier diffraction theorem which can be applied in combination with the Rytov approximation by calculating \gls{ub} from the measured scattered field $\gls{us} \approx \gls{ur}$ using \hyref{equation}{eq:ubornfromrytov}.
\begin{align}
\gls{ub} & = \gls{u0} \ln\!\left( \frac{a_\mathrm{0}}{a} \left[ \frac{\gls{ur}}{\gls{u0}} +1\right]\right) \tag{\ref{eq:ubornfromrytov}}\\
&= \gls{u0} \gls{pcr} \tag{\ref{eq:ubornfromrytovphase}}
\end{align}
The Fourier transform of the equation above at the detector $\vc{r} \rightarrow \vc{r_D}$ yields
\begin{align}
\widehat{U}_{\mathrm{B},\phi_0}(\vc{k_D}) & = \frac{1}{2\pi}\iint \!\! d^2 r_D \, \gls{u0} \varphi_{\mathrm{R},\phi_0}(\vc{r_D}) \exp(-i\vc{k_Dr_D}) \\
&= \frac{a_\mathrm{0}}{2\pi}\iint \!\! d^2r_D \, \varphi_{\mathrm{R},\phi_0}(\vc{r_D}) \exp(-i(\vc{k_D}-\gls{km}\gls{s0})\vc{r_D}) \\
&= a_\mathrm{0} \, \widehat{\varphi}_{\mathrm{R},\phi_0}(\vc{k_D} - \gls{km}\gls{s0})
\end{align}
where $\widehat{\varphi}_{\mathrm{R},\phi_0}(\vc{k_D})$ is the two-dimensional Fourier transform of $\varphi_{\mathrm{R},\phi_0}(\vc{r_D})$. For the Rytov approximation, \hyref{equation}{eq:UfandFf} then reads\footnote{Note that $\vc{k_D}$ is in a plane as $\vc{r_D}$ is defined in the detector plane. Therefore, $\widehat{\varphi}_{\mathrm{R}, \phi_0}(\vc{k_D}- \gls{km}\gls{s0})$ is a two-dimensional Fourier transform. However, $\widehat{F}\left(\vc{k_D} -\gls{km}\gls{s0}\right)$ is three-dimensional because it contains the $k_\mathrm{z}$ component.}
\begin{align}
\widehat{F}\left(\vc{k_D} -\gls{km}\gls{s0}\right) = 
- \sqrt{\frac{2}{\pi}}i M \gls{km}     		  	
 \exp(-i\gls{km}M \gls{lD})					
\widehat{\varphi}_{\mathrm{R}, \phi_0}(\vc{k_D}- \gls{km}\gls{s0}).
\end{align}
By replacing $\vc{k_D'} = \vc{k_D}- \gls{km}\gls{s0}$ one can rewrite this equation such that the argument of \gls{F} does not anymore contain the shift in Fourier space.
\begin{align}
\widehat{F}\left(\vc{k_D'}\right) = 
- \sqrt{\frac{2}{\pi}}i M^*& \gls{km}     		  	
 \exp(-i\gls{km}M^* \gls{lD})					
\widehat{\varphi}_{\mathrm{R}, \phi_0}(\vc{k_D'})
\label{eq:UfandFfRytov} \\
M^*& = \sqrt{1-(p'+p_\mathrm{0})^2 - (q'+q_\mathrm{0})^2} \\
\vc{k_D'}& = \gls{km}(p',q',M') \\
& = \gls{km}(p-p_\mathrm{0},q-q_\mathrm{0},M-M_\mathrm{0})
\end{align}
The data are still distributed on the surface of a semi-sphere in Fourier space. This information is hidden in the unintuitive restraint from $M^*$
\begin{align}
(p'+p_\mathrm{0})^2 + (q'+q_\mathrm{0})^2 \leq 1.
\end{align}
This expression can be simplified by noting that for $\gls{phi0}=0$, $\vc{k_D}$ is parallel to $\gls{s0}$ - our measuring $z$-direction ($M_0=1$) and therefore $p_\mathrm{0}=q_\mathrm{0}=0$ and
\begin{align}
p'^2 + q'^2 &\leq 1 \\
M^* &= \sqrt{1-p'^2 - q'^2} \\
\vc{k_D'} &= \gls{km}(p,q,M-1).
\end{align}
Note that the magnitude of the vector $\vc{k_D'}$ is not \gls{km}
\begin{align}
\left| \vc{k_D'} \right| = 2\gls{km}(1-M) \neq \gls{km}
\end{align}
and thus the restrictions for the $k_\mathrm{z}$ component of $\vc{k_D'}$ are different. It is shifted by~$-\gls{km}$ in the~$k_\mathrm{z}$-direction.
The shift in Fourier space is not obvious, but it is included in the restrictions for the $k_\mathrm{z}$ component of $\vc{k_D'}$.

In the short wavelength limit $\gls{lambda}$ approaches zero and thus \gls{km}, the radius of the semi-spherical surfaces in Fourier space, goes to infinity. The result is a data distribution that is identical to that of the Fourier slice theorem.
Furthermore, short wavelengths imply $k_\mathrm{x},k_\mathrm{y} \ll \gls{km}$ and we may write $M^*\rightarrow1$ \cite{Devaney1981} to obtain
\begin{align}
\widehat{F}\left(\vc{k_D'}\right) &= 
 - \sqrt{\frac{2}{\pi}}i \gls{km}     		  	
 \exp(-i\gls{km} \gls{lD})					
 \widehat{\varphi}_{\mathrm{R}, \phi_0}(\vc{k_D'})
\label{eq:RytPh1} \\
 \widehat{\varphi}_{\mathrm{R}, \phi_0}(\vc{k_D'}) &=
 \frac{1}{2\pi} \iint \!\! dx_\mathrm{D} dy_\mathrm{D}  \,
 \varphi_{\mathrm{R}, \phi_0}(\vc{r_D}) \exp(-i\vc{k_D'r_D})
\label{eq:RytPh2} \\
\widehat{F}\left(\vc{k_D'}\right) &= \frac{1}{(2\pi)^{3/2}}  
 \iiint \!\! dx dy dz \,
 f(\vc{r}) \exp(-i\vc{k_D'r}).
\end{align}
If we perform a two-dimensional Fourier transform of $\widehat{F}\left(\vc{k_D'}\right)$ in the reciprocal detector plane after inserting \ref{eq:RytPh2} into \ref{eq:RytPh1}, we get a projection along the $z$-axis smeared out by a periodic exponential.
\begin{align}
\frac{1}{\sqrt{2\pi}}
\int_{-A}^{+A} \!\! dz \,
 f(\vc{r})  =
 - \sqrt{\frac{2}{\pi}} i \gls{km} \exp(-i\gls{km} \gls{lD}) \varphi_{\mathrm{R}, \phi_0}(\vc{r_D})
\end{align}
Where the interval $[-A,A]$ is the domain of \gls{f} along the $z$-direction. Here, we used $k_\mathrm{Dz} = M-1 \rightarrow 0$. The exponential factor on the right hand side only yields a phase offset in the detector plane. We may set \gls{lD} to zero and arrive at \cite{Devaney1981}
\begin{align}
\int_{-A}^{+A} \!\! dz \,
 f(\vc{r})  =
 - 2 i \gls{km} \varphi_{\mathrm{R}, \phi_0}(\vc{r_D}).  \label{eq:RytLonW1}
\end{align}
The right hand side contains the complex Rytov phase in the detector plane $\varphi_{\mathrm{R}, \phi_0}(\vc{r_D}) = i \Phi(\vc{r_D})$ (see \hyref{section}{sec:Rytov}). The left hand side contains the scattering potential from \hyref{equation}{eq:fperturbance}.
\begin{align}
& \gls{f}  =  \gls{km}^2\left[ \left(\frac{\gls{n}}{\gls{nm}}\right)^2 -1 \right] \tag{\ref{eq:fperturbance}} \\
& \hphantom{\gls{f}} = \gls{km}^2\left[ \left(1 + \frac{\gls{ndelt}}{\gls{nm}}\right)^2 -1 \right] \\
&\overset{\gls{ndelt}^2 \ll \gls{ndelt}}{\approx} \frac{2\gls{km}^2}{\gls{nm}} \cdot \gls{ndelt}
\end{align}
Where \gls{ndelt} is the local refractive index variation from the surrounding medium~\gls{nm}. By writing the right hand side of \hyref{equation}{eq:RytLonW1} as an integral over $d\Phi(\vc{r_D})$, we thus get
\begin{align}
\int_{-A}^{+A} \!\! 
2 \gls{km}^2  \gls{ndelt} \, \frac{dz}{\gls{nm}}  &=
\int 2 \gls{km} \, d\Phi(\vc{r_D}) \\
  2 \gls{km}^2  \gls{ndelt} \, \frac{dz}{\gls{nm}}   &=
 2 \gls{km} \, d\Phi(\vc{r_D}) \\
  \frac{\gls{ndelt}}{\lambda }   \, dz   &=
 \, \frac{1}{2\pi} d\Phi(\vc{r_D}).
\end{align}
This relation describes the phase change $d\Phi$ that occurs over a distance $dz$, as derived in \hyref{section}{sec:Born}, \hyref{equation}{eq:phasesimplea}. Thus, in the limit of high \gls{km}, the Fourier diffraction theorem with the Rytov approximation becomes the Fourier slice theorem, which requires that the data ($\int \!\! d\Phi(\vc{r_D})$) recorded at the detector \vc{r_D} are computed from line integrals ($\int \!\!dz$) through the sample \gls{ndelt}.

\subsection{Derivation with the Fourier Transform Approach}
\label{sec:FDThFourTransFAppr}
An alternative way to derive the Fourier diffraction theorem (\hyref{equation}{eq:UfandFf}) is the convolution approach in Fourier space. This approach uses the convolution theorem for Fourier transforms. We only briefly discuss this approach. For a thorough discussion, see~\cite{Banos1966}.
\begin{align}
\gls{ub} = \int \!\! d^3r' \, \gls{G} f(\vc{r'}) u_\mathrm{0}(\vc{r'}) = (G \ast fu_\mathrm{0}) (\vc{r}) \label{eq:FDTbornalt}
\end{align}
\hyref{Equation}{eq:FDTbornalt} describes the Born approximation \gls{ub} as a convolution ($\ast$) of $G(\vc{r})$ with $\gls{f}\gls{u0}$. In Fourier space, we may write \gls{Ub} as\footnote{The factor $(2\pi)^{3/2}$ originates from the unitary angular frequency Fourier transform. The non-unitary angular frequency and the unitary ordinary frequency Fourier transforms do not show this factor.}
\begin{align}
\gls{Ub} = (2\pi)^{3/2} \gls{Gf} \cdot \widehat{(fu_\mathrm{0})}(\vc{k}),
\end{align}
where $\widehat{(fu_\mathrm{0})}(\vc{k})$ denotes the Fourier transform of $\gls{f}\gls{u0}$, and \gls{Gf} is the Fourier transform of $G(\vc{r})$.  We find that
\begin{align}
\widehat{(fu_\mathrm{0})}(\vc{k}) =\frac{ a_\mathrm{0}}{(2\pi)^{3/2}} \int \!\! d^3r' f(\vc{r'}) \exp(- i \left[\vc{k} - \gls{km} \gls{s0}\right]\vc{r'}) = a_\mathrm{0} \, \widehat{F}(\vc{k} - \gls{km} \gls{s0})
\end{align}
is simply a shifted version of the Fourier transform of the inhomogeneity \gls{f}. In order to find \gls{Gf}, we calculate the Fourier transform of the inhomogeneous Helmholtz equation for the Green's function (\hyref{equation}{eq:greendelta}).
\begin{align}
\frac{1}{(2\pi)^{3/2}}\int \!\! d^3r' \, \nabla^2 G(\vc{r'}) \exp(i\vc{kr'}) + 
 \overbrace{ \frac{1}{(2\pi)^{3/2}} \int \!\! d^3r' \, \gls{km}^2 G(\vc{r'}) \exp(i\vc{kr'})}^{\gls{km}^2 \gls{Gf}} = \notag \\
= - \frac{1}{(2\pi)^{3/2}} \underbrace{\int \!\! d^3r' \, \delta(\vc{r'}) \exp(i\vc{kr'}) }_{1}
\end{align}
Using integration by parts for each Cartesian coordinate and by considering the asymptotic behavior
\begin{align}
\frac{\exp(ikr)}{4\pi r} \rightarrow \frac{\exp(ik\left|x\right|)}{4\pi \left|x\right|} \text{~as~} \left|x\right| \rightarrow \infty 
\end{align}
one can show that the first integral in the equation above becomes \cite{Banos1966}
\begin{align}
\frac{1}{(2\pi)^{3/2}} \int \!\! d^3r' \, \nabla^2 G(\vc{r'}) \exp(i\vc{kr'}) = - k^2 \gls{Gf}.
\end{align}
Thus, we obtain the Fourier transform of the Green's function 
\begin{align}
\gls{Gf} = \frac{1}{(2\pi)^{3/2}} \frac{1}{k^2 - \gls{km}^2}
 = \frac{1}{(2\pi)^{3/2}} \frac{1}{k_\mathrm{x}^2 + k_\mathrm{y}^2 + k_\mathrm{z}^2 - \gls{km}^2}.
\end{align}
We combine the equations above and obtain
\begin{align}
\gls{Ub} = \frac{a_\mathrm{0}}{k^2 - \gls{km}^2} \cdot \widehat{F}(\vc{k} - \gls{km} \gls{s0}).
\end{align}
This equation looks very much like \hyref{equation}{eq:UfandFf}. However, here the entire field \gls{Ub} is on the left hand side of the equation. Since we measure the field at the detector $\widehat{U}_{\mathrm{B}, \phi_0}(\vc{k_D})$, we perform an inverse Fourier transform to normal space, then take the field \gls{ub} at the detector, and finally back-transform to two-dimensional Fourier space.
 We rotate our coordinate system such that \gls{s0} matches the $k_\mathrm{z}$-axis. The integral over $dk_\mathrm{z}$ can be evaluated using the residue theorem. We integrate around the singularity at $k_\mathrm{z} = \sqrt{\gls{km}^2 - k_\mathrm{x}^2 - k_\mathrm{y}^2}$.
\begin{align}
\gls{ub} &= \frac{a_\mathrm{0}}{(2\pi)^{3/2}} \iiint \!\!
dk_\mathrm{x}dk_\mathrm{y}dk_\mathrm{z} \, 
\frac{\exp(i\vc{kr})}{k^2 - \gls{km}^2} \cdot \widehat{F}(\vc{k} - \gls{km} \gls{s0}) \\
&= \frac{2 \pi i a_\mathrm{0}}{(2\pi)^{3/2}} \iint \!\!dk_\mathrm{x}dk_\mathrm{y}
\frac{
\exp\!\left(i\left[k_\mathrm{x}x +k_\mathrm{y}y + \sqrt{\gls{km}^2 - k_\mathrm{x}^2 - k_\mathrm{y}^2}z \right]\right)
}{
2 \sqrt{\gls{km}^2 - k_\mathrm{x}^2 - k_\mathrm{y}^2}
} 
\cdot \widehat{F}(\vc{k} - \gls{km} \gls{s0})
\end{align}
Here, we used the following partial fraction expansion to obtain the singularities.
\begin{align} 
\frac{1}{k_\mathrm{z}^2 - k_\mathrm{m}'^2}& = \frac{1}{2k_\mathrm{m}'}\left(
\frac{1}{k_\mathrm{z}-k_\mathrm{m}'}-\frac{1}{k_\mathrm{z}+k_\mathrm{m}'}
 \right) \\
 k_\mathrm{m}'^2& = \gls{km}^2 -  k_\mathrm{x}^2 - k_\mathrm{y}^2
\end{align}
We identify $\vc{k}$ as $\vc{k_D}=(x_\mathrm{D}, y_\mathrm{D}, z_\mathrm{D}=\gls{lD}$) (at $\gls{phi0} = 0$) and perform the two-dimensional Fourier transform at the detector. We obtain delta functions for $x$ and $y$ that we use to solve the integral for~$k_\mathrm{x}$ and~$k_\mathrm{y}$\footnote{Factors: the two-dimensional Fourier transform adds a factor of $1/(2\pi)$ and the delta functions contribute with $2\pi$ each.}. With $ M =\sqrt{\gls{km}^2 - k_\mathrm{Dx}^2 - k_\mathrm{Dy}^2}$ we arrive at \hyref{equation}{eq:UfandFf}.
\begin{align}
\widehat{U}_{\mathrm{B}, \phi_0}(\vc{k_D}) = 
2\pi \frac{i a_\mathrm{0} }{(2\pi)^{1/2} \cdot 2}	
\frac{											
 \exp(i\gls{km}M \gls{lD})
 }{												
 \gls{km}M
 }    						
\widehat{F}\left(\vc{k_D} -\gls{km}\gls{s0}\right)
\end{align}

\subsection{Backpropagation Algorithm}
\label{chap:backprop}
In this section we derive the three-dimensional analog to the two-dimensional backpropagation algorithm as described by Devaney \cite{Devaney1982336} and \hyref{section}{sec:2DBackprop} in this script. Our goal is to solve \hyref{equation}{eq:UfandFf} for \gls{f} such that we do not need any interpolation in Fourier space. In order to do that, we need to perform a change of coordinates. The coordinate system has its origin on the rotating axis $y$. We define our new coordinate system in the Fourier domain analogous to the two-dimensional case in three steps.

First, we construct our coordinate system such that we can eliminate any reciprocal distances and express any position of the recorded data in terms of angles. This is possible, because the Fourier diffraction theorem states that every point in Fourier space is placed on a semi-sphere with radius \gls{km} centered at~$-\gls{km}\gls{s0}$.
Second, we define the projection angle \gls{phi0} as the angle that corresponds to the angular sample position from which the projection was recorded. The angle \gls{phi0} coincides with the direction of the incoming plane wave \gls{s0}. Note that \gls{phi0} is a two-dimensional angle. However, if the sample is rotated only about one axis, the angle \gls{phi0} will become one-dimensional. We only consider the rotation about the $y$-axis as shown in \hyref{figure}{fig:CoordTrafoPhi0}.
\hyref{Figure}{fig:CoordTrafoThetaPsi} depicts the two angles which define the semi-spherical surface in spherical coordinates relative to the rotated coordinate system (\gls{phi0}). 
Third, we set the polar angle as \gls{theta} and the azimuthal angle as~\gls{psi}. For $\gls{phi0}=0$, the non-shifted semi-sphere lies within the half-space $k_\mathrm{z} \geq 0 $. The polar angle~\gls{theta} is then measured from the positive $k_\mathrm{z}$-axis and the azimuthal angle \gls{psi} is measured from the $k_\mathrm{x}$-axis in positive direction of $k_\mathrm{y}$. The corresponding variable domains are
\begin{align*}
\gls{theta} &\in \left[-\frac{\pi}{2},+\frac{\pi}{2}\right] \\
\gls{psi} &\in  \left[0,\pi\right].
\end{align*}

\begin{figure}[!t]
\centering
\subfloat[width=\linewidth/2][rotation through \gls{phi0}]
{\includegraphics[]{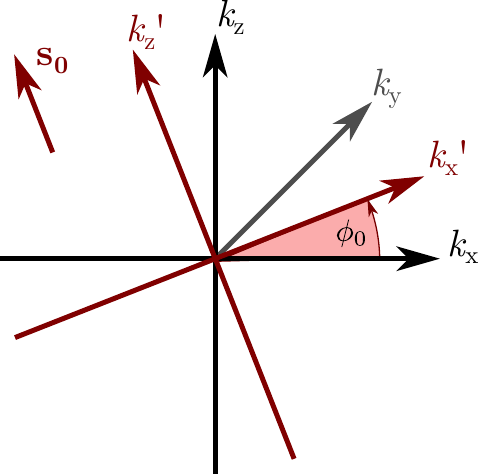} \label{fig:CoordTrafoPhi0}} \qquad
\subfloat[width=\linewidth/2][spherical coordinates\\ in the rotated system]
{\includegraphics[]{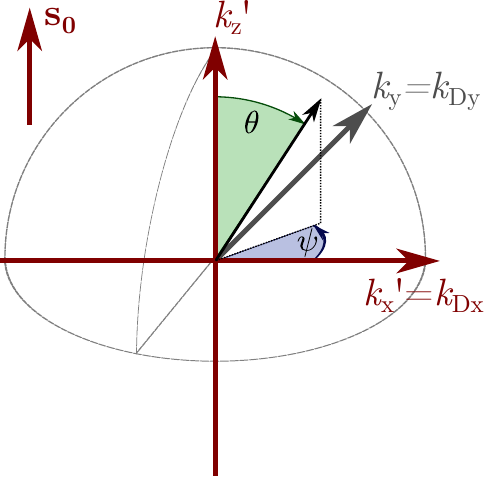}\label{fig:CoordTrafoThetaPsi}} \qquad
  \mycaption{Coordinate transforms for 3D backpropagation}{\textbf{a)} The coordinate system of the detector (red) is rotated about the $y$-axis which corresponds to a rotation about the $k_\mathrm{y}$-axis in Fourier space. \textbf{b)} The position on the semi-spherical surface with the radius \gls{km} is described in a spherical coordinate system with the polar angle~\gls{theta} and the azimuthal angle \gls{psi}.
\label{fig:CoordTrafo}}
\end{figure}

\noindent Let us first write the rotation matrix $D_{\gls{phi0}}$ for the rotation of the sample about the $y$-axis.
\begin{equation}
D_{\gls{phi0}} = 
\left(\begin{array}{ccc}
\cos\gls{phi0} & 0 & -\sin\gls{phi0} \\
0 & 1 & 0 \\
\sin\gls{phi0} & 0 & \cos\gls{phi0}
\end{array}\right)
\end{equation}
The matrix $D_{\gls{phi0}}$ rotates a vector \vc{s'} counter-clockwise through the angle \gls{phi0} about the $k_\mathrm{y}$-axis. The vector that describes the semi-spherical surface in spherical coordinates is given by
\begin{equation}
\vc{s'} = \left(\sin\theta \cos\psi,\,\sin\psi \sin\theta,\,\cos\theta\right)^\top
\end{equation}
where $()^\top$ denotes the transpose of the vector, i.e. \vc{s'} is a column vector.
We transform this vector into the rotated coordinate system by calculation of
\begin{align}
\vc{s} = D_{\gls{phi0}} \cdot \vc{s'} = 
\left(\begin{array}{c}
\sin\gls{theta} \cos\gls{phi0} \cos\gls{psi} - \sin\gls{phi0} \cos\gls{theta}  \\
\sin\gls{psi} \sin\gls{theta} \\
\sin\gls{phi0} \sin\gls{theta} \cos\gls{psi} + \cos\gls{phi0}\cos\gls{theta}
\end{array}\right).
\end{align}
We use the change of coordinates in Fourier space to rewrite the inverse Fourier integral for \gls{f} (\hyref{eq.}{eq:UfandFf})
\begin{align}
\widehat{F}\left(\vc{k_D} -\gls{km}\gls{s0}\right) &= 
- \sqrt{\frac{2}{\pi}} \frac{i M \gls{km}}{ a_\mathrm{0} }  	
 \exp(-i\gls{km}M \gls{lD})					
\widehat{U}_\mathrm{B}(\vc{k_D})
\tag{\ref{eq:UfandFf}} \\
\gls{f} &=
\frac{1}{(2\pi)^{3/2}} \iiint \!\! d^3k \,         
\widehat{F}\left(\vc{k}\right)					  
\exp(i\vc{kr})
\label{eq:fintF}
\end{align}
where we used
\begin{align}
\vc{k} &= \gls{km}(\vc{s} - \gls{s0}) \\
\vc{k_D} &= \gls{km} \vc{s} \\
\gls{s0} &= \left(-\sin\gls{phi0},\,0,\,\cos\gls{phi0}\right)^\top.
\end{align}
The integral over $d^3k$ can be expressed as an integral over \gls{phi0}, \gls{theta}, and \gls{psi} by means of the Jacobian matrix $J$.
\begin{align}
dk_\mathrm{x}dk_\mathrm{y}dk_\mathrm{z} &= \left| \det(J) \right| d\gls{phi0} d\gls{theta} d\gls{psi} \\
J &= \frac{\partial k_\mathrm{x} \partial k_\mathrm{y} \partial k_\mathrm{z}}
{\partial \gls{phi0} \partial \gls{theta} \partial \gls{psi}}
\end{align}
The Jacobian matrix is the matrix of all first-order partial derivatives of the vector
\begin{align}
\left(\begin{array}{c}
k_\mathrm{x} \\ k_\mathrm{y} \\ k_\mathrm{z} \\
\end{array}\right) =
\gls{km} (\vc{s} - \gls{s0}) = \gls{km}
\left(\begin{array}{c}
\sin\gls{theta} \cos\gls{phi0} \cos\gls{psi} - \sin\gls{phi0} \cos\gls{theta} +\sin\gls{phi0}  \\
\sin\gls{psi} \sin\gls{theta} \\
\sin\gls{phi0} \sin\gls{theta} \cos\gls{psi} + \cos\gls{phi0}\cos\gls{theta} - \cos\gls{phi0}
\end{array}\right)
\end{align}
and computes to (column wise):
\begin{align}
\frac{\partial k_\mathrm{x} \partial k_\mathrm{y} \partial k_\mathrm{z}}
{\partial \gls{phi0}} &= \gls{km}
\left(\begin{array}{c}
-\sin\gls{phi0} \sin\gls{theta} \cos\gls{psi} - \cos\gls{phi0} \cos\gls{theta} + \cos\gls{phi0}  \\
0   \\
\sin\gls{theta} \cos\gls{phi0} \cos\gls{psi} - \sin\gls{phi0} \cos\gls{theta} + \sin\gls{phi0}  \\
\end{array}\right) \\
\frac{\partial k_\mathrm{x} \partial k_\mathrm{y} \partial 
k_\mathrm{z}}
{\partial \gls{theta}} &= \gls{km}
\left(\begin{array}{c}
 \cos\gls{phi0} \cos\gls{psi} \cos\gls{theta} + \sin\gls{phi0} \sin\gls{theta} \\
\sin\gls{psi} \cos\gls{theta} \\
\sin\gls{phi0} \cos\gls{psi} \cos\gls{theta} - \sin\gls{theta} \cos\gls{phi0} \\
\end{array}\right) \\
\frac{\partial k_\mathrm{x} \partial k_\mathrm{y} \partial k_\mathrm{z}}
{\partial \gls{psi}} &= \gls{km}
\left(\begin{array}{c}
 -\sin\gls{psi} \sin\gls{theta} \cos\gls{phi0} \\
\sin\gls{theta} \cos\gls{psi} \\
-\sin\gls{phi0} \sin\gls{psi} \sin\gls{theta}
\end{array}\right)
\end{align}
For the integration by substitution we thus need to consider the following factor for the integral\footnote{This determinant can easily be calculated using SAGE (\url{http://www.sagemath.org/}).}
\begin{align}
\left| \det(J) \right| &= \left| -\gls{km}^3 (\sin\gls{theta})^2 \cos\gls{psi} \right|
\end{align}
and we may rewrite the integral for \gls{f} as
\begin{align}
\gls{f} &= \frac{1}{2}
\frac{1}{(2\pi)^{3/2}} \int_0^{2\pi} \!\! d\gls{phi0} 
\int_0^\pi \!\! d\gls{theta} \int_{-\pi/2}^{+\pi/2} \!\! d\gls{psi} \,         
\gls{km}^3 \left| (\sin\gls{theta})^2 \cos\gls{psi} \right| 
\widehat{F}\left(\vc{k}\right)					  
\exp(i\vc{kr})
\end{align}
where we have introduced a factor of $1/2$ to compensate for the double-\gls{phi0} coverage in Fourier space as we rotate the semi-sphere about the $k_\mathrm{y}$ axis from $0$ to $2\pi$. Note that this factor only holds for equidistant coverage over $2\pi$. In other cases, weighting factors need to be introduced for each angle~\gls{phi0}. We have to perform one more coordinate transform to arrive at an integral over the coordinates $k_\mathrm{Dx}$ and $k_\mathrm{Dy}$. These are the coordinates in Fourier space that correspond to direct Fourier transforms of the two-dimensional recorded images at angles \gls{phi0}. 
This coordinate transform is also depicted in \hyref{figure}{fig:CoordTrafo} and reads
\begin{align}
k_\mathrm{Dx} &= \gls{km} \cos\gls{psi} \sin\gls{theta} \label{eq:CoordTrafoDx} \\
k_\mathrm{Dy} &= \gls{km} \sin\gls{psi} \sin\gls{theta}.
\label{eq:CoordTrafoDy}
\end{align}
For $\gls{phi0}=0$, the $k_\mathrm{Dx}$-axis coincides with $k_\mathrm{x}$ and the $k_\mathrm{Dy}$-axis with $k_\mathrm{y}$. The Jacobian determinant $J_D$ for this coordinate transform at the detector then computes to
\begin{align}
\left| \det(J_D) \right| =&
 \frac{\partial k_\mathrm{Dx} \partial k_\mathrm{Dy}}{\partial \gls{theta} \partial \gls{psi}} \\
=& \gls{km}^2 | \cos\gls{theta} \sin\gls{theta} | \\
\overset{\gls{theta} \in \left[-\pi/2,+\pi/2\right]}{=}  & \gls{km}^2 \cos\gls{theta} | \sin\gls{theta} |.
\end{align}
In the second variables change of integral for \gls{f} we identify $\gls{km} \cos\theta = \gls{km} M$.
\begin{align}
d\gls{theta} d\gls{psi} &= \frac{1}{
\gls{km}^2 \cos\gls{theta} |\sin\gls{theta}|} dk_\mathrm{Dx} dk_\mathrm{Dy} \\
&=\frac{1}{
\gls{km}^2 M |\sin\gls{theta}|} dk_\mathrm{Dx} dk_\mathrm{Dy}
\end{align}

\noindent By applying these transforms, one obtains an integral for \gls{f} over the rotation angle \gls{phi0} and the coordinates of the Fourier-transformed image $k_\mathrm{Dx}$ and $k_\mathrm{Dy}$.
\begin{align}
\gls{f} &= \frac{1}{2}
\frac{1}{(2\pi)^{3/2}} \int_0^{2\pi} \!\! d\gls{phi0}  
\int_{-\gls{km}}^{\gls{km}} \!\! dk_\mathrm{Dx} \int_{-\gls{km}}^{\gls{km}} \!\! dk_\mathrm{Dy} \,         
\frac{\left| k_\mathrm{Dx} \right|}{
      M}                
\widehat{F}\left(\vc{k}\right)					  
\exp(i\vc{kr})
\end{align}
The $k_\mathrm{Dz}$-component of the vector \vc{k_D} can be expressed by $k_\mathrm{Dx}$, $k_\mathrm{Dy}$, and \gls{phi0}. We express \vc{k} in terms of \vc{k_D} and \gls{s0}, which can be separated into lateral (\vc{t_\perp}) and axial (\gls{s0}) components.

\begin{align}
\vc{k} &= \vc{k_D} - \gls{km} \gls{s0}  = \gls{km}(\vc{s}-\gls{s0})  \\
 &= k_\mathrm{Dx} \, \vc{t_\perp} + \gls{km}(M - 1) \, \gls{s0} \\
\vc{t_\perp} &= \left(\cos\gls{phi0}, \,
                \frac{k_\mathrm{Dy}}{k_\mathrm{Dx}}, \,
                \sin\gls{phi0} \right)^\top \\
\vc{s_0} &= \left(-\sin\gls{phi0}, \,0 ,\, \cos\gls{phi0} \right)^\top \\
\vc{s_0} \cdot \vc{t_\perp} &= 0
\end{align}

\begin{landscape}
\noindent By inserting the relations above, we may may express \gls{f} as
\begin{align}
\gls{f} &= \frac{1}{2}
\frac{1}{(2\pi)^{3/2}} \int_0^{2\pi} \!\! d\gls{phi0}  
\int_{-\gls{km}}^{\gls{km}} \!\! dk_\mathrm{Dx} \int_{-\gls{km}}^{\gls{km}} \!\! dk_\mathrm{Dy} \,         
\frac{\left| k_\mathrm{Dx} \right|}{ M }             
\widehat{F}\left(k_\mathrm{Dx} \, \vc{t_\perp} + \gls{km}(M - 1) \, \gls{s0}\right)					  
\exp[i(k_\mathrm{Dx} \, \vc{t_\perp} + \gls{km}(M - 1) \, \gls{s0})\vc{r}].
\end{align}
Now we insert \hyref{equation}{eq:UfandFf}. Note that the two-dimensional function $\widehat{U}_{\mathrm{B},\gls{phi0}}(\vc{k_D})$ was measured in the detector plane and thus we write $\widehat{U}_{\mathrm{B},\gls{phi0}}(k_\mathrm{Dx},k_\mathrm{Dy})$. \hyref{Equation}{eq:UfandFf} describes the Fourier-space distribution of the measurement data from one single projection at an angle \gls{phi0}.
\begin{align}
\widehat{F}\left(\vc{k_D} -\gls{km}\gls{s0}\right) = 
- \sqrt{\frac{2}{\pi}} \frac{i M \gls{km}}{ a_\mathrm{0} }  	
 \exp(-i\gls{km}M \gls{lD})					
\widehat{U}_{\mathrm{B},\gls{phi0}}(k_\mathrm{Dx},k_\mathrm{Dy})
\tag{\ref{eq:UfandFf}}
\end{align}
The integral over \gls{phi0} depends on the recorded data $U_{\mathrm{B},\gls{phi0}}$. Furthermore, our choice of coordinates requires the exponential factor in the Fourier transform to be dependent on \gls{phi0}.
\begin{align}
\gls{f} &= \frac{-i\gls{km}}{(2\pi)^{2}a_\mathrm0} 
\int_0^{2\pi} \!\! d\gls{phi0}  						  
\int_{-\gls{km}}^{\gls{km}} \!\! dk_\mathrm{Dx}				  
\int_{-\gls{km}}^{\gls{km}} \!\! dk_\mathrm{Dy} \,              
\left| k_\mathrm{Dx} \right| 
\widehat{U}_{\mathrm{B},\gls{phi0}}(k_\mathrm{Dx},k_\mathrm{Dy})
\exp(-i\gls{km}M \gls{lD}) 
\exp[i(k_\mathrm{Dx} \, \vc{t_\perp} + \gls{km}(M - 1) \, \gls{s0})\vc{r}] \label{eq:3DBackprop}
\end{align}
with
\begin{align}
\gls{km}M = \sqrt{\gls{km}^2 - k_\mathrm{Dx}^2 - k_\mathrm{Dy}^2}.
\end{align}
\hyref{Equation}{eq:3DBackprop} is the three-dimensional backpropagation algorithm of uniaxially rotated samples.
An application of the 3D backpropagation algorithm is illustrated in
\hyref{figure}{fig:FDT3D} for a centered sphere from computations that are based on Mie theory. Note that, as could be seen for the 2D case in \hyref{figure}{fig:FDT2D}, the reconstruction with the Born approximation breaks down and that the reconstruction quality with the Rytov approximation is dependent on the total number of projections that were acquired for the sinogram. Details of the simulation and reconstruction process can be found in \hyref{figure}{fig:FDT3D_cross-section}.
\end{landscape}

\begin{figure}[ht]
\centering
  \subfloat[][phantom, \gls{ndelt}]
  {\includegraphics[width=.285\linewidth]{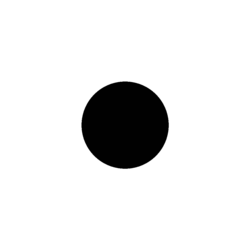}}
  \subfloat[][detector amplitude] 
  {\includegraphics[width=.285\linewidth]{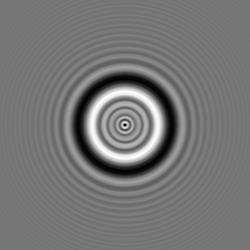}}
  \subfloat[][detector phase]
  {\includegraphics[width=.285\linewidth]{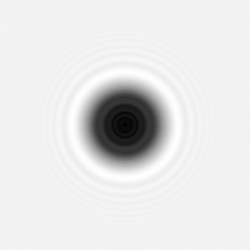}} \\
  \subfloat[][Born $x=0$ \\ \phantom{(a)~} (200 projections)]
  {\includegraphics[width=.285\linewidth]{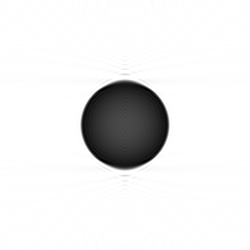}} 
  \subfloat[][Rytov $x=0$ \\ \phantom{(a)~} (50 projections)]
  {\includegraphics[width=.285\linewidth]{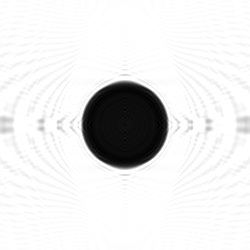}}
  \subfloat[][Rytov $x=0$ \\ \phantom{(a)~} (20 projections)]
  {\includegraphics[width=.285\linewidth]{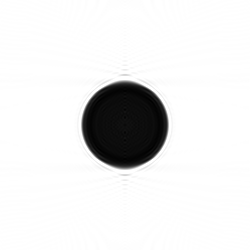}} \\
  \subfloat[][Born $y=0$ \\ \phantom{(a)~} (200 projections)]
  {\includegraphics[width=.285\linewidth]{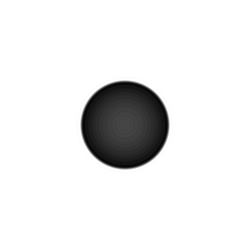}} 
  \subfloat[][Rytov $y=0$ \\ \phantom{(a)~} (50 projections)]
  {\includegraphics[width=.285\linewidth]{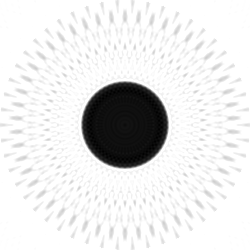}}
  \subfloat[][Rytov $y=0$ \\ \phantom{(a)~} (200 projections)]
  {\includegraphics[width=.285\linewidth]{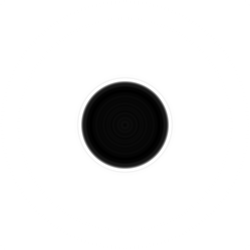}} \\
\mycaption{3D Backpropagation}{Image (a) shows the refractive index distribution of a dielectric sphere. White indicates the refractive index of the medium and black the higher refractive index of the sphere. The forward-scattering process was computed according to Mie theory. Amplitude and phase of the scattered near-field are shown in (b) and~(c). Because the sphere is rotated about its center, the amplitude and phase are identical for every slice of the sinogram. Images (d), (e), and (f) show the reconstruction of the refractive index distribution using the Born and Rytov approximations in the $x=0$ cross-section. Images (g), (h), and (i) show the corresponding reconstructions at the $y=0$ cross-section.
The reconstruction in (e) and (h) were performed with 50 projections only. See \hyref{figure}{fig:FDT3D_cross-section} for details. 
\label{fig:FDT3D}}
\end{figure}

\begin{figure}[ht]
\centering
  \includegraphics[width=\linewidth]{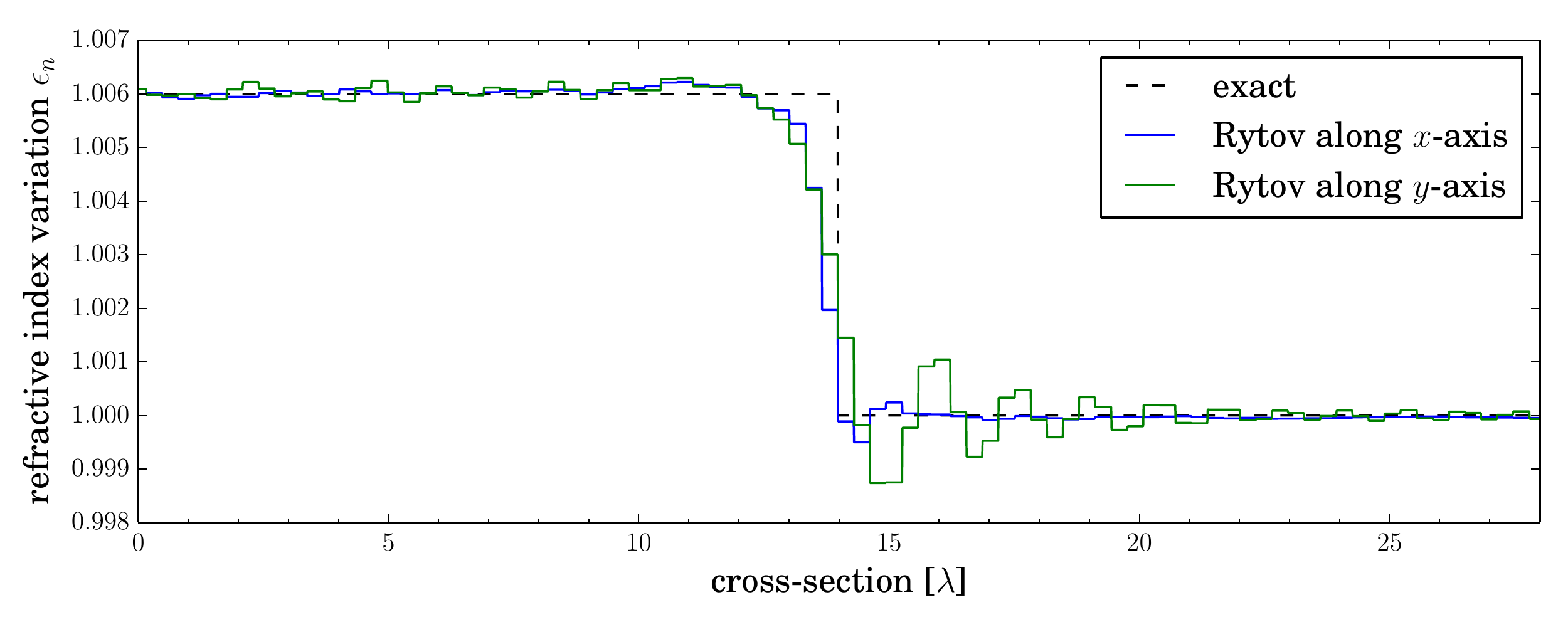}
\mycaption{3D Backpropagation cross-sections}{The refractive index of the medium is $\gls{nm} = 1.0$ and the local variation inside the sphere is $\gls{ndelt} = \gls{n} - \gls{nm} = 0.006$.
The radius of the sphere is $14 \gls{lambda}$ (vacuum wavelength \gls{lambda}).
The scattered wave is computed at an optical distance of $l_\mathrm {D} = 20 \gls{lambda}$ from the center of the cylinder and sampled at $\gls{lambda}/3.1125$ over 250 pixels using the software GMM-FIELD \cite{ringler08}. The refractive index map is reconstructed on a grid of $250\times 250$ pixels from 200 projections over $2 \pi$ using the software ODTbrain \cite{Mueller15}. The reconstruction with the Born approximation (not shown) follows the behavior as seen in \hyref{figure}{fig:FDT2D_cross-section}. The 2D cross-sections are shown in \hyref{figure}{fig:FDT3D}. Artifacts due to the incomplete $2\pi$-coverage are visible along the $y$-axis, as briefly discussed in \hyref{section}{sec:FourierDiffrRelevance1axis}.
\label{fig:FDT3D_cross-section}}
\end{figure}

\FloatBarrier

\subsubsection{Comparison to the Two-dimensional Backpropagation Algorithm}
\hyref{Table}{tab:2D3DProp} summarizes the similarities and differences between the 2D and the 3D backpropagation algorithms with respect to the dimension of the problem.
\begin{table}[ht]
\centering
\begin{tabular}{l|c|c|}
 & \multicolumn{2}{c}{\parbox{0.80\linewidth}{
    \begin{align*}
       f(\vc{r}) = -\frac{i \gls{km}}{2\pi a_0} 
       \left( \int  \!\! dK_\mathrm{D} \right)
       \int_0^{2 \pi} \!\!  d\gls{phi0} \,
		\exp(iB)
       \left| k_\mathrm{Dx} \right |  
       \widehat{U}_{\mathrm{B}, \phi_0}(\vc{k_D})
	\end{align*}
	}} \\
 & \textbf{2D} (\hyref{equation}{eq:2DBackprop}) & \textbf{3D} (\hyref{equation}{eq:3DBackprop}) \\ 
\hline 
\parbox{0.12\linewidth}{
    Sinogram \\     \gls{Ubphi0}
    } &
\parbox[][5em][c]{0.35\linewidth}{\hspace*{.5em}\centering
    1D Fourier transform of~complex~scattered~waves \\
    $\widehat{U}_{\mathrm{B},\phi_0}(k_\mathrm{Dx})$} &
\parbox{0.35\linewidth}{\centering
    2D Fourier transform of~complex~scattered~waves \\
    $\widehat{U}_{\mathrm{B},\phi_0}(k_\mathrm{Dx}, k_\mathrm{Dy})$

    }
 \\ 
\hline 
\parbox{0.12\linewidth}{
	Integral \\ $dK_\mathrm{D}$} &
\parbox{0.25\linewidth}{
	\begin{align*}
	\left( \int  \!\! dK_\mathrm{D} \right) =
	\frac{1}{\sqrt{2\pi}}
    \int \!\! dk_\mathrm{Dx}
    \end{align*}
     } & 
\parbox{0.25\linewidth}{
	\begin{align*}
	\left( \int  \!\! dK_\mathrm{D} \right) =
	\frac{1}{2\pi}
    \iint \!\! dk_\mathrm{Dx} \, dk_\mathrm{Dy}
    \end{align*}
    } \\
\hline
\parbox{0.12\linewidth}{
	Exponent \\ $B$} &
	 \multicolumn{2}{c|}{\parbox{0.80\linewidth}{
	\begin{align*}
	  B &= - \gls{km} M \gls{lD} + k_\mathrm{Dx}\vc{t_\perp r} + \gls{km}(M - 1)\gls{s0}\vc{r}
	\end{align*}
	}} \\
\parbox{0.12\linewidth}{
	} &
\parbox{0.25\linewidth}{ 
	\begin{align*}
	  M = \frac{1}{\gls{km}} \sqrt{\gls{km}^2 - k_\mathrm{Dx}^2}
	\end{align*}
	} &
\parbox{0.25\linewidth}{ 
	\begin{align*}
	  M = \frac{1}{\gls{km}} \sqrt{\gls{km}^2 - k_\mathrm{Dx}^2 - k_\mathrm{Dy}^2}		
	\end{align*}
	} \\
\hline
\parbox{0.12\linewidth}{
	Vectors \\ \vc{r}, \gls{s0}, \vc{t_\perp} } &	 
\parbox{0.25\linewidth}{ 
	\begin{align*}
	  \vc{r} &= (x,z) \\
	  \gls{s0} &= (-\sin \gls{phi0}, \, \cos \gls{phi0}) \\
	  \vc{t_\perp} &= (\cos \gls{phi0}, \, \sin \gls{phi0})
	\end{align*}
	} &
\parbox{0.25\linewidth}{ 
	\begin{align*}
	  \vc{r} &= (x,y,z) \\
	  \vc{s_0} &= \left(-\sin\gls{phi0}, \,0 ,\, \cos\gls{phi0} \right) \\
	  \vc{t_\perp} &= \left(\cos\gls{phi0}, \,
                \frac{k_\mathrm{Dy}}{k_\mathrm{Dx}}, \,
                \sin\gls{phi0} \right)
	\end{align*}
	} \\
\hline
\end{tabular} 
\mycaption{Comparison of 2D and 3D backpropagation}{The table compares the 2D and 3D backpropagation algorithms. The generalized backpropagation formula above the table shows that the structure of the algorithms are similar in 2D and 3D, as previously noted in \hyref{table}{tab:2D3DDiffract}. The only differences originate from the different numbers of dimensions.
See \hyref{table}{tab:2DProjProp} for a comparison between the  backpropagation algorithm and the backprojection algorithm. \label{tab:2D3DProp}}
\end{table}

%% file: chapter6_implementation.tex
\section{Implementation}
\label{sec:Implementation}

The actual algorithms for backprojection and backpropagation use a trick for the reconstruction. We do not need to numerically integrate the equations for the backprojection (\ref{eq:Backproj}), the two-dimensional (\ref{eq:2DBackprop}), and the three-dimensional (\ref{eq:3DBackprop}) backpropagation algorithms. Instead, we identify the Fourier transform for the reciprocal vector \vc{k_D} and perform the reconstruction projection-wise for each rotational position \gls{phi0}.
In this section, we subsequently derive the actual reconstruction algorithms from the equations of the previous sections.

\subsection{Backprojection}
The backprojection algorithm, as introduced in \hyref{section}{sec:NormalTomo}, inverts the Radon transform, i.e. it is used to reconstruct an object from equidistant projections. \hyref{Figure}{fig:FBP} depicts the process from image acquisition to image reconstruction with the backprojection algorithm. The backprojection algorithm is based on the Fourier slice theorem \cite{Bracewell1956, Brooks1976} and runs as follows:
\begin{itemize}
\item[1.] Each projection is filtered with a ramp filter ($\left| k_\mathrm{Dx} \right|$).
\item[2.] The filtered projection is backprojected onto the image volume according to its acquisition angle.
\item[3.] The sum of all backprojections constitutes the reconstructed image.
\end{itemize}

We start from the formula for the inverse Radon transform.
\begin{align}
f(x,z) = 
\frac{1}{2\pi}
\int \!\! dk_\mathrm{Dx}
\int_{0}^{\pi} \! \! d\gls{phi0} \,
\left| k_\mathrm{Dx} \right|
\frac{\widehat{P}_{\phi_0}(k_\mathrm{Dx})}{\sqrt{2\pi}}
\exp[i k_\mathrm{Dx}(x\cos \gls{phi0} + z \sin \gls{phi0} )]
\tag{\ref{eq:Backproj}}
\end{align} 
The data in real space at $\vc{r} = (x,y)$ are computed from integrals over $k_\mathrm{Dx}$ and \gls{phi0}. We can introduce a coordinate transform $D_{-\phi_0}$ that rotates \vc{r} through the angle $-\gls{phi0}$ along the $y$-axis, such that $x_{\phi_0} = x\cos \gls{phi0} + z \sin \gls{phi0}$. In the resulting equation we identify a one-dimensional inverse Fourier transform 
\begin{align}
f(x,z) = 
\frac{1}{2\pi}
\int_{0}^{\pi} \! \! d\gls{phi0} \
D_{-\phi_0} \! \! 
\underbrace{
 \left\lbrace
  \int \!\! dk_\mathrm{Dx}
  \left| k_\mathrm{Dx} \right|
  \frac{\widehat{P}_{\phi_0}(k_\mathrm{Dx})}{\sqrt{2\pi}}
  \exp[i k_\mathrm{Dx}x_{\phi_0}]
  \right\rbrace}_{
    \text{FFT}^{-1}_{\mathrm{1D}}
    \left\lbrace
     \left| k_\mathrm{Dx} \right|
     \widehat{P}_{\phi_0}(k_\mathrm{Dx})
    \right\rbrace
}.
\end{align} 
We have effectively replaced the integral over $k_\mathrm{Dx}$ by a one-dimensional inverse fast Fourier transform ($\text{FFT}^{-1}_{\mathrm{1D}}$) and a rotation in real space ($D_{-\phi_0}$). We replace the remaining integral over \gls{phi0} by a discrete sum over $N$ equidistant projections and obtain
\begin{align}
f(x,z) = 
\frac{1}{2\pi} 
\sum_{j=1}^{N} \! \Delta \gls{phi0} \,
D_{-\phi_j} \! \! 
\left\lbrace
 \text{FFT}^{-1}_{\mathrm{1D}}
 \left\lbrace
  \left| k_\mathrm{Dx} \right|
  \widehat{P}_{\phi_j}(k_\mathrm{Dx})
 \right\rbrace
\right\rbrace
\label{eq:alg.backproj}
\end{align} 
with the discrete angular distance $\Delta \gls{phi0} = \pi/N$ and the discrete angles $\phi_j = j\cdot \Delta \gls{phi0}$ (${j = 1,2,\dots,N}$). A numerical method that implements \hyref{equation}{eq:alg.backproj} is much faster than the direct computation of \hyref{equation}{eq:Backproj}, because it can make use of the fast Fourier transform. The common name \textit{filtered} backprojection algorithm comes from an interpretation of \hyref{equation}{eq:alg.backproj}. The one-dimensional detector data are ramp-filtered and then projected over the entire two-dimensional reconstruction plane according to the angle they were acquired.

\subsubsection*{A Note on the Ramp Filter}
The ramp filter  $\left| k_\mathrm{Dx} \right|$, derived in \hyref{section}{sec:2DInvRadon} can also be motivated by the  point spread function of the backprojection process, which is $\nicefrac{1}{\left| \vc{r} \right|}$. By filtering each projection with the ramp filter $\left|  k_\mathrm{Dx} \right|$, the effect of the point spread function can be reversed. To illustrate this, let us consider a reconstructed image that has not been filtered ($f_0$). Then we may write a formula for the filtered image $f$ by means of the convolution
\begin{align}
f_0 = f \ast \frac{1}{\left| \vc{r} \right|}.
\end{align}
Inverting this formula using the Fourier transform $\mathcal{F}$, leads to
\begin{align}
f_0 &= f \ast \frac{1}{\left| \vc{r} \right|} = \mathcal{F}^{-1}\left( 
\mathcal{F}(f)  \cdot
\underbrace{\mathcal{F}\left(\frac{1}{\left| \vc{r} \right|}\right)}_{2\pi/\left|\vc{k}\right|} \right) \\ 
f &= \mathcal{F}^{-1} \left( \mathcal{F}(f_0) \cdot \left| \vc{k} \right|/2\pi \right) = f \ast
 \mathcal{F}^{-1}\left( \underbrace{
  \mathcal{F}\left( \frac{1}{\left| \vc{r} \right|} \right) \cdot \frac{\left| \vc{k} \right|}{2\pi} 
  }_1 \right).
\end{align}
The derivation shows, that the image in real space $f$ can be computed from the non-filtered backprojection $f_0$ by multiplication with a two-dimensional ramp filter $|\vc{k}|$. Filtering each projection with $|k_\mathrm{Dx}|$ is mathematically equivalent and yields better results when working with discrete data sets.
With regard to detector symmetries or sample type, other filters (e.g. cosine, Hamming, Hann) have been developed. These filters are usually heuristic and have different effects on image contrast or noise.

\subsection{Two-dimensional Backpropagation}
The backpropagation algorithm in two dimensions ($\vc{r} = (x,z))$ can be derived analogous to the backprojection algorithm. An additional difficulty is introduced by the fact that the projection is now multiplied by a phase factor which is dependent on the distance from the center of the object. Let us consider  \hyref{equation}{eq:2DBackprop}.
\begin{align}
\gls{f} = 
- \frac{i \gls{km}}{a_0 (2\pi)^{3/2}}
\int \!\! dk_\mathrm{Dx} \int_0^{2 \pi} \!\!  d\gls{phi0} \,
\left| k_\mathrm{Dx} \right |  
\widehat{U}_{\mathrm{B},\phi_0}(k_\mathrm{Dx})
\exp( -i \gls{km} M \gls{lD} ) \times \notag \\
\times \exp \! \left[i (k_\mathrm{Dx} \, \vc{t_\perp} + \gls{km}(M - 1) \, \gls{s0})\vc{r}\right] 
\tag{\ref{eq:2DBackprop}}
\end{align}
\begin{align}
\vc{t_\perp} &=  (\cos\gls{phi0}, \, \sin\gls{phi0}) \notag \\
\vc{s_0} &= (-\sin\gls{phi0}, \, \cos\gls{phi0}) \notag
\end{align}
We begin by introducing the rotation $D_{-\phi_0}$ through $-\gls{phi0}$ along the $y$-axis that transforms \vc{r} to \vc{r_{\phi_0}}.
\begin{align}
\vc{r_{\phi_0}} &= (x_{\phi_0}, z_{\phi_0})  \\
x_{\phi_0} &= x\cos \gls{phi0} + z \sin \gls{phi0} \\
z_{\phi_0} &= -x\sin \gls{phi0} + z \cos \gls{phi0} \\
\vc{t_\perp} \cdot \vc{r} &=  x_{\phi_0} \\
\vc{s_0} \cdot \vc{r} &= z_{\phi_0}
\end{align}

\begin{align}
\gls{f} & =
- \frac{i \gls{km}}{a_0 (2\pi)^{3/2}}
\int_0^{2 \pi} \!\!  d\gls{phi0}  \times \notag \\
\times & D_{-\phi_0} \!\!
\left\lbrace
 \int \!\! dk_\mathrm{Dx} 
 \left| k_\mathrm{Dx} \right |  
 \widehat{U}_{\mathrm{B},\phi_0}(k_\mathrm{Dx})
 \exp( -i \gls{km} M \gls{lD} )
  \exp \! \left[i (k_\mathrm{Dx} x_{\phi_0} + \gls{km}(M - 1) z_{\phi_0})\right] 
\right\rbrace
\end{align}
Because of the factor $\gls{km}(M - 1) \, z_{\phi_0}$ in the integral, we cannot proceed exactly as we did in the previous section. The inverse Fourier transform needs to be performed for all coordinates $z_{\phi_0}$ before the rotation.
\begin{align}
\gls{f} = &
- \frac{i \gls{km}}{2\pi \cdot a_0}
\int_0^{2 \pi} \!\!  d\gls{phi0}  \times \notag \\
& \times  D_{-\phi_0} \!\!
\left\lbrace
 \text{FFT}^{-1}_{\mathrm{1D}}
 \left\lbrace
  \left| k_\mathrm{Dx} \right |  
  \widehat{U}_{\mathrm{B},\phi_0}(k_\mathrm{Dx})
  \exp( -i \gls{km} M \gls{lD} )
  \exp \! \left[i \gls{km}(M - 1) z_{\phi_0}\right]
 \right\rbrace 
\right\rbrace
\end{align}
In practice, the measured scattered field \gls{us}, which is approximated by the Born approximation \gls{ub}, is already background corrected. We insert the incident plane wave $u_0(\gls{lD}) = a_0 \exp(i\gls{km}\gls{lD})$ and thus, obtain an additional factor $\exp(+i\gls{km}\gls{lD})$ in the Fourier filter.
\begin{align*}
\frac{\widehat{U}_{\mathrm{B}}(k_\mathrm{Dx})}{a_0} = \frac{\text{FFT}_{\mathrm{1D}} \left\lbrace u_\mathrm{B}(x) \right\rbrace }{a_0} = \text{FFT}_{\mathrm{1D}} \left\lbrace \frac{u_\mathrm{B}(x)}{u_0(\gls{lD})} \cdot \exp(i\gls{km}\gls{lD}) \right\rbrace
= \frac{\widehat{U}_{\mathrm{B}}(k_\mathrm{Dx}) \cdot \exp(i\gls{km}\gls{lD})}{u_0(\gls{lD})}
\end{align*}
Using this substitution and the discretization of the integral over \gls{phi0} according to the previous section, we obtain the backpropagation algorithm in two dimensions
\begin{align}
\gls{f} = 
- \frac{i \gls{km}}{2\pi}
\sum_{j=1}^{N} \! \Delta \gls{phi0}   D_{-\phi_j} \!\!
\left\lbrace
 \text{FFT}^{-1}_{\mathrm{1D}}
 \left\lbrace
  \left| k_\mathrm{Dx} \right |  
  \frac{
  		\widehat{U}_{\mathrm{B},\phi_j}(k_\mathrm{Dx})
		}
		{u_0(\gls{lD})
		}
  \exp \! \left[i \gls{km}(M - 1) \cdot (z_{\phi_j}-\gls{lD}) \right]
 \right\rbrace 
\right\rbrace
\label{eq:alg.backprop2d}
\end{align}
with the discrete angular distance $\Delta \gls{phi0} = 2\pi/N$ and the discrete angles ${\phi_j = j\cdot \Delta \gls{phi0}}$ ($j = 1,2,\dots,N$).
When comparing this equation with the backprojection algorithm from \hyref{equation}{eq:alg.backproj}, we can see one major difference besides the different filter. The inverse Fourier transform needs to be calculated separately for every distance $z_{\phi_j}$. 
In practice, one first calculates the one-dimensional signal $\widehat{U}_{\mathrm{B},\phi_j}(k_\mathrm{Dx}) / u_0(\gls{lD}) \cdot \exp[ -i \gls{km} (M-1) \gls{lD} ]$ and then expands the signal by one dimension through the application of the second filter $\exp \! \left[i \gls{km}(M - 1) \, z_{\phi_j}\right]$. The inverse Fourier transform is then computed along the axis with constant $z_{\phi_j}$. The resulting two-dimensional data are rotated by $\phi_j$ and added to the reconstruction plane. The name \textit{backpropagation} comes from an interpretation of the {$z_{\phi_j}$-exponential}, which looks like a propagation in $z_{\phi_j}$-direction.

\subsection{Three-dimensional Backpropagation}
With the previously described two-dimensional algorithm, it is now straight-forward to derive the three-dimensional analog. We start again from the integral representation.
\begin{align}
\gls{f} &= \frac{-i\gls{km}}{(2\pi)^{2}a_\mathrm0} 
\int_0^{2\pi} \!\! d\gls{phi0}  						  
\int_{-\gls{km}}^{\gls{km}} \!\! dk_\mathrm{Dx}				  
\int_{-\gls{km}}^{\gls{km}} \!\! dk_\mathrm{Dy} \,              
\left| k_\mathrm{Dx} \right| \times \notag \\
&\times \widehat{U}_{\mathrm{B},\gls{phi0}}(k_\mathrm{Dx},k_\mathrm{Dy})
\exp(-i\gls{km}M \gls{lD}) 
\exp[i(k_\mathrm{Dx} \, \vc{t_\perp} + \gls{km}(M - 1) \, \gls{s0})\vc{r}] \tag{\ref{eq:3DBackprop}}
\end{align}
\begin{align}
\vc{t_\perp} &= \left(\cos\gls{phi0}, \,
                \frac{k_\mathrm{Dy}}{k_\mathrm{Dx}}, \,
                \sin\gls{phi0} \right)^\top \notag \\
\vc{s_0} &= \left(-\sin\gls{phi0}, \,0 ,\, \cos\gls{phi0} \right)^\top \notag
\end{align}
In this section, we only consider the rotation of the sample along the $y$-axis. This introduces artifacts, as briefly discussed in \hyref{section}{sec:FourierDiffrRelevance1axis}. The rotation through $-\gls{phi0}$ is described by the rotation operator $D_{-\phi_0}$ that transforms \vc{r_{\phi_0}} to \vc{r}.
\begin{align}
\vc{r} &= (x, y, z)^\top  \\
\vc{r_{\phi_0}} &= (x_{\phi_0}, y_{\phi_0}, z_{\phi_0})^\top  \\
x_{\phi_0} &= x\cos \gls{phi0} + z \sin \gls{phi0} \\
y_{\phi_0} &= y \\
z_{\phi_0} &= -x\sin \gls{phi0} + z \cos \gls{phi0} \\
\vc{t_\perp} \cdot \vc{r} &=  x_{\phi_0} + \frac{k_\mathrm{Dy}}{k_\mathrm{Dx}} \cdot y_{\phi_0} \\
\vc{s_0} \cdot \vc{r} &= z_{\phi_0}
\end{align}

\begin{align}
\gls{f} &= \frac{-i\gls{km}}{(2\pi)^{2}a_\mathrm0} 
\int_0^{2\pi} \!\! d\gls{phi0}  	\					  
D_{-\phi_0} \!\!
\left\lbrace
 \int_{-\gls{km}}^{\gls{km}} \!\! dk_\mathrm{Dx}				  
 \int_{-\gls{km}}^{\gls{km}} \!\! dk_\mathrm{Dy} \,              
 \left| k_\mathrm{Dx} \right| \times \right. \notag \\
 & \left. \times \widehat{U}_{\mathrm{B},\gls{phi0}}(k_\mathrm{Dx},k_\mathrm{Dy})
 \exp(-i\gls{km}M \gls{lD}) 
 \exp[i(k_\mathrm{Dx} x_{\phi_0} + k_\mathrm{Dy} y_{\phi_0} + \gls{km}(M - 1) z_{\phi_0})]
\right\rbrace 
\end{align}
Here, we identify the two-dimensional Fourier transform over $k_\mathrm{Dx}$ and $k_\mathrm{Dy}$ and simplify the equation.
\begin{align}
\gls{f} &= \frac{-i\gls{km}}{2\pi \cdot a_\mathrm0} 
\int_0^{2\pi} \!\! d\gls{phi0} \ \times	\notag \\
& \times D_{-\phi_0} \!\!
\left\lbrace
 \text{FFT}^{-1}_\mathrm{2D}
 \left\lbrace
  \widehat{U}_{\mathrm{B},\gls{phi0}}(k_\mathrm{Dx},k_\mathrm{Dy})
  \left| k_\mathrm{Dx} \right|
  \exp(-i\gls{km}M \gls{lD}) 
  \exp[i \gls{km}(M - 1) z_{\phi_0}]
 \right\rbrace
\right\rbrace 
\end{align}
Following the derivation in two dimensions, we identify the incident plane wave $u_0(\gls{lD})$ and discretize the \gls{phi0}-integral
\begin{align}
\gls{f} &= \frac{-i\gls{km}}{2\pi} 
\sum_{j=1}^{N} \! \Delta \gls{phi0}  \times \notag \\
& \times D_{-\phi_j} \!\!
\left\lbrace
 \text{FFT}^{-1}_\mathrm{2D}
 \left\lbrace
  \frac{\widehat{U}_{\mathrm{B},\phi_j}(k_\mathrm{Dx},k_\mathrm{Dy})}
  		{u_0(\gls{lD})}
  \left| k_\mathrm{Dx} \right|
  \exp[i \gls{km}(M - 1) \cdot (z_{\phi_j} - \gls{lD})]
 \right\rbrace
\right\rbrace
\label{eq:alg.backprop3d}
\end{align}
with the discrete angular distance $\Delta \gls{phi0} = 2\pi/N$ and the discrete angles ${\phi_j = j\cdot \Delta \gls{phi0}}$ (${j = 1,\dots,N}$). \hyref{Equation}{eq:alg.backprop3d} and \hyref{equation}{eq:alg.backprop2d} only differ in the number of dimensions. In the three-dimensional case, the acquired data are two-dimensional projections through the sample. The fast Fourier transform is used again to speed up the computation. The subsequent rotations are performed for three-dimensional volumes which are then added to the reconstruction volume in the sum over \gls{phi0}.

%% file: conclusion.tex
\section*{Conclusions}
This manuscript reviewed the theory of diffraction tomography. We discussed the wave equation with the Born and Rytov approximations. Next, we compared the mathematical formalism of two-dimensional diffraction tomography with the formalism of classical tomography and pointed out the major differences between them. Subsequently, we derived the full three-dimensional equivalent of diffraction tomography and discussed the implications for biological samples. Finally, we gave details on how to implement the backpropagation algorithm in software. This script is intended as a literature review, establishing the mathematical foundations of diffraction tomography with a coherent notation. The reconstruction methods described are applicable not only to optical diffraction tomography but also to ultrasonic diffraction tomography whose underlying principle is the scalar wave equation. In general, diffraction tomography opens new ways for three-dimensional tissue imaging and with growing computational power at hand, will become a feasible and flexible tool in modern imaging.

%% file: appendix_notation_in_literature.tex
\section{Notations in Literature}
\subsection{Two-dimensional Diffraction Tomography}
\label{sec:2DDiffractLiterature}
Devaney and Slaney derived the Fourier diffraction theorem with a slightly different notation \cite{Devaney1982336, Slaney1984}. This section illustrates the main differences in notation and translates between them. Besides different variable names and a differing definition of the object function/scattering potential \gls{f}, the Devaney and Slaney use the non-unitary angular frequency Fourier transform, whereas this script uses the unitary angular frequency Fourier transform. The Fourier diffraction theorem from \hyref{equation}{eq:2DFDTF} reads
\begin{align}
\widehat{F}(\gls{km}(\vc{s} - \gls{s0})) = &
- \sqrt{\frac{2}{\pi}} 
\frac{i \gls{km}}{a_0} M
\widehat{U}_{\mathrm{B},\phi_0}(k_\mathrm{Dx})
\exp \! \left(-i \gls{km} M \gls{lD}\right) \tag{\ref{eq:2DFDTF}}.
\end{align}
Using the notation of Slaney, we start with the Fourier transform of the scattered wave $\psi_\mathrm{s}$ (\cite{Slaney1984}, eq. (62)\footnote{The factor $\frac{1}{2\pi}$ is missing in \cite{Slaney1984}: The derivation with the convolution theorem does not seem to be correct. Equation (47) in \cite{Slaney1984} should read 
$\widetilde{\psi}_s(\vec{\Lambda}) = \widetilde{G}(\vec{\Lambda}) * \widetilde{H}(\vec{\Lambda})$, where $\widetilde{H}(\vec{\Lambda})$ is the Fourier transform of  $H(\vec{r}) = \{ O(\vec{r}) \psi_0(\vec{r}) \} $
. The additional factor of $2\pi$ seems to originate from a delta function in equation (50). The full three-dimensional Fourier transform approach is derived in \hyref{section}{sec:FDThFourTransFAppr}.}).
\begin{align*}
\int \psi_\mathrm{s}(x,y=l) 
e^{-i\alpha x} dx = 
\frac{1}{2\pi} & 
\widetilde{\Gamma}(\alpha,l) \\
& \widetilde{\Gamma}(\alpha',l) = 
 \int e^{-i \alpha' x} \int \Gamma_1 (\alpha,l) e^{i \alpha x} d \alpha dx
\end{align*}
Note that Slaney uses the non-unitary angular frequency Fourier transform.
With the identity of the delta function
\begin{align*}
\delta (\alpha - \alpha') = \frac{1}{2\pi} \int e^{i (\alpha-\alpha') x} dx
\end{align*}
we may write 
\begin{align*}
 \widetilde{\Gamma}(\alpha',l) = 
 \int 2\pi \delta (\alpha - \alpha') 
   \Gamma_1 (\alpha',l) d\alpha  = 2\pi \Gamma_1 (\alpha',l).
\end{align*}
Using equation 60 from \cite{Slaney1984}, we get\footnote{The exclamation mark (!) in this formula points at the sign of equation (60) and (61) in \cite{Slaney1984}, which should be switched. This follows from the path integration described in the paper.}
\begin{align*}
\int \psi_\mathrm{s}(x,y=l) e^{-i\alpha x} dx \overset{!}{=}
 \frac{i \widetilde{O}(\alpha, \sqrt{k_\mathrm{o}^2 - \alpha^2} - k_\mathrm{o} )}
 	  {2 \sqrt{k_\mathrm{o}^2 - \alpha^2}}
 e^{i\sqrt{k_\mathrm{o}^2 - \alpha^2}l}. \tag{\cite{Slaney1984}}
\end{align*}
Note that in Devaney et al. \cite{Devaney1982336}, equation (20) is equivalent to the equation above, when replacing $ \widetilde{O}$ with $-k^2 \widetilde{O}$ and performing a one-dimensional Fourier transform. Now we perform the following translations to get from  Slaney's notation to the notation in this script.
\begin{align*}
\text{detector position:}& \\
(x,y=l) &\rightarrow (x_\mathrm{D},z_\mathrm{D}=\gls{lD}) \\
l &\rightarrow \gls{lD} \\
\text{correlating factor:}& \\
\sqrt{k_\mathrm{o}^2 - \alpha^2} &\rightarrow \gls{km}M  \\
\text{Fourier transformed object:}& \\
 \widetilde{O}(\alpha, \sqrt{k_\mathrm{o}^2 - \alpha^2} - k_\mathrm{o} ) &\rightarrow  2\pi \widehat{F}\left(\gls{km}(\vc{s} - \gls{s0})\right) \\
& \text{(object function defined at \cite{Slaney1984}, eq. (10))} \\
   \int \psi_\mathrm{s}(x,y=l) e^{-i\alpha x} dx &\rightarrow
  \frac{(2\pi)^{1/2}}{a_\mathrm{0}} \widehat{U}_{\mathrm{B},\phi_0}(k_\mathrm{Dx})
\end{align*}
By replacing these things, we get the equivalent to \hyref{equation}{eq:2DFDTF}.
\begin{align*}
\frac{(2\pi)^{1/2}}{a_\mathrm{0}}\widehat{U}_{\mathrm{B},\phi_0}(k_\mathrm{Dx}) =
\frac{2 \pi i \widehat{F}(\gls{km}(\vc{s} - \gls{s0}))}{
	  2 \gls{km}M}
\exp(i\gls{km}M\gls{lD})
\end{align*}
The backpropagation formula (\hyref{equation}{eq:2DBackprop}) can be translated via the exact same way. Using the previously mentioned publication by Devaney \cite{Devaney1982336} and combining the equations (35), (36) and (29a) therein, the backpropagation formula reads
\begin{align*}
O_\text{LD}(\vc{r}) = \frac{1}{(2\pi)^2} 
			   \int_{-\pi}^\pi \! \! d\phi_0 \int_{-k}^k \!\! d\kappa \,
			   |\kappa|
			   &\underbrace{\widetilde{\Gamma}_{\phi_0}(\kappa, \omega)}_{\Downarrow}
			   e^{i (\kappa \xi + ( \gamma -k)(\eta-l_\mathrm{o})}. \\
\int_{-\infty}^\infty \!\! d \xi \,
        & \underbrace{\Gamma_{\phi_0}(\xi', \omega)}_{\Downarrow}
		e^{-i \kappa \xi'} \\
&\frac{ie^{-ikl_\mathrm{o}}}{kU_\mathrm{o}(\omega)}
		U_{\phi_0}^{\mathrm{(s)}} (\xi', \eta=l_\mathrm{o}, \omega) \\
\end{align*}	
By inserting all of these translations we find the Fourier transform 
$\widehat{U}_{\phi_0}^{\mathrm{(s)}}(\kappa, \omega)$ of the scattered wave
$U_{\phi_0}^{\mathrm{(s)}} (\xi, \eta=l_\mathrm{o}, \omega)$.
\begin{align*}
O_\text{LD}(\vc{r}) = \frac{1}{(2\pi)^2} 
			   \int_{-\pi}^\pi \! \! d\phi_0 \int_{-k}^k \!\! d\kappa \,
			   |\kappa| \frac{i}{k U_\mathrm{o}(\omega)}
			   &\underbrace{
			   		\int \!\! d\xi' \,
			   		U_{\phi_0}^{\mathrm{(s)}} (\xi', \eta=l_\mathrm{o}, \omega)
			   		e^{-i \kappa \xi'}
			   }
				e^{i(\kappa\xi + (\gamma - k)\eta)}
				e^{-i\gamma l_\mathrm{o}}			   
			    \\
			   & 		\hphantom{\int \!\! d\xi' \, U_{\phi_0}^{\mathrm{(s)}}}
			   \widehat{U}_{\phi_0}^{\mathrm{(s)}}(\kappa, \omega)
\end{align*}
We now perform the following translations.
\begin{align*}
\text{variables:}& \\
l_\mathrm{o} &\rightarrow \gls{lD} \\
k &\rightarrow \gls{km} \\
\gamma &\rightarrow \gls{km} M \\
\text{factors (notation):}& \\
O_\text{LD}(\vc{r}) &\rightarrow - \frac{1}{\gls{km}^2} f(\vc{r})  \\
& \text{ (sign and factor $\gls{km}^2$ defined at \cite{Devaney1982336}, eq. (12))} \\
\widehat{U}_{\phi_0}^{\mathrm{(s)}}(\kappa, \omega) &\rightarrow \sqrt{2\pi} \, \widehat{U}_{\mathrm{B},\phi_0}(k_\mathrm{Dx}) \\
& \text{ (non-unitary to unitary Fourier transform)} \\
\text{amplitude:}& \\
U_\mathrm{o}(\omega) &\rightarrow  a_\mathrm{0} \\
\text{vectors:}& \\
\xi &\rightarrow \vc{t_\perp} \\
\eta &\rightarrow \gls{s0}
\end{align*}
We arrive at the previously derived backpropagation formula. 
\begin{align}
\gls{f} = 
- \frac{i \gls{km}}{a_0 (2\pi)^{3/2}}
\int \!\! dk_\mathrm{Dx} \int_0^{2 \pi} \!\!  d\gls{phi0} \,
\left| k_\mathrm{Dx} \right |  
\widehat{U}_{\mathrm{B},\phi_0}(k_\mathrm{Dx})
\exp( -i \gls{km} M \gls{lD} ) \times \notag \\
\times \exp \! \left[i (k_\mathrm{Dx} \, \vc{t_\perp} + \gls{km}(M - 1) \, \gls{s0})\vc{r}\right] 
\tag{\ref{eq:2DBackprop}}
\end{align}

\clearpage
\subsection{Three-dimensional Diffraction Tomography}
In \hyref{section}{sec:2DDiffractLiterature} we looked at two notations in literature that describe the two-dimensional case. The three-dimensional case is briefly discussed by e.g. Devaney~\cite{Devaney1982336}. We show the equivalence of the notations by performing a Fourier transform of \hyref{equation}{eq:FfandUf}.
\begin{align*}
u_{B,\phi_0}(\vc{r_D}) = 
\iint \!\! \frac{dk_\mathrm{Dx}  dk_\mathrm{Dy}}{2\pi} \, \exp(i\vc{k_Dr_D})
\frac{i \pi a_\mathrm{0} }{(2\pi)^{1/2}}	
\frac{											
 \exp(i\gls{km}M \gls{lD})
 }{												
 \gls{km}M
 }    						
\widehat{F}\left(\vc{k_D} -\gls{km}\gls{s0}\right)
\end{align*}
The following translations have to be performed to translate between the notation of Devaney and ours.
\begin{align*}
\text{detector position:}& \\
(x_\mathrm{D},y_\mathrm{D},z_\mathrm{D}=\gls{lD}) &\rightarrow (x,z,y=l_\mathrm{o}) \\
\gls{lD} &\rightarrow l_\mathrm{o} \\
u_{B,\phi_0}(\vc{r_D}) &\rightarrow  U_\mathrm{B}^\mathrm{(s)}(x,y=l_\mathrm{o},z)\\
\text{correlating factors:}& \\
a_\mathrm{0} &\rightarrow U_\mathrm{o}(\omega) \\
\gls{km}M &\rightarrow \gamma \\
\text{Fourier transformed object:}& \\
\widehat{F}\left(\vc{k_D} -\gls{km}\gls{s0}\right) &\rightarrow -\frac{k^2}{(2\pi)^{3/2}} \widetilde{O}(K_\mathrm{x}, \gamma-k, K_\mathrm{z}) \\
& \text{ (sign and factor $k^2$ defined at \cite{Devaney1982336}, eq. (12))} \\
\end{align*}
We then arrive at equation (46) in \cite{Devaney1982336}.
\begin{align*}
U_\mathrm{B}^\mathrm{(s)}(x,y=l_\mathrm{o},z) &=
 -i \frac{k^2}{8\pi^2} U_\mathrm{o}(\omega) 
 \iint \!\! \frac{dK_\mathrm{x} dK_\mathrm{z}}{\gamma}
 e^{i\gamma l_\mathrm{o}} \widetilde{O}(K_\mathrm{x}, \gamma-k, K_\mathrm{z}) e^{i[K_\mathrm{x}x + K_\mathrm{z}z}] \tag{\cite{Devaney1982336}, equation (46)}
\end{align*}

%% file: appendix_changes.tex
\section{Changes to this Document}
\subsection*{[v2] March 2016}
\begin{itemize}
\item[-] errata:
	\begin{itemize}
	\item[-] glossary entry \gls{pc}: removed amplitude $a$ as it is part of the complex phase
    \item[-] replaced $i$ with $j$ in text after equations 6.2, 6.13, and 6.23
	\end{itemize}
\item[-] updated references:
	\begin{itemize}
    \item[-] implementation of the backpropagation algorithm (updated \cite{Mueller15})
    \item[-] clarified missing-apple-core problem (added \cite{Vertu09})
    \item[-] removed journal abbreviations
    \end{itemize}
\item[-] replaced 'object function' by 'scattering potential' for more consistency with literature on scattering theory
\item[-] deepened discussion on validity of Rytov approximation
\item[-] changed equations for backpropagation algorithms 6.13 and 6.23 to include background correction
\end{itemize}

\subsection*{[v3] October 2016}
\begin{itemize}
\item[-] errata:
	\begin{itemize}
	\item[-] glossary entry \gls{pc}: added logarithm of amplitude $a$ to complex phase
	\item[-] added missing negation in the discussion of the Rytov approximation
	\item[-] corrected equation 4.1: The inhomogeneous Helmhotz equation contains the scattering potential and not the delta distribution.
	\item[-] replaced equation 4.2: In 2D, the Green's function is not $\propto \exp(r)/r$; replaced with the definition of the Green's function.
	\end{itemize}
\end{itemize}